\pdfoutput=1
\documentclass[12pt,a4paper,twoside]{report}
\DeclareMathAlphabet{\mathpzc}{OT1}{pzc}{m}{it} %
\usepackage[english, science, titlepage]{ku-frontpage}
\usepackage[utf8]{inputenc}
\usepackage{hyperref}
\usepackage{physics}
\usepackage{graphicx}
\usepackage{subcaption}
\usepackage{wrapfig}
\usepackage{tikz}
\usepackage{float}
\usepackage{fancyhdr}
\usepackage{caption}
\usepackage{cite}
\usetikzlibrary{positioning}
\assignment{Master's Thesis}
\author{Gustav Uhre Jakobsen}
\title{General Relativity from Quantum Field Theory}
\subtitle{} 
\date{Handed in: July 1, 2020}
\advisor{Academic advisor: Poul Henrik Damgaard}
\coadvisor{Academic co-advisor: N. Emil J. Bjerrum-Bohr}
\frontpageimage{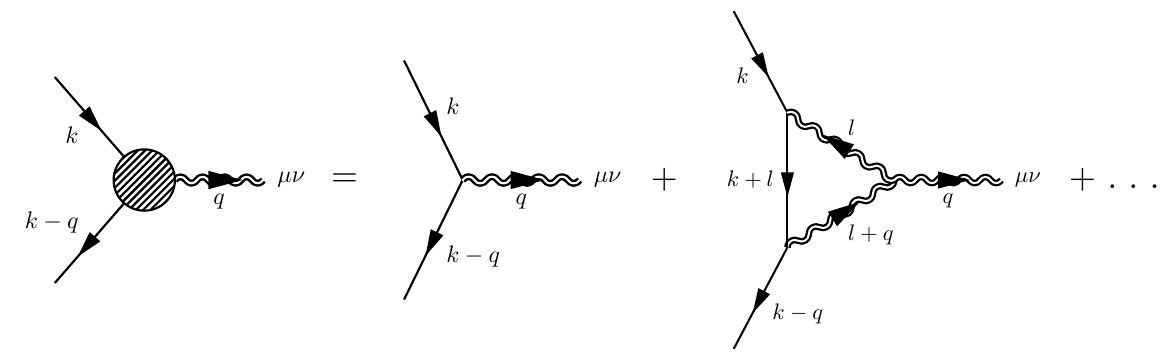}
\heja{
  Niels Bohr International Academy, The Niels Bohr Institute,  University of Copenhagen
}
\hejb{
  Blegdamsvej 17, DK-2100 Copenhagen, Denmark
}

\newcommand*{\Xmnb}[2]{\frac{\ x^\bot_{#1} x^\bot_{#2}}{\ x_\bot^2}}
\newcommand\dmt{n}
\newcommand{\fcR}{\frac{\mu}{R^{\dmt}}}
\newcommand{\fcr}{\frac{\mu}{r^{\dmt}}}
\newcommand{\fcrs}{\Big(\fcr\Big)^2}

\newcommand*{\Xrhob}[1]{\frac{\ x^\bot_{#1}}{r}}

\newcommand\prl{\parallel}
\newcommand\sqg{\sqrt{-g}}
\newcommand\mn{{\mu\nu}}

\newcommand\pe{{\phi\epsilon}}
\newcommand\ab{{\alpha\beta}}

\newcommand\gd{{\gamma\delta}}

\newcommand\rs{{\rho\sigma}}
\newcommand\sr{{\sigma\rho}}

\newcommand*{\absvec}[1]{\abs{\vct{#1}}}
\newcommand*{\vct}[1]{\boldsymbol{#1}}
\newcommand\Delc{{\Delta_\tecl}}
\newcommand\Delgf{{\Delta_\tegf}}

\newcommand*{\deltb}[1]{\eta^{\prl}_{#1}}
\newcommand*{\delrb}[1]{\eta^{\bot}_{#1}}
\newcommand*{\etat}[1]{\eta^{\parallel}_{#1}}
\newcommand*{\etar}[1]{\eta^{\bot}_{#1}}
\newcommand{\STM}{Schwarzschild-Tangherlini}

\newcommand\maP{\mathcal{P}}
\newcommand\maPi{{\mathcal{P}^{-1}}}
\newcommand\maM{\mathcal{M}}
\newcommand\teol{\text{1-loop}}
\newcommand\tenl{\text{non-linear}}
\newcommand\hka{\mathpzc{h}}
\newcommand\thka{\tilde {\mathpzc{h}}}
\newcommand\thmn{{\tilde h}}
\newcommand\tHmn{{\tilde H}}
\newcommand\tGmn{{\tilde G}}
\newcommand\ttau{{\tilde \tau}}
\newcommand\tfmn{{\tilde f}}

\newcommand\Gpz{\mathpzc{G}}
\newcommand\Hpz{\mathpzc{H}}
\newcommand*{\Ghn}[2]{\Gpz_{h^{#1}}^{#2}}
\newcommand*{\tGhn}[2]{\tilde\Gpz_{h^{#1}}^{#2}}
\newcommand*{\Hhn}[2]{\Hpz_{h^{#1}}^{#2}}
\newcommand*{\tHhn}[2]{\tilde\Hpz_{h^{#1}}^{#2}}

\newcommand\ddx{d^D x}
\newcommand*{\oov}[1]{\frac{1}{#1}}
\newcommand\stdim{D}
\newcommand\dDx{d^\stdim x}
\newcommand*\dDp[1]{\frac{d^D#1}{(2\pi)^D}}
\newcommand*\ddp[1]{\frac{d^{D-1}#1}{(2\pi)^{D-1}}}

\newcommand*{\labelt}[1]{\label{eqn:#1}}
\newcommand*{\flabel}[1]{}
\newcommand*{\ffig}[1]{}
\newcommand*{\labelx}[1]{\label{#1}}
\newcommand*{\eqreft}[1]{\eqref{eqn:#1}}
\newcommand\dDo{de Donder}
\newcommand\cGno{G}
\newcommand*{\cGn}[1]{G^{#1}}
\newcommand\chno{h}
\newcommand*{\chn}[1]{h^{#1}}
\newcommand\maJ{\mathcal{J}}
\newcommand*{\qbmnu}[2]{\frac{q_\bot^{#1} q_\bot^{#2}}{q_\bot^2}}
\newcommand*{\qbmnd}[2]{\frac{q^\bot_{#1} q^\bot_{#2}}{q_\bot^2}}
\newcommand*{\hhatu}[1]{\hat h^{#1}}

\newcommand\tecc{\text{(c)}}
\newcommand\CEEq{\cgE\ equation}
\newcommand\teze{{(0)}}
\newcommand\teon{{(1)}}
\newcommand\tetw{{(2)}}
\newcommand\tetr{{(3)}}
\newcommand\tenn{{(n)}}
\newcommand\teve{\text{vertex}}
\newcommand\tenp{\text{n-loop}}
\newcommand\EHA{{\text{Einstein-Hilbert}}}
\newcommand\tegh{{\text{(gh)}}}
\newcommand\tecl{{\text{(c)}}}
\newcommand\tegf{{\text{(gf)}}}
\newcommand\teEH{{\text{EH}}}

\newcommand\teee{{\text{tree}}}
\newcommand\nzi{{{n=0..\infty}}}
\newcommand\noi{{{n=1..\infty}}}
\newcommand\gff{gauge-fixing function}
\newcommand\ellipsis{.\ .\ .}
\newcommand\IVe{V}
\newcommand\cgE{covariant gauge}

\newcommand\Mbo{{M^\bot}}
\newcommand\blankpage{
    \newpage
    \null
    \thispagestyle{empty}%
    \addtocounter{page}{-1}%
    \newpage
}
\begin{document}
\pagenumbering{roman}
\maketitle
\blankpage
\chapter*{Abstract}
The quantum field theoretic description of general relativity is a modern approach to gravity where gravitational force is carried by spin-2 gravitons.
In the classical limit of this theory, general relativity as described by the Einstein field equations is obtained.
This limit, where classical general relativity is derived from quantum field theory is the topic of this thesis.
\par
The \STM\ metric, which describes the gravitational field of an inertial point particle in arbitrary space-time dimensions, $D$, is analyzed.
The metric is related to the exact three-point vertex function of a massive scalar interacting with a graviton to all orders in $G_N$, and the one-loop contribution to this amplitude is computed from which the leading-order self-interaction contribution to the metric is derived.
\par
To understand the gauge-dependence of the metric, covariant gauge (i.e. $R_\xi$-gauge) is used which introduces the arbitrary parameter, $\xi$, and the gauge-fixing function $G_\sigma$.
In the classical limit, the gauge-fixing function turns out to be the coordinate condition, $G_\sigma=0$.
As gauge-fixing function a novel family of gauges, which depends on an arbitrary parameter $\alpha$ and includes both harmonic and \dDo\ gauge, is used.
\par
Feynman rules for the graviton field are derived and important results are the graviton propagator in covariant \dDo-gauge and a general formula for the n-graviton vertex in terms of the Einstein tensor.
The Feynman rules are used both in deriving the \STM\ metric from amplitudes and in the computation of the one-loop correction to the metric.
\par
The one-loop correction to the metric is independent of the covariant gauge parameter, $\xi$, and satisfies the gauge condition $G_\sigma=0$ where $G_\sigma$ is the family of gauges depending on $\alpha$.
It is compared to the literature for particular values of $\alpha$ and $D$ and, also, to an independent derivation using only methods from classical general relativity.
In space-time $D=5$ a logarithm appears in position space and this phenomena is analyzed in terms of redundant gauge freedom.


\tableofcontents

\chapter{Introduction}
\labelx{sec:Introduction}
\pagenumbering{arabic}
The modern approach to general relativity using the framework of quantum field theory has successfully been used to derive several exciting results in the theory of gravity.
These include the description of classical binary systems as well as predictions on quantum corrections to gravitational processes at low energies.
For example, the analytic tools from quantum field theory are essential to improve the accuracy of theoretic predictions on gravitational waves.
Here, modern methods of quantum field theory such as on-shell scattering amplitudes and generalized unitarity are helpful.
The predictions on quantum corrections to gravity, being at the moment insignificant to experimental observations, are of great interest for investigations into the quantum theory of gravity.
\par
Well-known physicists have worked on the quantum field theoretic description of gravity.
In Refs.~\cite{Feynman:1996kb,Feynman:1963ax}, Feynman introduced spin-2 particles to describe gravitational interactions.
From these and similar investigations it was realized that it is necessary to include ``ghosts'' in the Feynman rules, which is now understood from the Faddeev-Popov gauge-fixing procedure~\cite{DeWitt:1967ub,DeWitt:1967uc,Faddeev:1973zb,Faddeev:1967fc}.
In Ref.~\cite{tHooft:1974toh}, 't Hooft and Veltman analyzed the one-loop divergencies of quantum gravity.
At one-loop order, they found that ``pure gravity'', that is gravity described by the \EHA\ action, is renormalizable.
Later investigations~\cite{Goroff:1985th} have shown that this fails at two-loop order so that ``pure gravity'' is non-renormalizable.
This is not surprising considering the negative mass dimension of the gravitational constant.
In Weinberg's comprehensive treatment of general relativity, Ref.~\cite{Weinberg:1972kfs}, the field theoretic point of view was consistently used in favor of the geometric description of gravity.
Additional significant contributions include Refs.~\cite{Schwinger:1963re,Schwinger:1968rh,Weinberg:1964ew,Iwasaki:1971vb}.
\par
An important contribution to the quantum field theoretic description of gravity was that of Donoghue who used an effective field theoretic approach which made it possible to deal rigorously with the non-renormalizability of quantum gravity and to compute quantum corrections to gravity at low energies \cite{Donoghue:1995cz,Donoghue:1994dn,Donoghue:1993eb,BjerrumBohr:2002kt,Bjerrum-Bohr:cand}.
In one line of work, quantum corrections to the metric was computed~\cite{Donoghue:2001qc,BjerrumBohr:2002ks}.
\par
Today, methods from quantum field theory have been used to derive a number of results in classical general relativity~\cite{Duff:1973zz,Buonanno:1998gg,Goldberger:2004jt,Holstein:2008sx,Neill:2013wsa,Akhoury:2013yua,Vaidya:2014kza,Damour:2017zjx,Damour:2019lcq,Guevara:2017csg,Cachazo:2017jef,KoemansCollado:2019ggb,Bjerrum-Bohr:2018xdl,Cheung:2018wkq,Bern:2019nnu,Bern:2019crd,Chung:2019yfs,Cheung:2020gyp,Kosower:2018adc,Bjerrum-Bohr:2019kec,Cristofoli:2019neg,Jakobsen:2020ksu,Cristofoli:2020uzm,Cristofoli:2020hnk}.
These include the post-Minkowskian expansion with which the binary system of two gravitationally interacting objects have been analyzed.
Here, scattering amplitudes are used to derive Hamiltonians, potentials and scattering angles of the two classical, interacting objects.
\par
Instead of the four space-time dimensions that we usually associate with the physical world, the field theoretic description of gravity is easily generalized to arbitrary space-time dimensions, $D$.
This is also relevant when the dimensional regularization scheme is used.
Studies of gravity in arbitrary dimensions can be found in Refs.~\cite{Cristofoli:2020uzm,Emparan:2008eg,Collado:2018isu,Cristofoli:2020hnk}.
In arbitrary dimensions, the metric of an inertial point particle is generalized from the Schwarzschild metric to the \STM\ metric.
The derivation of this metric from the quantum field theoretic approach to gravity is a main topic of this thesis.
\par
In quantum field theory, gauge theories of spin-1 particles are a great success and play a major role in the Standard Model of particle physics.
Examples are Yang-Mills theory, quantum chromodynamics and quantum electrodynamics.
In the quantum field theoretic approach to general relativity it is found that gravitational interactions can be described by spin-2 particles.
These spin-2 gravitons carry the gravitational force in analogy to e.g. the strong force carried by spin-1 gluons from quantum chromodynamics.
These similarities invite for a new interpretation of the general covariance of general relativity where the graviton field is treated as a gauge particle in analogy to the gluon field.
An exciting discovery is the double copy nature of quantum gravity in terms of Yang-Mills theory~\cite{Bern:2019crd,Bern:2010ue,Cheung:2016say}.
For example, tree amplitudes of gravitons are expanded in terms of products of tree amplitudes of gluons and gravity is said to be the square of Yang-Mills.
\par
In general, the gauge theory of spin-2 gravitons is much more complicated than that of spin-1 particles.
In $D=4$, the negative mass dimension of the gravitational constant is in contrast to the dimensionless coupling constant of Yang-Mills theory.
Also, the Einstein field equations are highly non-linear which usually means that the Feynman rules of quantum gravity include vertices with an arbitrary number of gravitons.
This makes the \STM\ metric very different from the Coulomb potential of quantum electrodynamics.
While the Coulomb potential is exact at tree level, the \STM\ metric gets corrections from diagrams with an arbitrary number of loops.
This is an exciting fact about the classical limit of quantum field theory that, in general, diagrams with any number of loops contribute.
\par
In this thesis, the Feynman diagram expansion of the \STM\ metric is analyzed from the quantum field theoretic approach to general relativity.
Such metric expansions were first analyzed by Duff in Ref.~\cite{Duff:1973zz} and have later been studied in Refs.~\cite{Bjerrum-Bohr:2018xdl,BjerrumBohr:2002ks,Galusha:cand,Cristofoli:2020hnk,Chung:2019yfs,Donoghue:2001qc,Jakobsen:2020ksu}.
In Ref.~\cite{Donoghue:2001qc,BjerrumBohr:2002ks} quantum corrections to the classical metric was considered while the reduction of triangle n-loop integrals in the classical limit was discussed in Refs.~\cite{Bjerrum-Bohr:2018xdl,Galusha:cand}.
\par
An interesting aspect of the \STM\ metric derived from amplitudes is its gauge/coordinate dependence.
In what coordinates is the \STM\ metric when computed from amplitudes?
What kind of coordinate conditions is it possible to use in the gauge-fixed action?
We will use the quantum field theoretic point of view and instead of coordinates, we will speak of gauge choices and conditions.
The question is then how the quantum gauge-fixing procedure is related to gauge conditions in the classical limit.
To this aim it is advantageous to use the path integral method and Feynman rules where gauge dependence is explicit.
\par
As a whole, general relativity from quantum field theory is an exciting field which incorporates gravity into the modern description of quantum particles.
It has found importance in providing analytic tools for the analysis of gravitational waves and is of interest for the development of quantum gravity.
To this end, a better understanding of the relation of the \STM\ metric to this framework as well as its gauge dependence is desirable.
%
%
\par
The thesis is structured as follows.
In Ch.~\ref{sec:Background} we briefly go through relevant aspects of classical general relativity and quantum field theory.
This serves as a starting point from which the two theories can be combined and also, importantly, we state several conventions on signs and notation in this chapter.
\par
The next chapters are concerned with the quantum field theoretic description of gravity.
In Ch.~\ref{sec:GaugeDependence} we analyze the gauge theory of spin-2 gravitons and how it is related to general covariance.
Then, in Ch.~\ref{sec:ExpansionsAround} we expand the objects of general relativity in the graviton field and in the gravitational constant around flat space-time.
In Ch.~\ref{sec:FeynmanRules} we use these results to derive Feynman rules for the graviton.
\par
We choose to work in covariant gauge (i.e. $R_\xi$-gauge) with the arbitrary covariant gauge parameter, $\xi$, and as gauge-fixing function, $G_\sigma$, we choose a family of gauge conditions depending on an arbitrary parameter $\alpha$.
The expansions in Ch.~\ref{sec:ExpansionsAround} are used for Feynman rules and also later in Ch.~\ref{sec:STM} to analyze the classical equations of motion.
We have not found a detailed treatment of quantum gravity in covariant gauge in the literature and the Feynman rules in Ch.~\ref{sec:FeynmanRules} present new results such as the graviton propagator in covariant \dDo-gauge.
Another interesting result of Ch.~\ref{sec:ExpansionsAround} and~\ref{sec:FeynmanRules} is an expression of the n-graviton vertex in terms of the Einstein tensor and an analogous tensor $H^\mn$.
\par
The later chapters are concerned with the derivation of the \STM\ metric from amplitudes.
In Ch.~\ref{sec:STM}, general formulas relating the all-order metric from general relativity to amplitudes from quantum field theory are derived.
This includes a detailed analysis of the gauge-fixed classical equations of motion which are solved in a perturbative expansion analogous to a Feynman diagram expansion.
In particular, the \STM\ metric is derived from the exact three-point function of a massive scalar interacting with a graviton.
Then, in Ch.~\ref{sec:PerturbativeExpansion2} we specialize to the one-loop diagram contribution to the metric.
This is the first correction to the metric due to graviton self-interaction and depends on the three-graviton vertex.
\par
Both chapters~\ref{sec:STM} and~\ref{sec:PerturbativeExpansion2} present exciting results.
The Feynman rules for the general n-graviton vertex derived in earlier chapters clearly shows how the Feynman diagram expansion of the three-point vertex function is related to the perturbative solution of the classical equations of motion.
The one-loop contribution to the metric is a very general result depending on the arbitrary dimension $D$ and the gauge parameter $\alpha$ from the gauge-fixing function $G_\sigma$.
This result is compared with the literature.
An interesting phenomenon occurs in the metric in $D=5$ where a logarithmic dependence on the radial coordinate appears.
Also, we present a classical derivation of the metric contribution which confirms the results from the amplitude computation.
\par
Finally, in Ch.~\ref{sec:Conclusion} we summarize the results of this thesis and suggest directions for further research.

\chapter{Background}
\labelx{sec:Background}
We briefly discuss the classical theory of general relativity and the path integral approach to quantum field theory.
We show how the action of general relativity can be included in the path integral in a minimal way.
Finally we consider how to interpret the classical limit $\hbar\rightarrow0$ of quantum field theory.
\par
We work with the mostly negative metric and with units ($c=\hbar=1$).
We will use the comma-notation for partial derivatives.

\section{General Relativity}
\labelx{sec:GeneralRelativity}
In the traditional approach to general relativity, gravity is described by the metric tensor which measures the geometry of space-time.
Physical equations are required to be general covariant, that is, invariant under general coordinate transformations.
To this end, tensor fields are introduced which obey definite transformation laws.
In the modern field theoretic approach, it is recognized that general covariance can be thought of as the gauge symmetry of spin-2 gravitons.
In this section we will follow the traditional approach and later in Ch.~\ref{sec:GaugeDependence} we will discuss the description of general relativity in terms of spin-2 gravitons.
\par
We have found the treatments of classical general relativity of Weinberg~\cite{Weinberg:1972kfs} and Dirac~\cite{Dirac:GR} useful.
In particular, we use the same conventions as Dirac~\cite{Dirac:GR} which also coincide with the conventions of Refs.~\cite{Donoghue:1995cz,Bjerrum-Bohr:cand,BjerrumBohr:2002ks}.
These conventions include the mostly negative metric and how the Ricci tensor is defined in terms of the curvature tensor.
We work all the time in arbitrary space-time dimensions $D$.
Investigations in gravity in arbitrary space-time dimensions can be found in Refs.~\cite{Cristofoli:2020uzm,Emparan:2008eg,Collado:2018isu,Cristofoli:2020hnk}.
\par
The strength of gravitational interactions are described by the gravitational constant, $G_N$.
As mentioned in the introduction, in $D=4$ the gravitational constant is dimensionful in contrast to e.g. Yukawa and Yang-Mills couplings.
In general the mass dimension of $G_N$ is:
\begin{align}
  [G_N] = [\text{mass}]^{-(D-2)}
  \ .
  \labelt{nn21}
\end{align}
In $D=2$ and $D=3$ general relativity behaves very differently from $D\geq4$.
In this thesis we will only consider $D\geq4$ and by arbitrary dimensions $D$ we always assume $D\geq4$.
Instead of working with $G_N$, we will often use $\kappa$ where:
\begin{equation}
  \kappa^2 = 32 \pi G_N
  \ .
  \labelt{nn22}
\end{equation}
This definition of $\kappa$ agrees with Refs.~\cite{Donoghue:1995cz,Bjerrum-Bohr:cand,BjerrumBohr:2002ks}
\par
The curvature tensor is:
\begin{equation}
  R_{\mn\rs}
  =
  \Gamma_{\mn\sigma,\rho} - \Gamma_{\mn\rho,\sigma}
  -\Gamma_{\beta\mu\rho}\Gamma^\beta_{\nu\sigma} + \Gamma_{\beta\mu\sigma}\Gamma^\beta_{\nu\rho}
  \ .
  \labelt{nn23}
\end{equation}
We use the comma-notation to denote partial derivatives so that e.g. $g_{\mn,\rho}=\partial_\rho g_\mn$.
\par
The Christoffel symbols are
\begin{equation}
  \Gamma_{\rho\mn}
  =
  \frac{1}{2}
  \big(
  g_{\rho\mu,\nu} + g_{\rho\nu,\mu} - g_{\mn,\rho}
  \big)
  \ ,
  \labelt{nn24}
\end{equation}
and $\Gamma^\rho_{\mn}=g^\rs \Gamma_{\sigma\mn}$.
\par
From the curvature tensor we get the Ricci tensor, $R_\mn$, and scalar, $R$:
\begin{align}
  &R_\mn = g^\rs R_{\rho\mn\sigma}
  \ ,
  \labelt{nn25}
  \\
  &R = g^\mn R_\mn
  \ .
  \labelt{nn26}
\end{align}
The Einstein-Hilbert action can be defined in terms of the Ricci scalar,
\begin{equation}
  S_{EH} = 
  \int \dDx
  \sqrt{-g}
  \ R
  \ ,
  \labelt{nn27}
\end{equation}
where $g$ is the determinant of $g_\mn$.
If we have a Lagrangian, $\mathcal{L}_\phi$, describing matter in special relativity, we can add a matter term to the Einstein-Hilbert action,
\begin{align}
  S_\phi = \int \dDx \sqrt{-g} \
  \mathcal{L}_\phi
  \ ,
  \labelt{nn28}
\end{align}
where now contractions in $\mathcal{L}_\phi$ should be made with $g_\mn$ and $g^\mn$.
In our case, matter will be described by massive scalar fields and hence:
\begin{align}
  \mathcal{L}_\phi
  =
  \frac{1}{2}
  \Big(
  g^\mn \phi_{,\mu} \phi_{,\nu}
  -m^2\phi^2
  \Big)
  \ .
\end{align}
The Einstein field equations can be derived from the variational principle $\delta S_\tecl=0$ where:
\begin{align}
  S_\tecl = \frac{2}{\kappa^2} S_{EH} + S_\phi
  \ .
  \labelt{nn29}
\end{align}
The subscript on $S_\tecl$ can be read as ``classical''.
\par
The Einstein field equations are
\begin{equation}
  G^\mn = -\frac{\kappa^2}{4} T^\mn
  \ ,
  \labelt{n210}
\end{equation}
where $G^\mn$ is the Einstein tensor
\begin{equation}
  G^\mn = R^\mn - \frac{1}{2} R\  g^\mn
  \ ,
  \labelt{n211}
\end{equation}
and $T^\mn$ is the energy-momentum tensor of matter.
\par
The Einstein tensor appears when we vary the Einstein-Hilbert action:
\begin{equation}
  \delta S_{EH}
  =
  -
  \int \dDx
  \sqrt{-g}
  \ G^\mn \delta g_\mn
  \ .
  \labelt{ein1}
\end{equation}
It obeys
\begin{equation}
  D_\mu G^\mn = 0
  \labelt{n213}
\end{equation}
where $D_\mu$ is the covariant derivative.
\par
Similarly $T^\mn$ appears when we vary the matter action
\begin{equation}
  \delta S_\phi = -\frac{1}{2}
  \int \dDx \sqg
  \ T^\mn \delta g_\mn
  \ ,
  \labelt{n214}
\end{equation}
and it, too, obeys $D_\mu T^\mn=0$.
\par
In $D=4$ the well-known Schwarzschild metric describes the gravitational field of an inertial, non-spinning point particle.
In arbitrary space-time dimensions it is generalized to the \STM\ metric, which in spherical coordinates is given by:
\begin{align}
  d\tau^2 =
  (1-\frac{\mu}{r^\dmt}) dt^2
  -\frac{1}{1-\frac{\mu}{r^\dmt}} dr^2
  - r^2 d\Omega^2_{D-2}
  \labelt{n215}
\end{align}
Here, $n=D-3$ and $\mu$ is the \STM\ parameter:
\begin{align}
  \mu = \frac{16 \pi G_N M}{(D-2) \Omega_{D-2}}
  \labelt{mud1}
\end{align}
In this equation $\Omega_{d-1}$ is the surface area of a sphere in d-dimensional space and is given explicitly by:
\begin{align}
  \Omega_{d} = \frac{2\sqrt{\pi}^{d+1}}{\Gamma((d+1)/2)}
  \ .
  \labelt{n217}
\end{align}
The \STM\ metric in Eq.~\eqreft{n215} can be found in e.g. \cite{Emparan:2008eg}.
It solves the Einstein field equations in vacuum with a point particle singularity at centrum.
\section{Quantum Field Theory}
\labelx{sec:QuantumField}
Quantum field theory unites special relativity and quantum mechanics.
The Standard Model of particle physics is formulated in its framework.
Here, non-abelian gauge theories play a dominant role.
The treatments of quantum field theory of Srednicki~\cite{Srednicki:2007qs} and of Schwartz~\cite{Schwartz:2013pla} have been useful.
In particular, we have used the same approach to path integral quantization as in Srednicki.
\par
Naively, we will treat the gravitational field, $g_\mn$, as any other quantum field and use the \EHA\ action minimally coupled to a scalar field in the path integral.
As already mentioned, this leads to a non-renormalizable quantum theory.
However, in the classical, low-energy limit we can ignore the divergent terms.
\par
Later, we will expand the action in the ``graviton field'' $h_\mn=g_\mn-\eta_\mn$.
This field, $h_\mn$, will describe spin-2 gravitons.
The general covariance of the gravitational action translates into gauge theory of the spin-2 particles.
As with spin-1 particles, it is necessary to ``fix a gauge'' in the path integral.
We will briefly indicate how this is done with the Faddeev-Popov method following Srednicki~\cite{Srednicki:2007qs}.
\par
Our action is that of Eq.~\eqreft{nn29} which is:
\begin{equation}
  S_c = \int \dDx \sqrt{-g}  
  \bigg(
  \frac{2R}{\kappa^2} \
  +\ 
  \frac{1}{2}(g^{\mu \nu}
  \partial_\mu \phi \partial_\nu \phi - m^2 \phi^2)
  \bigg)
  \ .
  \labelt{act1}
\end{equation}
The partition function is then given by:
\begin{equation}
  Z_\omega = \int \mathcal{D} g_{\mu \nu} \ \mathcal{D} \phi \ \det(\frac{\delta G}{\delta \epsilon}) \
  \delta ( G_\sigma - \omega_\sigma ) \   e^{iS_c}
  \ .
  \labelt{n219}
\end{equation}
Here we have used the Faddeev-Popov gauge-fixing procedure.
The $\delta$-function picks out a specific gauge-choice so that the path integral extends only over independent field configurations.
The gauge-fixing function $G_\sigma$ breaks the general covariance of the Einstein Hilbert action.
The Jacobian determinant is expanded by introducing ghosts.
The arbitrary field $\omega_\sigma$ on which $Z_\omega$ depends is integrated out with a Gaussian weight function.
\par
This gives the final expression for the partition function $Z$ in covariant gauge with gauge-fixing function $G_\sigma$:
\begin{align}
  Z &= \int
  \mathcal{D} \omega_\sigma  \ Z_\omega \ 
  \exp(i\int \dDx \ \frac{1}{\kappa^2 \xi} \eta^{\sigma \rho} \omega_\sigma \omega_\rho)
  \labelt{n220}
  \\
  &= \int \mathcal{D} g_{\mu \nu} \  \mathcal{D} \phi \  \mathcal{D}c \mathcal{D}\bar{c} \
  e^{iS_\tecl + i\frac{2}{\kappa^2}S_\tegf + iS_\tegh}
  \ .
  \nonumber
\end{align}
There are three types of fields.
The gravitational field $g_\mn$, the scalar field $\phi$ and the ghost fields $c$ and $\bar c$.
\par
In addition to $S_\tecl$ two new terms appear in the action.
The gauge-fixing term, $S_\tegf$, which comes from the Gaussian weight function and the ghost term $S_\tegh$ which comes from the expansion of the Jacobian determinant.
In this work, we will mainly be concerned with the classical limit of quantum gravity and in this limit we can ignore the ghosts.
The gauge-fixing term, $S_\tegf$, is given by:
\begin{equation}
  S_{gf} =
  \frac{1}{2 \xi}
  \int \dDx\ 
  \eta^{\sigma \rho} G_\sigma G_\rho
  \ ,
  \labelt{gau1}
\end{equation}
Neglecting the ghost term, we get a final expression for the gauge-fixed action:
\begin{align}
  S &= S_\tecl + \frac{2}{\kappa^2} S_\tegf
  \labelt{act2}
  \\
  &= \frac{2}{\kappa^2}
  \big(
  S_\teEH + S_\tegf
  \big)
  +S_\phi
  \ .
  \nonumber
\end{align}
This action is relevant for the classical limit of quantum gravity.
We get the Feynman rules after we expand the action in $h_\mn$.
It is however more convenient to use the normalization $\hka_\mn=\frac{1}{\kappa}h_\mn$ which will clearly show, at which order in $G_N$, the different vertices contribute.
This will be relevant in Ch.~\ref{sec:FeynmanRules} when the Feynman rules are derived.
\par
Let us briefly discuss how the Feynman rules are derived from a given action.
The quadratic terms of the action give rise to propagators.
There will be a second derivative operator between the two fields, which when inverted gives the corresponding propagator.
This is most easily done in momentum space, which we will also use in this work.
\par
The vertex rules come from terms with more than two fields.
In our case we will have two types of vertices, namely a scalar meeting an arbitrary number of gravitons or an arbitrary number of gravitons meeting in a single self-interaction vertex.
We will denote them $\phi^2 h^n$ and $h^n$.
For the \STM\ metric computation, the self-interaction vertices are essential.
\par
To derive the vertex rule in momentum space we transform our fields:
\begin{subequations}
  \label{n223}
  \begin{align}
    &\tilde \phi(q)
    =
    \int \dDx e^{iqx} \phi
    \ ,
    \labelt{n223a}
    \\
    &\tilde \hka(q)
    =
    \int \dDx e^{iqx} \hka
    \ .
    \labelt{n223b}
  \end{align}
\end{subequations}
When we change to momentum space, the space integration removes the exponential factors and introduces a $\delta$-function which describes conservation of momentum.
A term in the action which results in a $\phi^2 h^2$ vertex would look like:
\begin{align}
  &\kappa^2
  \int \dDx\ 
  \eta^{\mu\alpha}\eta^{\nu\beta} \hka_\mn  \hka_\ab  \eta^\rs  \phi_{,\rho}  \phi_{,\sigma}
  =
  -\kappa^2
  \int
  \dDp{l_\teon}\dDp{l_\tetw}\dDp{p}\dDp{q}
  \labelt{exa2}
  \\
  &\qquad\qquad\qquad\qquad\times
  (2\pi)^D \delta^D\big(l_\teon+l_\tetw+p+q\big)
  \eta^{\mu\alpha}\eta^{\nu\beta}
  \tilde \hka_\mn(l_1) \tilde \hka_\ab(l_2)
  \tilde \phi(p) \tilde \phi(q)
  \eta^\rs p_\sigma q_\rho
  \nonumber{}
\end{align}
To get the corresponding vertex rule we would take the integrand without the $\delta$-function.
We would then remove the fields and multiply by $i$ and $2!2!$.
The two factorial factors come from the two factors of $h$ and the two factors of $\phi$.
For a $h^5$ vertex it would be $5!$ and for a $\phi^2h^3$ it would be $2!3!$.
When we remove the fields, we have to make sure that the leftover vertex is symmetric in the fields.
In the case of Eq.~\eqreft{exa2} this is easily done.
Thus, from the example in Eq.~\eqreft{exa2} we get the following vertex rule, $V^{\mn\ \ab}$:
\begin{align}
  V^{\mn\ \ab}
  =
  -i4\kappa^2 \frac{1}{2}\big(\eta^{\mu\alpha}\eta^{\nu\beta} + \eta^{\mu\beta}\eta^{\nu\alpha}\big) p_\sigma q^\sigma
  \labelt{n225}
\end{align}
Here, $\mn$ and $\ab$ are graviton indices of the two graviton lines and $p$ and $q$ are the incoming (or outgoing) momenta of the two scalar lines.

\subsection{The Classical Limit}
\labelx{Sec:TheClassical}
The classical limit of quantum field theory can be defined as the limit where $\hbar\rightarrow0$.
The conventional interpretation of this limit is that even slight variations from classical field configurations makes the integrand oscillate greatly so that only configurations where $\delta S=0$ are significant.
The equations $\delta S=0$ are the classical equations of motion.
The classical limit of quantum field theory is analyzed in detail in Ref.~\cite{Kosower:2018adc} and for the particular case of gravity in Refs.~\cite{Bjerrum-Bohr:2018xdl,Cheung:2020gyp,Bjerrum-Bohr:2019kec}.
\par
In general, Feynman diagrams with any number of loops still contribute in the classical limit.
This is so due to several cancellations of $\hbar$.
An important distinction in classical physics is that between waves and particles which in the quantum theory is blurred.
Thus, in quantum theory, wavenumbers, $l^\mu$, and particle momenta, $p^\mu$, are related through the formula:
\begin{align}
  p^\mu = \hbar l^\mu
  \ .
  \labelt{me31}
\end{align}
If we wish to introduce $\hbar$ explicitly as a dimensionful quantity in a Feynman diagram we should consider, for each (quantum) momentum, whether it belongs to a particle or wave in the classical limit.
If we assume that, by default, any momentum variable in the Feynman diagram represents a particle momentum, then a change to wavenumber introduces a factor $\hbar$.
The importance of this distinction is that in the classical limit a particle has a finite particle momentum while a wave has a finite wavenumber.
For a given quantum particle the question is whether $p^\mu$ or $l^\mu$ in Eq.~\eqreft{me31} should stay finite.
\par
We will work with units $\hbar=1$ which seems contradictory to the limit $\hbar\rightarrow0$.
In this setup, the classical limit rather means that the action $S$ is much larger than $\hbar$, that is much larger than unity.
After we have extracted results in the classical limit, they should be independent of $\hbar$ and we can forget about the initial choice $\hbar=1$.
If we assume that momenta are by default ``particle momenta'', we see from Eq.~\eqreft{me31} that the momentum of classical particles should stay finite while the momentum of classical waves should be sent to zero.
Thus, if $q^\mu$ is the momentum of a wave-like particle in a Feynman diagram, then $q^\mu=\hbar l^\mu$ where $l^\mu$ is the wavenumber which should stay finite.
Then $q^\mu$ must be small in comparison.
\par
In our case, we have two types of particles, massive scalars and gravitons.
The massive scalars are interpreted as point particles and we let their momenta stay finite.
On the other hand, gravitons behave like waves in the classical theory and their momenta are sent to zero.
These conclusions apply both to external momenta and internal loop momenta.
In our case the classical limit is to some extend equivalent to a long-range limit.
Thus, from the theory of Fourier transforms we learn that small wavenumbers in momentum space are related to long distances in position space.
A rigorous discussion of these conclusions is found in Ref.~\cite{Kosower:2018adc}.
\par
In the classical limit massive scalars can be interpreted as point particles.
Also, we can think of these as an effective description of larger extended objects.
In Ref.~\cite{Goldberger:2004jt} finite size effects are described by including non-minimal terms in the action.
\par
In our work we will explore the \STM\ metric which is generated by a single particle.
This should be compared to the Coulomb potential from electrodynamics.
While the Coulomb potential is exact at tree-level the \STM\ metric gets corrections from diagrams with an arbitrary number of loops.
This is a consequence of the fact that gravitons interact with themselves and makes the investigation of the \STM\ metric interesting.

\chapter{Gauge Theory of Gravity in Quantum Field Theory}
\labelx{sec:GaugeDependence}

Gauge symmetry is an important concept in modern physics.
Successful gauge theories are Yang Mills theory, quantum chromodynamics and electroweak theory.
It is an exciting idea, that general covariance in general relativity can be considered as gauge symmetry of spin-2 particles.
A related insight of modern physics is that of describing gravity as a double copy of Yang-Mills theory~\cite{Bern:2019crd,Bern:2010ue,Cheung:2016say}.
\par
In Sec.~\ref{sec:GeneralRelativity2}, we discuss the description of gravity in terms of spin-2 gravitons on a flat background.
Then, in Sec.~\ref{sec:GaugeTransformations} we derive the transformation properties of the graviton field under gauge transformations.
Finally, in Sec.~\ref{sec:GaugeFixing} we discuss the freedom in choosing a parameterization for the graviton field as well as gauge-fixing functions.
\section{General Relativity in Lorentz Covariant Quantum Field Theory}
\labelx{sec:GeneralRelativity2}
In our approach, the action of general relativity is expanded around flat space-time and the dynamical field is chosen to be this perturbation, $h_\mn$.
This makes the graviton field, $h_\mn$, look very similar to any other quantum field on a flat space-time.
In the long-range limit of gravity, where space-time is approximately flat, it is possible to change completely to a quantum field theoretic point of view.
We then describe spin-2 particles and their gauge symmetry on a flat space-time instead of general covariant objects on an arbitrary space-time.
From this point of view, general coordinate transformations are instead interpreted as gauge transformations and choosing a coordinate system is translated to ``fixing the gauge''.
This is e.g. the point of view developed in~\cite{Feynman:1996kb}.
\par
In general relativity, general covariant tensors are often taken as fundamental quantities.
Instead, we will mostly focus on Lorentz covariant tensors.
Thus, when speaking of tensors, we will mostly mean Lorentz covariant tensors.
Such tensors are defined with respect to the nearly flat space-time far away from matter.
The indices on Lorentz covariant tensors are raised and lowered with the flat space metric.
Thus, the indices on most tensors in this work are raised and lowered with $\eta^\mn$ and $\eta_\mn$.
\par
We will often need to change between position and momentum space.
For this purpose the conventions of relativistic quantum field theory will be used.
Generally, we will use a tilde to denote objects in momentum space.
For example, we have the graviton field in position space $h_\mn(x)$ and in momentum space $\tilde h_\mn (q)$ and their relation is:
\begin{subequations}
  \label{nn31}
  \begin{align}
    &h_\mn(x)
    =
    \int \dDp{q}
    \ 
    e^{-iqx}
    \ 
    \tilde h_\mn(q)
    \labelt{nn31a}
    \\
    &\tilde h_\mn(q)
    =
    \int \dDx
    \ 
    e^{iqx}
    \ 
    h_\mn(x)
    \labelt{nn31b}
  \end{align}
\end{subequations}
These transformations are then meant, when we speak of momentum or position space, or Fourier transforms.
Often, we will leave out the explicit dependence on $x$ or $q$ and simply write $h_\mn$ or $\tilde h_\mn$ when we feel no confusion can arise.
\par
Later, in Ch.~\ref{sec:STM} and Ch.~\ref{sec:PerturbativeExpansion2} we will focus in great detail on the \STM\ metric.
There, we will keep all our equations Lorentz covariant which is natural when working with amplitudes and Feynman diagrams.
Here, we will develop a notation which makes these expressions clear from a physical, and mathematical, point of view.
\par
The \STM\ metric describes the gravitational field of an inertial point particle.
This particle will have momentum which we take as $k^\mu$ and a mass $m = \sqrt{k^2}$.
We then introduce the following projection operators:
\begin{subequations}
  \label{nn32}
  \begin{align}
    &\etat{\mn}=\frac{k_\mu k_\nu}{m^2}
    \ ,
    \labelt{nn32a}
    \\
    &\etar{\mn}=\eta_\mn-\frac{k_\mu k_\nu}{m^2}
    \ .
    \labelt{nn32b}
  \end{align}
\end{subequations}
These are projection operators, i.e. their sum is $\eta_\mn$, they are both idempotent, and they are orthogonal to each other.
The tensor $\etat{\mn}$ projects tensors parallel to $k^\mu$ and $\etar{\mn}$ orthogonal to $k^\mu$.
Alternatively, $\etat{\mn}$ projects tensors along the worldline of the particle, while $\etar{\mn}$ projects tensors into the orthogonal space.
We will then use similar symbols on tensors to denote projection.
For example, if we have a vector $q^\mu$ we will write:
\begin{subequations}
  \labelt{nn33}
  \begin{align}
    &q^\mu_\prl = {\eta_\prl}^\mu_\nu q^\nu
    \ ,
    \labelt{nn33a}
    \\
    &q^\mu_\bot = {\eta_\bot}^\mu_\nu q^\nu
    \ .
    \labelt{nn33b}
  \end{align}
\end{subequations}
Note that $q^\mu_\prl$ is time-like so that $q^\mu_\prl q_\mu^\prl>0$ and in some sense, it is 1-dimensional.
Similarly $q^\mu_\bot$ is space-like so that $q^\mu_\bot q_\mu^\bot<0$ and in some sense, it is $(D-1)$-dimensional.
We will define the short-hand:
\begin{align}
  q_\prl = \sqrt{q^\mu_\prl q_\mu^\prl}
  \ .
  \labelt{nn34}
\end{align}
Then $k_\prl=m$ and $k_\mu q^\mu = m q_\prl$.
\par
It is particularly simple to work in the inertial frame of $k^\mu$.
Here $\eta_\prl^\mn$ and $\eta_\bot^\mn$ are both diagonal and represent the time and space components of $\eta^\mn$ respectively.
Then $\eta_\prl^{00}=1$ and zero otherwise and $\eta_\bot^{ij} = -\delta^{ij}$ and zero otherwise.
Also, $q_\bot^\mu$ is the space components of $q^\mu$ and $q_\prl=q^0$.
Hence, in this special frame, $q_\bot^2 = - \absvec{q}^2$.

\section{Gauge Transformations of the Graviton Field}
\labelx{sec:GaugeTransformations}
In this section we analyze gauge transformations of $h_\mn$.
In general, all fields transform under gravitational gauge transformations.
This is clear from the traditional point of view, since a gravitational gauge transformation, that is a coordinate transformation, changes the functional dependence of every field on the coordinates.
On the other hand, it is sometimes argued, that this does not constitute a real change, so that a scalar field is left unchanged under a coordinate transformation.
\par
In this work, the gauge transformations of the fields will not play an essential role.
However, since we have not found detailed accounts of the gauge transformations of $h_\mn$ in other sources, we will include this brief discussion on its gauge transformations.
On the other hand, the formulas for gauge transformations in linearized gravity are well known.
We will see, that these formulas are consequences of the more general equations in this section.
We will use part of these results later in Sec.~\ref{sec:AppearanceOf}.
\par
We derive the formulas from the point of view, that gravitons are spin-2 fields defined on a flat space-time background.
As mentioned, this point of view only works at long distances, when the space-time is approximately flat.
\par
We start with formulas from the traditional framework of general relativity and rewrite them in terms of $h_\mn$.
Under a general transformation of coordinates to $\hat x^\mu(x)$ we have:
\begin{align}
  \hat g_\mn (\hat x)
  = g_\ab (x)
  \frac{\partial x^\alpha}{\partial \hat x^\mu}
  \frac{\partial x^\beta}{\partial \hat x^\nu}
  \ .
  \labelt{gtr1}
\end{align}
Choose a coordinate transformation according to:
\begin{align}
  x^\mu = \hat x^\mu + \hat \epsilon^\mu(\hat x)
  \labelt{rel1}
  \ .
\end{align}
Let us also write the transformation in another symmetric way:
\begin{align}
  \hat x^\mu = x^\mu + \epsilon^\mu(x)
  \labelt{rel3}
  \ .
\end{align}
Thus $\epsilon^\mu(x)$ relates the new coordinates to the old ones as a function of the old coordinates.
In contrast $\hat \epsilon^\mu(\hat x)$ relates the new coordinates to the old ones as a function of the new coordinates.
Often it would be most natural to use $\epsilon^\mu(x)$ to define the new coordinates.
The analysis is, however, simpler when we use $\hat \epsilon(\hat x)$.
They obey the equation
\begin{align}
  \epsilon(x)
  + \hat\epsilon(\hat x) = 0
  \ ,
\end{align}
from which they can be related to each other by using Eqs.~\eqreft{rel1} and~\eqreft{rel3} and using Taylor expansions.
\par
Let us insert the coordinate transformation Eq.~\eqreft{rel1} into the formula for transforming $g_\mn$ Eq.~\eqreft{gtr1}.
First, we compute the partial derivative of $x^\alpha$:
\begin{align}
  \frac{\partial x^\alpha}{\partial \hat x^\mu}
  = \delta^\alpha_\mu
  +\frac{\partial \hat \epsilon^\alpha(\hat x)}{\partial \hat x^\mu}
\end{align}
We insert this into Eq.~\eqreft{gtr1}:
\begin{align}
  \hat g_\mn (\hat x)
  &= g_\mn (x)
  + g_{\alpha\nu} (x)
  \frac{\hat\partial \hat \epsilon^\alpha(\hat x)}{\hat\partial \hat x^\mu}
  + g_{\mu\beta} (x)
  \frac{\hat\partial \hat \epsilon^\beta(\hat x)}{\hat\partial \hat x^\nu}
  + g_{\ab} (x)
  \frac{\hat\partial \hat \epsilon^\alpha(\hat x)}{\hat\partial \hat x^\mu}
  \frac{\hat\partial \hat \epsilon^\beta(\hat x)}{\hat\partial \hat x^\nu}
  \labelt{gmn1}
  \\
  &=
  g_\ab(x)
  \Big(
  \delta^\alpha_\mu \delta^\beta_\nu
  +
  \frac{\hat\partial \hat \epsilon^\alpha(\hat x)}{\hat\partial \hat x^\mu}
  \delta^\beta_\nu
  +
  \delta^\alpha_\mu
  \frac{\hat\partial \hat \epsilon^\beta(\hat x)}{\hat\partial \hat x^\nu}
  +
  \frac{\hat\partial \hat \epsilon^\alpha(\hat x)}{\hat\partial \hat x^\mu}
  \frac{\hat\partial \hat \epsilon^\beta(\hat x)}{\hat\partial \hat x^\nu}
  \Big)
  \nonumber{}
\end{align}
The fields are evaluated at different coordinates, either $x$ or $\hat x$.
We want all fields to be evaluated at the same coordinate and we choose $\hat x$.
The only occurrence of $x$ is in $g_\mn(x)$.
We use Eq.~\eqref{eqn:rel1} to relate $x$ to $\hat x$ in $g_\ab(x)$ and make a Taylor expansion:
\begin{subequations}
  \label{n312}
  \begin{align}
    g_\mn(x)
    &= g_\mn(\hat x + \hat \epsilon(\hat x))
    \labelt{n312a}
    \\
    &=
    \sum_{\nzi}\oov{n!}
    \hat \epsilon^\sigma(\hat x)
    \ .\ .\ .\ \hat\partial_\sigma\ .\ .\ .\  g_\mn(\hat x)
    \labelt{n312b}
    \\
    &=
    g_\mn(\hat x)
    +\hat \epsilon^\sigma(\hat x) \hat\partial_\sigma g_\mn(\hat x)
    +\oov{2}\hat \epsilon^\sigma(\hat x) \hat \epsilon^\rho(\hat x)
    \hat\partial_\sigma \hat\partial_\rho
    g_\mn(\hat x)
    +\ .\ .\ .
    \labelt{n312c}
    \\
    &=
    \sum_{n=0..\infty}
    \oov{n!}
    \Big(
    \hat \epsilon^\sigma(\hat x) \hat\partial_\sigma^{(g)}
    \Big)^n
    g_\mn(\hat x)
    \ .
    \labelt{n312d}
  \end{align}
\end{subequations}
This is a complicated formula, and it is not easily written without developing some notation.
In Eq.~\eqreft{n312c} the expansion is written out explicitly and in Eq.~\eqreft{n312d} the superscript on $\partial^{(g)}_\sigma$ means that the partial derivative only hits $g_\mn$ and ignores any $\hat \epsilon$.
Now that we can express both sides of equation~\eqref{eqn:gmn1} in terms of the same coordinate we ignore the dependence on coordinates:
\begin{align}
  \hat g_\mn = 
  \Big(
  \delta^\alpha_\mu \delta^\beta_\nu
  +
  \partial_\mu \hat\epsilon^\alpha
  \delta^\beta_\nu
  +
  \delta^\alpha_\mu
  \partial_\nu \hat\epsilon^\beta
  +
  \partial_\mu \hat\epsilon^\alpha
  \partial_\nu \hat\epsilon^\beta
  \Big)
  \sum_{n=0..\infty}
  \oov{n!}
  \Big(
  \hat \epsilon^\sigma \partial_\sigma^{(g)}
  \Big)^n
  g_\ab
  \labelt{gmn2}
\end{align}
We can insert the definitions of $g_\mn$ in terms of $h_\mn$ to arrive at:
\begin{align}
  \hspace*{-0.3cm}
  \hat h_\mn =
  \partial_\mu \hat \epsilon_\nu
  +
  \partial_\nu \hat \epsilon_\mu
  +
  \partial_\mu \hat \epsilon_\alpha
  \partial_\nu \hat \epsilon^\alpha
  +
  \Big(
  \delta^\alpha_\mu \delta^\beta_\nu
  +
  \partial_\mu \hat\epsilon^\alpha
  \delta^\beta_\nu
  +
  \delta^\alpha_\mu
  \partial_\nu \hat\epsilon^\beta
  +
  \partial_\mu \hat\epsilon^\alpha
  \partial_\nu \hat\epsilon^\beta
  \Big)
  \sum_{n=0..\infty}
  \oov{n!}
  \Big(
  \hat \epsilon^\sigma \partial_\sigma^{(g)}
  \Big)^n
  h_\ab
  \labelt{gmn3}
\end{align}
This is the transformation law of the graviton field under a transformation with gauge parameter $\epsilon^\mu$.
For example we can get the well known linear transformations of linearized gravity if we assume $\hat \epsilon$ and $h_\mn$ to be small of the same order:
\begin{align}
  \hat h_\mn \approx 
  h_\mn
  +
  \partial_\nu\hat \epsilon_\mu
  +
  \partial_\mu\hat \epsilon_\nu
  \ .
\end{align}
This equation is reminiscent of the gauge transformations of the vector potential in electrodynamics.

\section{Gauge-Fixing Functions and Coordinate Conditions}
\labelx{sec:GaugeFixing}     
In this work we use covariant gauge with an arbitrary covariant parameter $\xi$.
This results in the gauge-fixed action from Eq.~\eqreft{act2}:
\begin{align}
  S =
  \int \dDx \sqrt{-g}  
  \bigg(
  \frac{2R}{\kappa^2} \
  +\ 
  \mathcal{L}_\phi
  \bigg)
  +
  \int \dDx
  \frac{1}{\kappa^2 \xi} \eta^{\sigma \rho} G_\sigma G_\rho
  \ .
  \labelt{act3}
\end{align}
What is the classical limit of this action?
In Sec.~\ref{sec:GaugeFixed} we will analyze the classical equations of motion of this action in detail.
The result is that the classical limit of this action is general relativity described by the Einstein field equations together with the coordinate condition $G_\sigma=0$.
\par
In this section we will discuss possible choices of $G_\sigma$ and alternative parameterizations of the graviton field $h_\mn$.
Let us first mention some coordinate choices from the traditional approach to general relativity.
In the study of black holes spherical or cylindrical-type coordinates are often used.
However, these are not well suited for expansions around flat space-time.
Another well-known coordinate condition is harmonic gauge:
\begin{align}
  g^\mn \Gamma_\mn^\sigma = 0
  \labelt{har1}
\end{align}
The coordinates in this gauge are cartesian-like and well suited for expansions around flat space-time.
The linearized version of the harmonic gauge condition is familiar from linearized gravity:
\begin{align}
  \partial_\mu (h^\mu_\sigma - \frac{1}{2} \eta^\mu_\sigma h^\nu_\nu) = 0
  \labelt{ddo1}
\end{align}
However, it is rarely used as an exact coordinate-condition and we do not know of any exact metrics in the linear gauge of Eq.~\eqreft{ddo1}.
\par
The study of gravity from the quantum field theoretic point of view has initiated new investigations into the gauge theory of gravity.
For example, in Refs.~\cite{Cheung:2020gyp,Cheung:2016say} very general choices of gauge functions and parameterizations of the graviton field are studied.
Instead of
\begin{equation}
  g_\mn = \eta_\mn + h_\mn
  \ ,
  \labelt{par2}
\end{equation}
we can use non-linear parameterizations such as:
\begin{equation}
  g_\mn =
  e^{\pi_\mn}
  \ .
  \labelt{par3}
\end{equation}
In this equation the exponential function should be evaluated as though $\pi$ is a matrix and contractions should be made with $\eta_\mn$.
With this parameterization, the inverse metric is simply:
\begin{align}
  g^\mn = e^{-\pi^\mn}
  \ .
  \labelt{par4}
\end{align}
Other choices are:
\begin{align}
  \sqrt{-g} g_\mn = \eta_\mn + h'_\mn
  \labelt{par5}
  \ .
\end{align}
In Ref.~\cite{Cheung:2020gyp}, the most general parameterization to second order in $h_\mn$ is considered.
However, in our work we will only consider the simple parameterization of Eq.~\eqreft{par2}.
\par
As with the choice of parameterization, there is a large freedom in the choice of the gauge-fixing function.
It would be interesting to investigate this freedom from the perspective of traditional general relativity.
In our work we will use the following \gff:
\begin{equation}
  G_\sigma =
  (1-\alpha) \ \partial_\mu (h^\mu_\sigma - \frac{1}{2} \eta^\mu_\sigma h_\nu^\nu)
  + \alpha\  g^\mn \Gamma_{\sigma\mn}
  \ .
  \labelt{gau3}
\end{equation}
It is an interpolation between the two gauge choices of Eq.~\eqreft{har1} and Eq.~\eqreft{ddo1}, that is between harmonic and \dDo\ gauge.
Here, we use the same terminology as~\cite{Cheung:2020gyp}, that is harmonic gauge is $g^\mn \Gamma_\mn^\sigma=0$ and \dDo\ gauge is
\begin{align}
  \eta^\mn \partial_\mu
  g_{\sigma\nu}
  -\frac{1}{2}
  \partial_\sigma
  \eta^\mn
  g_\mn
  =0
  \ ,
  \labelt{vv42}
\end{align}
where we have written the condition in terms of $g_\mn$ instead of $h_\mn$ to stress that it is meant as an exact constraint rather than an approximate constraint in linearized gravity.
\par
Let us look into the details of the generalized \dDo-type gauge function Eq.~\eqreft{gau3}.
As mentioned, it combines harmonic and \dDo\ gauge so that when $\alpha=0$ we have \dDo\ gauge and when $\alpha=1$ we have harmonic gauge.
However, any choice of $\alpha$ is valid and corresponds to some gauge condition.
The dependence on $\alpha$ is chosen such that when $G_\sigma$ is expanded in $h_\mn$ the linear term is independent of $\alpha$ and the non-linear terms are scaled by $\alpha$.
In particular, this means that the graviton propagator will be independent of $\alpha$.
Because the gauge function agrees with \dDo\ gauge at linear order, we will speak of the gauge choice as being of \dDo-type.
\chapter{Expansions Around Flat Space-Time}
\labelx{sec:ExpansionsAround}
It will be convenient to develop notation and concepts to facilitate expansions of the objects from general relativity in the graviton field and in the gravitational constant $G_N$.
The expansions in the graviton field are used to derive Feynman rules for quantum gravity in Ch.~\ref{sec:FeynmanRules}.
The expansions in $G_N$ are used in the analysis of the classical equations of motion in Ch.~\ref{sec:STM}.
In Sec.~\ref{sec:ExpansionsIn} we will distinguish the two types of expansions, namely in $h_\mn$ and in $G_N$.
In Sec.~\ref{sec:ExpansionOf}, we will expand the Einstein tensor, $G^\mn$, and the action, $S$.
We will compute the expansion of $S$ explicitly to third order in $h_\mn$ from which we can derive the graviton propagator and three-graviton vertex in covariant \dDo-type gauge.
We will then relate the expansion of $S$ to that of $G^\mn$.

\par
In the following sections we will work with a multitude of tensors.
Two important ones are $I^\mn_\ab$ and $\maP^\mn_\ab$:
\begin{subequations}
  \label{eqn:imt1}
  \begin{align}
    &I^\mn_\ab = \frac{1}{2}
    \Big(
    \delta^\mu_\alpha\delta^\nu_\beta
    +
    \delta^\mu_\beta\delta^\nu_\alpha
    \Big)
    \ ,
    \labelt{imt2}
    \\
    &\maP^\mn_\ab = I^\mn_\ab - \frac{1}{2}\eta^\mn\eta_\ab
    \ .
    \labelt{imt3}
  \end{align}
\end{subequations}
These tensors, as well as most other tensors in this chapter, are considered as Lorentz covariant tensors, and thus indices are raised and lowered with the flat space metric.
The definitions in Eqs.~\eqreft{imt1} agree with those of Refs.~\cite{Donoghue:1994dn,Bjerrum-Bohr:cand,BjerrumBohr:2002kt}.

\section{Expansions in $h_\mn$ and $G_N$}
\labelx{sec:ExpansionsIn}
We can expand the objects of general relativity in two different, though slightly related, ways.
First, we can expand in the graviton field, $h_\mn = g_\mn - \eta_\mn$.
Second, we can expand in the gravitational constant $G_N$.
The expansions in $h_\mn$ are important for deriving the Feynman rules.
The expansions in $G_N$ are useful when we analyze the classical equations of motion.
Also, the expansions in $G_N$ can be related to the expansions in $h_\mn$.
\par
If we have an object from general relativity such as the Ricci scalar we can expand it in $h_\mn$,
\begin{subequations}
  \label{eqn:exp5}
  \begin{equation}
    R = \sum_{\noi} R_{\chn{n}}
    \ ,
    \labelt{exp5a}
  \end{equation}
  or in $G_N$,
  \begin{equation}
    R = \sum_{\noi} R_{\cGn{n}}
    \ .
    \labelt{exp5b}
  \end{equation}
\end{subequations}
We will use this kind of notation in this and later chapters.
Thus a subscript or superscript with $\chn{n}$ or $\cGn{n}$ denotes the $n$'th term in the expansion in $h_\mn$ or $G_N$ respectively.
\par
Let us start with two simple, but important, examples of expansions in $h_\mn$.
These are $g^\mn=(\eta_\mn+h_\mn)^{-1}$ and $\sqrt{-g}$.
In principle, this allows us to expand the action to any order in $h_\mn$.
\par
For $g^\mn$, we find:
\begin{align}
  g^\mn
  &=
  \eta^\mn -  h^\mn +  h^\mu_\rho h^{\rho\nu}
  -  h^{\mu}_{\rho}h^\rho_\sigma h^{\sigma\nu}
  + .\ .\ .
  \labelt{exp3}
\end{align}
This expansion should be compared with the geometric series:
\begin{equation}
  \frac{1}{1+x} = \sum_{\nzi} (-x)^n
  =
  1 - x + x^2 - x^3 + \ellipsis
  \labelt{exp4}
\end{equation}
Eq.~\eqreft{exp3} can be derived by introducing:
\begin{equation}
  \hat h^\mn = g^\mn - \eta^\mn
  \ .
  \labelt{nn45}
\end{equation}
Using the equation $g^\ab g_{\beta\gamma} = \delta^\alpha_\gamma$ we get an equation between $h_\mn$ and $\hat h_\mn$:
\begin{equation}
  \hat h_\mn = - h_\mn - \hat h^\sigma_\mu h_{\sigma\nu}
  \ .
  \labelt{nn46}
\end{equation}
This is solved inductively by inserting the equation into itself repeatedly.
\par
The expansion of $g^\mn$ in $h_\mn$ is then rather straightforward and follows the structure of the geometric series.
Using the notation introduced in Eqs.~\eqreft{exp5} we would e.g. write:
\begin{equation}
  (g^{\mn})_{\chn{3}} =   -  h^\mu_{\rho}h^\rho_\sigma h^{\sigma\nu}
  \ .
  \labelt{nn47}
\end{equation}
This is the $\chn{3}$ term of the expansion of $g^\mn$.
\par
Let us turn to $\sqg$.
Here, we use the trace log expansion of the determinant:
\begin{subequations}
  \label{eqn:nn48}
  \begin{align}
    \sqrt{-g} &= \exp(\frac{1}{2}\tr\ln(\eta^\mu_\nu+h^\mu_\nu))
    \labelt{nn48a}
    \\
    &=\exp(\frac{1}{2}\big(
    h^\mu_\mu
    - \frac{1}{2} h^\mu_\nu h^\nu_\mu
    + \frac{1}{3} h^\mu_\nu h^\nu_\rho h^\rho_\mu
    - \frac{1}{4} h^\mu_\nu h^\nu_\rho h^\rho_\sigma h^\sigma_\mu
    + .\ .\ .
    \big))
    \labelt{nn48b}
    \\
    &=
    1
    + \frac{1}{2} h
    - \frac{1}{4} \maP^\rs_\mn h^\mn h_\rs
    + .\ .\ .
    \labelt{nn48c}
  \end{align}
\end{subequations}
This series is less straightforward and it would be interesting to look into methods to derive the terms more effectively.
In the third line, we have computed terms to second order in $h_\mn$.
In case of the graviton propagator and the three-graviton vertex it is sufficient to know the linear term.
\par
Again, using the notation of Eqs.~\eqreft{exp5} we can e.g. write:
\begin{equation}
  \sqrt{-g}
  _{\chn{2}} = - \frac{1}{4} \maP^\rs_\mn h^\mn h_\rs
  \ .
  \labelt{nn49}
\end{equation}
In general, the $h^n$ term in the $h_\mn$ expansion will be a function of $n$ factors of $h_\mn$.
For example in Eq.~\eqreft{nn49} we have the $h^2$ term of the $\sqg$-expansion which is a quadratic function of $h_\mn$.
Sometimes it will be useful to show the dependence on $h_\mn$ explicitly so that we e.g. write $\sqrt{-g}_{\chn{2}}(h,h)$ to denote that $\sqrt{-g}_{\chn{2}}$ is a quadratic function of $h_\mn$.
It is then possible to evaluate $\sqrt{-g}_{\chn{2}}$ with different arguments than $h_\mn$.
\par
For example, we can evaluate $\sqg_{\chn{2}}(h,h)$ in $h^{(a)}$ and $h^{(b)}$ which we think of as some given tensors:
\begin{align}
  \sqrt{-g}_{\chn{2}}(h^{(a)}_\mn,h^{(b)}_\mn) = - \frac{1}{4} \maP^\rs_\mn h^{(a)\mn} h^{(b)}_\rs
  \ .
  \labelt{n410}
\end{align}
This idea is uniquely defined as long as we demand the functions to be symmetric in the factors of $h_\mn$.
\par
Let us now consider expansions in $G_N$.
We will relate these to the expansions in $h_\mn$.
As an example, we will consider the Einstein tensor, $G^\mn$, and relate its expansion in $G_N$ to its expansion in $h_\mn$.
\par
By definition, we have
\begin{equation}
  G^\mn
  =
  \sum_{n=1..\infty}
  G_{\chn{n}}^\mn
  \labelt{n411}
\end{equation}
where we have used that $G^\mn$ does not have any $h^0$ terms.
We assume, that $h_\mn$ is at least of first order in $G_N$.
Then $G_{\chn{n}}^\mn$ is at least of $n$'th order in $G_N$.
\par
We will expand the terms $G_{\chn{n}}^\mn$ in $G_N$ separately.
We will do the cases $G_{\chno}^\mn$ and $G_{\chn{2}}^\mn$ explicitly after which the general case can be inferred.
\par
For the linear case:
\begin{subequations}
  \label{n412}
  \begin{align}
    G^\mn_{\chno}(h_\mn)
    &=
    G^\mn_{\chno}\big(\sum_{\nzi}h^{\cGn{n}}_\mn\big)
    \labelt{n412a}
    \\
    &=
    \sum_{\noi} G^\mn_{\chno}\big(h^{\cGn{n}}_\mn\big)
    \labelt{n412b}
  \end{align}
\end{subequations}
In the first line we inserted the expansion of $h_\mn$ in terms of $G_N$.
Then, in the second line we used that $G^\mn_{\chno}(h_\mn)$ is a linear function of $h_\mn$.
The term $G^\mn_{\chno}\big(h^{\cGn{n}}_\mn\big)$ is of n'th order in $G_N$.
\par
We can expand the quadratic term similarly:
\begin{subequations}
  \label{eqn:n413}
  \begin{align}
    G^\mn_{\chn{2}}(h_\mn,h_\mn)
    &=
    G^\mn_{\chn{2}}
    \big(
    \sum_{n=0..\infty}h^{\cGn{n}}_\mn
    ,
    \sum_{m=0..\infty}h^{\cGn{m}}_\mn
    \big)
    \labelt{n413a}
    \\
    &=
    \sum_{n=1..\infty}
    \sum_{m=1..\infty}
    G^\mn_{\chn{2}}
    \big(
    h^{\cGn{n}}_\mn
    ,
    h^{\cGn{m}}_\mn
    \big)
    \labelt{n413b}
    \\
    &=
    G^\mn_{\chn{2}}
    \big(
    h^{\cGno}_\mn
    ,
    h^{\cGno}_\mn
    \big)
    +
    2
    G^\mn_{\chn{2}}
    \big(
    h^{\cGno}_\mn
    ,
    h^{\cGn{2}}_\mn
    \big)
    +...
    \labelt{n413c}
  \end{align}
\end{subequations}
Again, in the first line we inserted the expansions of $h_\mn$ in terms of $G_N$.
In the second line we used that $G^\mn_{\chn{2}}(h,h)$ is linear in both of its arguments.
In the third line we have written out terms explicitly to third order in $G_N$.
For example $G^\mn_{\chn{2}} \big( h^{\cGno}_\mn , h^{\cGn{2}}_\mn \big)$ is of third order in $G_N$.

\par
These expansions can be compared to the expansion of the following polynomial:
\begin{align}
  (x_1 + x_2 + \ellipsis + x_n)^n
  \ .
  \labelt{n414}
\end{align}
For the case $n=3$ there would e.g. be a term $6G_{\chn{3}}^\mn(h^{\cGn{1}},h^{\cGn{2}},h^{\cGn{3}})$ which should be compared to $6x_1 x_2 x_3$ in the expansion of Eq.~\eqreft{n414}.
\par
We can now write down the explicit expansion of $G^\mn$ to third order in $G_N$ in terms of the functions $G_{\chn{n}}^\mn$.
\begin{align}
  G^\mn
  \approx&
  \
  G^\mn_{\chno}(h^{\cGno}_\mn)
  \labelt{exp1}
  \\
  &+
  G^\mn_{\chno}(h^{\cGn{2}}_\mn)
  +
  G^\mn_{\chn{2}}
  \big(
  h^{\cGno}_\mn
  ,
  h^{\cGno}_\mn
  \big)
  \nonumber
  \\
  &+
  G^\mn_{\chno}(h^{\cGn{3}}_\mn)
  +
  2 G^\mn_{\chn{2}}
  \big(
  h^{\cGno}_\mn
  ,
  h^{\cGn{2}}_\mn
  \big)
  +
  G^\mn_{\chn{3}}
  \big(
  h^{\cGno}_\mn
  ,
  h^{\cGno}_\mn
  ,
  h^{\cGno}_\mn
  \big)
  \nonumber
\end{align}
where in the first line we have the linear $G_N$ term, in the second line the $(G_N)^2$ terms and in the third line the $(G_N)^3$ terms.
Thus, when we know the expansion of $G^\mn$ in terms of $h_\mn$ we can find the expansion of $G^\mn$ in terms of $G_N$ as well.
\section{Action and Einstein Tensor Expanded in the Graviton Field}
\labelx{sec:ExpansionOf}
It is now the goal to expand the action $S$ and the Einstein tensor $G^\mn$ in the graviton field $h_\mn$.
First, we will focus on the gravitational part of the action, that is $S_\teEH+S_\tegf$, which from Eqs.~\eqreft{nn27} and~\eqreft{gau1} is given by:
\begin{equation}
  S_\teEH + S_\tegf
  =
  \int \dDx
  \Big(
  \sqg\ R + \frac{1}{2\xi} \eta^\rs G_\rho G_\sigma
  \Big)
  \labelt{act5}
\end{equation}
For the gauge-fixing function, $G_\sigma$, the \dDo-type function from Eq.~\eqreft{gau3} will be used.
\par
We will use two different expansions of the action.
First, with partial integrations the action can be rewritten so that it depends only on first derivatives of the metric.
We will then expand this form of the action explicitly to third order in $h_\mn$ from which we can derive the three-graviton vertex.
Second, we will write a general expansion in terms of tensor functions $\Ghn{n}{\mn}$ and $\Hhn{n}{\mn}$ which will be related to $G^\mn$ and an analogous tensor $H^\mn$ respectively.
\subsection{Action in terms of First Derivatives}
\labelx{sec:ActionIn}
The idea to rewrite the \EHA\ action in terms of first derivatives of the metric can e.g. be found in Dirac~\cite{Dirac:GR}.
We get:
\begin{subequations}
  \begin{align}
    S_{EH}
    &=
    \int \dDx
    \sqrt{-g} \ 
    R
    \labelt{dir1}
    \\
    &=
    \int \dDx
    \sqrt{-g}
    \ 
    g^\mn
    \Big(
    \Gamma^\rho_{\rho\mu,\nu} - \Gamma^\rho_{\mn,\rho}
    -
    \Gamma^\rho_\mn
    \Gamma^\sigma_{\rho\sigma}
    +
    \Gamma^\rho_{\mu\sigma}
    \Gamma^\sigma_{\nu\rho}
    \Big)
    \labelt{dir2}
    \\
    &=
    \int \dDx
    \sqrt{-g}
    \ 
    g^\mn
    \Big(
    \Gamma^\rho_\mn
    \Gamma^\sigma_{\rho\sigma}
    -
    \Gamma^\rho_{\mu\sigma}
    \Gamma^\sigma_{\nu\rho}
    \Big)
    \labelt{dir3}
  \end{align}
\end{subequations}
In the second line we inserted the definition of the Ricci scalar $R$ and the third line follows after partial integrations.
The result in the third line is that the first two terms of Eq.~\eqref{eqn:dir2} are removed while the last two terms of Eq.~\eqref{eqn:dir2} change sign.
\par
We can write the \EHA\ action entirely in terms of $g_\mn$ and $g^\mn$ by inserting the definition of the Christoffel symbols:
\begin{equation}
  S_{EH} = \int \dDx \sqrt{-g}\
  \frac{1}{4}
  \Big(
  2 g^{\sigma\gamma} g^{\rho\delta} g^{\ab}
  - g^\gd g^\ab g^\rs
  - 2g^{\sigma\alpha}g^{\gamma\rho}g^{\delta\beta}
  + g^{\rs}g^{\alpha\gamma}g^{\beta\delta}
  \Big)
  g_{\ab,\rho} g_{\gd,\sigma}
  \labelt{act6}
\end{equation}
This expression conforms to the traditional idea of a Lagrangian as a function of the field and its first derivatives.
\par
Both $S_\teEH$ and $S_\tegf$ are now quadratic functions of $h_{\mn,\sigma}=g_{\mn,\sigma}$.
In case of $S_\tegf$ this is so, since $G_\sigma$ is linear in $h_{\mn,\sigma}$.
We can now expand everything in $h_\mn$ and collect orders in $h_\mn$.
The only necessary expansions are those of $g^\mn$ and $\sqrt{-g}$ which we know from Sec.~\ref{sec:ExpansionsIn}.
\par
The expansion will be done to third order in $h_\mn$.
Let us start with the gauge-fixing term, which is simpler than the \EHA\ action.
It will be necessary to know the gauge-fixing function $G_\sigma$ to second order in $h_\mn$.
This is found to be
\begin{equation}
  G_\sigma \approx
  \maP^\mn_\rs h_\mn^{,\rho}
  - \alpha\
  \Gamma^{\rho\ab}_{\sigma\mn} h^\mn h_{\ab,\rho}
  \ ,
  \labelt{n419}
\end{equation}
where we have used $\maP$ from Eq.~\eqreft{imt3} and introduced a new tensor $\Gamma^{\rho\ab}_{\sigma\mn}$.
This tensor will sometimes be useful in the following and is defined such that:
\begin{align}
  \Gamma_{\rho\mn} = \Gamma_{\rho\mn}^{\sigma\ab} g_{\ab,\sigma}
  \ .
  \labelt{n420}
\end{align}
It is given entirely in terms of $\delta^\mu_\nu$ by the formula:
\begin{align}
  \Gamma^{\rho\ab}_{\sigma\mn} = I^{\ab}_{\sigma\kappa} I^{\rho\kappa}_{\mn} - \frac{1}{2} I^{\ab}_{\mn} \delta^\rho_\sigma
  \labelt{gam2}
\end{align}
It is now straightforward to expand $G_\sigma G^\sigma$ to third order in $h_\mn$:
\begin{align}
  G_\sigma G^\sigma
  \approx
  h^{,\rho}_\mn \maP^\mn_{\rho\kappa} \maP^{\kappa\sigma}_\ab h^\ab_{,\sigma}
  - 2\alpha
  \maP_\gd^{\rho\kappa} h^\gd_{,\rho}
  \Gamma^{\sigma\ab}_{\kappa\mn}
  h^\mn h_{\ab,\sigma}
  \ .
  \labelt{act9}
\end{align}
This expression can now be inserted in the gauge-fixing term $S_\tegf$ of Eq.~\eqreft{act5}.
\par
The expansion of the \EHA\ action is more complicated than that of the gauge-fixing term.
Using Eq.~\eqreft{act6} we get the quadratic term by replacing $g^\mn$ by $\eta^\mn$ and $\sqg$ by unity since the two factors of $h_\mn$ come from $h_{\mn,\sigma}$.
We get:
\begin{align}
  \hspace*{-.2cm}
  (S_\teEH)_{\chn{2}}
  &=
  \int \dDx 
  \frac{1}{4}
  \Big(
  2 \eta^{\sigma\gamma} \eta^{\rho\delta} \eta^{\ab}
  - \eta^\gd \eta^\ab \eta^\rs
  - 2\eta^{\sigma\alpha}\eta^{\gamma\rho}\eta^{\delta\beta}
  + \eta^{\rs}\eta^{\alpha\gamma}\eta^{\beta\delta}
  \Big)
  h_{\ab,\rho} h_{\gd,\sigma}
  \labelt{act7}
  \\
  &=
  \int \dDx 
  \frac{1}{4}
  \Big(
  2 \eta^{\sigma\gamma} \eta^{\rho\delta} \eta^{\ab}
  - \eta^\gd \eta^\ab \eta^\rs
  - 2\eta^{\sigma\alpha}\eta^{\gamma\rho}\eta^{\delta\beta}
  + \eta^{\rs}\eta^{\alpha\gamma}\eta^{\beta\delta}
  \Big)
  h_{\ab,\sigma} h_{\gd,\rho}
  \ .
  \nonumber{}
\end{align}
In the second line we made partial integrations so that the partial derivatives on the two factors of $h_\mn$ were exchanged.
\par
Using
\begin{align}
  2\maP^\ab_{\kappa\rho} \maP_\gd^{\kappa\sigma} h_{\ab}^{,\rho} h^\gd_{,\sigma}
  =
  \Big(
  2\eta^{\sigma\alpha}\eta^{\gamma\rho}\eta^{\delta\beta}
  - 2\eta^{\sigma\gamma} \eta^{\rho\delta} \eta^{\ab}
  + \frac{1}{2} \eta^\gd \eta^\ab \eta^\rs
  \Big)
  h_{\ab,\sigma} h_{\gd,\rho}
  \ ,
  \labelt{n424}
\end{align}
we can rewrite the quadratic term, Eq.~\eqreft{act7}, as:
\begin{equation}
  (S_\teEH)_{\chn{2}} =
  \frac{1}{4}
  \int \dDx
  \ h_{\gd}^{,\rho}
  \Big(
  \delta^\rho_\sigma \mathcal{P}^\gd_\ab
  -2 \mathcal{P}^{\gd}_{\rho\kappa} \mathcal{P}_{\ab}^{\sigma\kappa}
  \Big)
  h^{\ab}_{,\sigma}
  \ .
  \labelt{n425}
\end{equation}
This is a rather simple result and the tensor structure of the second term in the quadratic operator is the same as the one that comes from the gauge-fixing term.
\par
The three-graviton term of the \EHA\ action is more involved than that of the gauge-fixing term.
In Eq.~\eqreft{act6} we should in turn replace one factor of $g^\mn$ with $-h^\mn$ and the rest with $\eta^\mn$.
Then we should also add the contribution from $\sqg \approx 1 + \frac{1}{2} h^\nu_\nu$.
Naively, this gives one term for each factor of $g^\mn$, that is 12 terms, and 4 additional terms from the contribution of $\sqg$ multiplied into the brackets.
However, some of these terms are equivalent.
\par
Computing all of these terms we have found that:
\begin{equation}
  (S_\teEH)_{\chn{3}} =
  \frac{1}{2} \int \dDx
  \ U_\tecl^{\mn\ \ab\rho\ \gd\sigma} h_\mn h_{\ab,\rho} h_{\gd,\sigma}
  \ ,
  \labelt{n426}
\end{equation}
where the three-graviton term can be written in the compact form:
\begin{align}
  U_\tecl^{\mn\ \ab\rho\ \gd\sigma} \ h_{\ab,\rho} h_{\gd,\sigma}
  =
  &
  2 I^\mn_\pe \maP^\ab_\rs \maP^{\sigma\phi}_{\gamma\delta} h^{,\epsilon}_\ab h^{\gd,\rho}
  - \maP^{\mu\rho}_\ab \maP^{\nu\sigma}_\gd \eta_\rs h^\ab_{,\kappa} h^{\gd,\kappa}
  \labelt{ute1}
  \\
  &
  + \maP^\mn_\rs
  \Big(
  h^{\rho\alpha}_{,\beta} h^{\sigma\beta}_{,\alpha}
  -\frac{1}{2} h^{\alpha,\rho}_{\beta} h^{\beta,\sigma}_{\alpha}
  - h^\rs_{,\alpha} h^\ab_{,\beta}
  \Big)
  \nonumber
  \ .
\end{align}
It can also be written in a less compact form, which is more easily compared to the action in Eq.~\eqreft{act6}:
\begin{align}
  U_\tecl^{\mn\ \ab\rho\ \gd\sigma} h_\mn h_{\ab,\rho} h_{\gd,\sigma}
  = &\frac{1}{2} h^\mu_\nu h_{,\mu} h^{,\nu}
  - \frac{1}{4} h  h_{,\rho} h^{,\rho}
  + h^\mu_\nu h^\nu_{\mu,\rho} h^{,\rho}
  - h^\mu_\nu h_\mu^{\nu,\sigma} h_{\sigma,\rho}^\rho
  +\frac{1}{4} h h^\mu_{\nu,\rho} h_\mu^{\nu,\rho}
  \nonumber{}
  \\
  &- h^\nu_\mu h^\mu_{\sigma,\nu} h^{,\sigma}
  - h^\mu_\nu h^{,\nu} h^\rho_{\mu,\rho}
  +\frac{1}{2} h h^\rho_{\sigma,\rho} h^{,\sigma}
  -h^\mu_\nu h^\rho_{\mu,\sigma} h_\rho^{\nu,\sigma}
  -\frac{1}{2} h h^\rho_{\nu,\mu} h_\rho^{\mu,\nu}
  \nonumber{}
  \\
  &+h^\mn h_{\mu,\rho}^\sigma h_{\nu,\sigma}^\rho
  -\frac{1}{2} h^\mu_\nu h^\rho_{\sigma,\mu} h_\rho^{\sigma,\nu}
  +2 h^\mu_\nu h_\rho^{\sigma,\nu} h^\rho_{\mu,\sigma}
  \labelt{ute4}
\end{align}
Here, $h=h_\nu^\nu$.
There are 13 terms instead of the 16 terms estimated from the naive counting.
\par
By analogy we define a $U_\tegf$:
\begin{equation}
  (S_\tegf)_{\chn{3}} =
  \frac{1}{2\xi} \int \dDx
  \ U_\tegf^{\mn\ \ab\rho\ \gd\sigma} h_\mn h_{\ab,\rho} h_{\gd,\sigma}
  \ .
  \labelt{act8}
\end{equation}
The tensor, $U_\tegf$, can easily be read off from Eq.~\eqreft{act9}:
\begin{subequations}
  \label{eqn:n430}
  \begin{align}
    U_{gf}^{\mn\ \ab\rho\ \gd\sigma} \ h_{\ab,\rho} h_{\gd,\sigma}
    &=
    -2 \alpha \maP^\ab_\rs h^{,\sigma}_\ab \Gamma^{\rho\mn}_{\kappa\gd} h^{\gd,\kappa}
    \labelt{n430a}
    \\
    &=
    \alpha
    \mathcal{P}^\rs_\ab
    h^\ab_{,\sigma}
    \Big(
    - h_{\rho}^{\mu,\nu}
    - h_{\rho}^{\nu,\mu}  
    + h^\mn_{,\rho}  
    \Big)
    \ .
    \labelt{n430b}
  \end{align}
\end{subequations}
In the second line we inserted the definition of $\Gamma^{\rho\mn}_{\kappa\gd}$.
\par
We define
\begin{equation}
  U^{\mn\ \ab\rho\ \gd\sigma} =
  U_\tecl^{\mn\ \ab\rho\ \gd\sigma}
  +
  \frac{1}{\xi}
  U_\tegf^{\mn\ \ab\rho\ \gd\sigma}
  \ ,
  \labelt{n431}
\end{equation}
and we get the final expression for the expansion of the gravitational action, Eq.~\eqreft{act5}, to third order in $h_\mn$:
\begin{align}
  S_\teEH + S_\tegf
  \approx
  &\ \frac{1}{4}
  \int \dDx
  \ h_{\mn}^{,\rho}
  \Big(
  \delta^\rho_\sigma \mathcal{P}^\mn_\ab
  -2(1-\frac{1}{\xi})
  \mathcal{P}^{\mn}_{\rho\kappa} \mathcal{P}_{\ab}^{\sigma\kappa}
  \Big)
  h^{\ab}_{,\sigma}
  \nonumber{}
  \\
  &+ \frac{1}{2} \int \dDx
  \ U^{\mn\ \ab\rho\ \gd\sigma} h_\mn h_{\ab,\rho} h_{\gd,\sigma}
  \ .
  \labelt{n432}
\end{align}
This is the main result of this section.
\par
The tensors $U^{\mn\ \ab\rho\ \gd\sigma}$, $U_\tecl$ and $U_\tegf$ are defined to be symmetric under exchange $\ab\rho \leftrightarrow \gd\sigma$.
With this symmetry as well as symmetry in $\mu\leftrightarrow \nu$, $\alpha \leftrightarrow \beta$ and $\gamma \leftrightarrow \delta$ the definitions of the $U$-tensors can be read off from the expressions in Eqs.~\eqreft{ute1},~\eqreft{ute4} and~\eqreft{n430}.
\par
Let us show how this is done with the simple $U_\tegf$ in Eq.~\eqreft{n430b}.
For example the last term of Eq.~\eqreft{n430b}:
\begin{align}
  \alpha
  \maP^\rs_\ab h^\ab_{,\sigma} h^\mn_{,\rho}
  &=
  \alpha \maP^{\ab\rs} I^{\gd\mn} h_{\ab,\rho}h_{\mn,\sigma}
  \labelt{n433}
  \\
  &=
  \frac{1}{2}
  \alpha
  \Big(
  \maP^{\ab\rs} I^{\gd\mn}
  +
  \maP^{\gd\rs} I^{\ab\mn}
  \Big)
  h_{\ab,\rho}h_{\mn,\sigma}
  \labelt{n434}
\end{align}
Hence this term contributes with
\begin{align}
  \frac{1}{2}
  \alpha
  \Big(
  \maP^{\ab\rs} I^{\gd\mn}
  +
  \maP^{\gd\rs} I^{\ab\mn}
  \Big)
  \labelt{n435}
\end{align}
to $U_\tegf$.
In general $U_\tegf$ is:
\begin{align}
  \hspace*{-1cm}
  U_\tegf^{\mn\ \ab\rho\ \gd\sigma}
  =
  -\alpha
  \Big(
  I^{\mn\rho\kappa}
  I^\ab_{\kappa\lambda}
  \maP^{\lambda\sigma\gd}
  +
  I^{\mn\sigma\kappa}
  I^\gd_{\kappa\lambda}
  \maP^{\lambda\rho\ab}
  \Big)
  +
  \frac{1}{2}
  \alpha
  \Big(
  \maP^{\ab\rs} I^{\gd\mn}
  +
  \maP^{\gd\rs} I^{\ab\mn}
  \Big)
  \labelt{n436}
\end{align}
A similar expression can be derived for $U_\tecl$.
\subsection{Action in terms of the Einstein Tensor}
We will now use a different approach for the expansion of the gravitational action which is suited for the metric computation in the classical limit.
We postulate an expansion of $S_\teEH$ to any order in $h_\mn$ in terms of tensor functions $\Ghn{n}{\mn}$:
\begin{align}
  S_\teEH
  =
  -
  \int \dDx
  \
  h_\mn
  \sum_{n=1..\infty} \oov{(n+1)}  \Ghn{n}{\mn}(h,h,\ellipsis,h)
  \ .
  \labelt{n437}
\end{align}
The functions $\Ghn{n}{\mn}$ will then be related to the Einstein tensor.
They are evaluated from $n$ factors of $h_\mn$.
We can make an analogous expansion of $S_\tegf$.
We postulate:
\begin{align}
  S_\tegf
  =
  -
  \frac{1}{\xi}
  \int \dDx
  \
  h_\mn
  \sum_{n=1..\infty} \oov{(n+1)}  \Hhn{n}{\mn}(h,h,\ellipsis,h)
  \ .
  \labelt{n438}
\end{align}
Here $\Hhn{n}{\mn}$ are analogous to $\Ghn{n}{\mn}$ and will be related to a tensor analogous to the Einstein tensor which we will call $H^\mn$.
\par
Let us focus on $S_\teEH$ first.
The case of $S_\tegf$ will be similar.
We require that
\begin{equation}
  \Ghn{n}{\mn}(h,h,\ellipsis,h)
  \ ,
  \labelt{n439}
\end{equation}
is symmetric in its n arguments, that is, it is symmetric in the $n$ factors of $h_\mn$.
In addition, we require that
\begin{equation}
  \int \dDx
  \
  h_\mn
  \oov{(n+1)}  \Ghn{n}{\mn}(h,h,\ellipsis,h)
  \ ,
  \labelt{n440}
\end{equation}
is symmetric in the $(n+1)$ factors of $h_\mn$.
Obviously, the integrand is not symmetric by itself, since the $h_\mn$ contracted to $\Ghn{n}{\mn}$ plays an asymmetrical role in comparison to the other $n$ factors of $h_\mn$.
However, the integral can still be symmetric in the $(n+1)$ factors of $h_\mn$ due to partial integrations.
\par
The last condition, that the integral Eq.~\eqreft{n440} is symmetric in its factors of $h_\mn$, means that it is straightforward to vary this integral in $h_\mn$:
\begin{align}
  \delta  \int \dDx
  \
  h_\mn
  \oov{(n+1)}  \Ghn{n}{\mn}(h,h,\ellipsis,h)
  =
  \int \dDx
  \
  \Ghn{n}{\mn}(h,h,\ellipsis,h)
  \ 
  \delta h_\mn
  \labelt{n441}
\end{align}
This makes it easy to relate the functions $\Ghn{n}{\mn}$ to the Einstein tensor.
\par
The requirements Eqs.~\eqreft{n439} and~\eqreft{n440} are always possible to fulfill and uniquely define the functions $\Ghn{n}{\mn}$.
As an example let us relate $\Ghn{2}{\mn}$ to $U^{\mn\ \ab\rho\ \gd\sigma}_c$:
\begin{subequations}
  \label{sht5}
  \begin{align}
    \hspace*{-1cm}
    (S_\teEH)_{\chn{3}}
    &= -\oov{3}\int\ddx\  h_\mn \Ghn{2}{\mn}(h,h)
    \labelt{sht1}
    \\
    &= \oov{2}\int\ddx\  h_\mn U_\tecl^{\mn\ \ab\rho\ \gd\sigma} h_{\ab,\rho}h_{\gd,\sigma}
    \labelt{sht2}
    \\
    &= \oov{6}\int\ddx\
    \Big(
    h_\mn U_\tecl^{\mn\ \ab\rho\ \gd\sigma} h_{\ab,\rho}h_{\gd,\sigma}
    +h_{\mn,\sigma} U_\tecl^{\ab\ \gd\rho\ \mn\sigma} h_{\ab}h_{\gd,\rho}
    +h_{\mn,\rho} U_\tecl^{\gd\ \mn\rho\ \ab\sigma} h_{\ab,\sigma}h_{\gd}
    \Big)
    \label{eqn:sht3}
    \\
    &= \oov{6}\int\ddx\
    h_\mn
    \Big(
    U_\tecl^{\mn\ \ab\rho\ \gd\sigma} h_{\ab,\rho}h_{\gd,\sigma}
    - U_\tecl^{\ab\ \gd\rho\ \mn\sigma} \partial_\sigma(h_{\ab}h_{\gd,\rho})
    - U_\tecl^{\gd\ \mn\rho\ \ab\sigma} \partial_\rho(h_{\ab,\sigma}h_{\gd})
    \Big)
    \label{eqn:sht4}
  \end{align}
\end{subequations}
First, in Eqs.~\eqreft{sht1} and~\eqreft{sht2} we use the definitions of $S_\teEH$ in terms of $\Ghn{n}{\mn}$ and $U_\tecl^{\mn\ \ab\rho\ \gd\sigma}$ respectively.
Then in Eq.~\eqreft{sht3} we pretend that the three factors of $h_\mn$ in the action are distinguished by their written order and rewrite the action in terms of $U_\tecl^{\mn\ \ab\rho\ \gd\sigma}$ so that it is symmetric in these three factors.
Then it certainly obeys an equation similar to~\eqreft{n441}.
After partial integrations which leave the action unchanged we rewrite the action in Eq.~\eqreft{sht4} in the same form as $\Ghn{n}{\mn}$ in Eq.~\eqreft{n437}.
After expanding the partial derivatives in Eq.~\eqreft{sht4} and using the symmetries of $U$, we find that $\Ghn{2}{\mn}$ is given in terms of $U_\tecl$ according to:
\begin{align}
  \hspace*{-1cm}
  \Ghn{2}{\mn}(h,h)
  =
  -\frac{1}{2}
  \Big(
  U_\tecl^{\mn\ \ab\rho\ \gd\sigma} h_{\ab,\rho}h_{\gd,\sigma}
  - 2U_\tecl^{\ab\ \gd\rho\ \mn\sigma} h_{\ab,\sigma}h_{\gd,\rho}
  - 2U_\tecl^{\ab\ \gd\rho\ \mn\sigma} h_{\ab}h_{\gd,\rho\sigma}
  \Big)
  \labelt{ute2}
\end{align}
By similar arguments any term in the action can uniquely be rearranged to obey the two conditions in Eqs.~\eqreft{n439} and~\eqreft{n440}.
\par
We will now relate the functions $\Ghn{n}{\mn}$ to the Einstein tensor.
Using Eq.~\eqreft{n441} we vary the Einstein-Hilbert action in the form of Eq.~\eqreft{n437} and get:
\begin{align}
  \delta S_{EH}
  =
  -
  \int \dDx\ 
  \delta h_\mn
  \sum_{n=1..\infty}  \Ghn{n}{\mn}(h,h,\ellipsis,h)
  \ .
  \labelt{n444}
\end{align}
This is easily compared to the expression for $\delta S$ in terms of $G^\mn$ Eq.~\eqreft{ein1}:
\begin{align}
  \delta S_{EH}
  = -  \int \ddx
  \ \sqrt{-g} \ G^\mn \delta h_\mn
  \labelt{n445}
\end{align}
In this way $\Ghn{n}{\mn}$ is related to the Einstein tensor:
\begin{align}
  \sum_{n=1..\infty}  \Ghn{n}{\mn}(h,h,\ellipsis,h)
  = \sqrt{-g}\ G^\mn
  \labelt{n446}
\end{align}
Using the notation introduced in Sec.~\ref{sec:ExpansionsIn} we can write the relation as:
\begin{subequations}
  \label{eqn:n447}
  \begin{align}
    \Ghn{n}{\mn}
    &=
    \big(
    \sqrt{-g} G^\mn
    \big)_{\chn{n}}
    \labelt{n447a}
    \\
    &= \sum_{j=0..n}
    G^\mn_{h^{n-j}} \ (\sqrt{-g})_{h^j}
    \labelt{n447b}
  \end{align}
\end{subequations}
And in particular:
\begin{subequations}
  \label{eqn:n448}
  \begin{align}
    &\Ghn{1}{\mn} =  G^\mn_{h^1}
    \labelt{n448a}
    \\
    &\Ghn{2}{\mn} =  G^\mn_{h^2} +\frac{1}{2} h\ G^\mn_{h^1}
    \labelt{n448b}
    \\
    &\Ghn{3}{\mn}
    = G^\mn_{h^3}
    +\frac{1}{2} h\ G^\mn_{h^2}
    - \frac{1}{4} \maP^\rs_\ab h^\ab h_\rs\ G^\mn_{h^1}
    \labelt{n448c}
  \end{align}
\end{subequations}
In general, it makes sense to define
\begin{equation}
  \Gpz^\mn = \sqg\ G^\mn
  \ ,
  \labelt{n449}
\end{equation}
which nicely summarizes the relation of the functions $\Ghn{n}{\mn}$ to the Einstein tensor.
\par
We will now discuss the similar result for $S_\tegf$.
Let us introduce the analogous tensor to the Einstein tensor by:
\begin{equation}
  \delta S_\tegf =
  \frac{1}{\xi}
  \int
  \dDx
  \
  \eta^\rs
  G_\rho
  \delta G_\sigma
  =
  -
  \frac{1}{\xi}
  \int \dDx \
  \sqrt{-g}
  H^\mn \delta h_\mn
  \ .
  \labelt{hmn1}
\end{equation}
In Ch.~\ref{sec:STM} the tensor $H^\mn$ will be analyzed in detail for the generalized \dDo-type gauge function introduced in Sec.~\ref{sec:GaugeFixing}.
\par
We require the functions $\Hhn{n}{\mn}$ to obey the same conditions as $\Ghn{n}{\mn}$ in Eqs.~\eqreft{n439} and~\eqreft{n440}.
By similar arguments as for $\Ghn{n}{\mn}$ we find that $\Hhn{n}{\mn}$ is related to $H^\mn$ by:
\begin{align}
  \Hhn{n}{\mn}
  &=
  \big(
  \sqrt{-g} H^\mn
  \big)_{\chn{n}}
  \ .
  \labelt{n451}
\end{align}
And we define also
\begin{align}
  \Hpz^\mn
  =
  \sqg \ H^\mn
  \labelt{n452}
\end{align}
which summarizes the relation between $\Hhn{n}{\mn}$ and $H^\mn$.
\par
In Sec.~\ref{sec:GravitonSelf} we will derive the vertex rules for the n-graviton self-interaction vertices in terms of $\Ghn{n}{\mn}$ and $\Hhn{n}{\mn}$.
This makes it possible to compare the n-graviton amplitude to the classical equations of motion which we will do in Ch.~\ref{sec:STM}.
\subsection{Einstein Tensor to Second Order in the Graviton Field}
\labelx{sec:EinsteinTensor}
We will now derive results for the expansion of the Einstein tensor $G^\mn$ to second order in $h_\mn$.
These results will not be used for explicit computations in this work.
However, since we have related the expansion of the \EHA\ action to the Einstein tensor, they can be used as an alternative definition of the three-graviton vertex instead of the $U$-tensor.
They are suited for the one-loop computation of the \STM\ metric if the triangle integrals are simplified appropriately.
Also, we will introduce a tensor $Q^{\mn\ \ab\ \gd}$ which describes the quadratic term in the \EHA\ action.
\par
The Einstein tensor can be written in the following way:
\begin{subequations}
  \begin{align}
    G^\mn
    &=
    \frac{1}{2}
    \Big(
    g^{\mu\alpha}g^{\nu\beta}
    +
    g^{\mu\beta}g^{\nu\alpha}
    - g^\mn g^\ab
    \Big)
    R_\ab
    \\
    &=
    \frac{1}{4}
    \Big(
    g^{\mu\alpha}g^{\nu\beta}
    +
    g^{\mu\beta}g^{\nu\alpha}
    - g^\mn g^\ab
    \Big)
    g^\gd
    F_{\ab\ \gd}^{\rs\ \pe}
    \Big(
    g_{\pe,\rs}
    +
    g^{\kappa\lambda}
    \Gamma_{\kappa\rs}
    \Gamma_{\lambda\pe}
    \Big)
    \labelt{ein2}
  \end{align}
\end{subequations}
Recall, that $\Gamma_{\sigma\mn}$ is linear in $h_{\mn,\sigma}$.
Here $F^{\rs\ \pe}_{\ab\ \gd}$ is a tensor introduced for convenience defined by:
\begin{align}
  F^{\ab\ \gd}_{\mn\ \rs}
  =
  I^\ab_\mn I^\gd_\rs
  +
  I^\ab_\rs I^\gd_\mn
  -2
  I^\ab_{\epsilon\zeta}
  I^{\zeta\eta}_{\mn}
  I^{\gd}_{\eta\theta}
  I^{\theta\epsilon}_\rs
  \ .
\end{align}
It is expressible entirely in terms of $\delta^\mu_\nu$.
\par
It can be helpful to separate $G^\mn$ into a second derivative part $G_a$ and a first derivative part $G_b$.
\begin{subequations}
\begin{align}
  &G^\mn_a
  =
  \frac{1}{4}
  \Big(
  g^{\mu\alpha}g^{\nu\beta}
  +
  g^{\mu\beta}g^{\nu\alpha}
  - g^\mn g^\ab
  \Big)
  g^\gd
  F_{\ab\ \gd}^{\rs\ \pe}
  \partial_\rho \partial_\sigma g_\pe
  \ ,
  \\
  &G^\mn_b
  =
  \frac{1}{4}
  \Big(
  g^{\mu\alpha}g^{\nu\beta}
  +
  g^{\mu\beta}g^{\nu\alpha}
  - g^\mn g^\ab
  \Big)
  g^\gd
  F_{\ab\ \gd}^{\rs\ \pe}
  g^{\kappa\lambda}
  \Gamma_{\kappa\rs}
  \Gamma_{\lambda\pe}
  \ .
\end{align}
\end{subequations}
Here, $G_a^\mn$ is at least of first order in $h_\mn$ and $G^\mn_b$ is at least of second order.
\par
For the linear term of $G^\mn$ in $h_\mn$ we expand $G_a^\mn$.
All instances of $g^\mn$ are replaced by $\eta^\mn$ and we get:
\begin{align}
  G^\mn_{\chno}
  = (G_{a}^\mn)_{\chno}
  &=
  \frac{1}{2} \maP^{\mn\ab} \eta^\gd
  F^{\rs\ \pe}_{\ab\ \gd}
  h_{\rs,\pe}
  \nonumber{}
  \\
  &=
  \frac{1}{2} Q^{\mn\ \ab\ \gd} h_{\ab,\gd}
  \labelt{n457}
\end{align}
The $Q$-tensor describes the tensor structure of $G^\mn_\chno$.
From $G^\mn_\chno$ together with the similar term of $H^\mn_\chno$ we would be able to derive the graviton propagator.
The $Q$-tensor is:
\begin{subequations}
  \label{eqn:n459}
  \begin{align}
    Q^{\mn\ \ab\ \gd}
    &=
    \maP^{\mn\rs} \eta^\pe
    F^{\ab\ \gd}_{\rs\ \pe}
    \labelt{n459a}
    \\
    &=
    \eta^\mn \maP^{\ab\gd}
    -2
    I^\mn_{\sigma\rho}
    \maP^{\rho\phi\ab}
    \eta_\pe
    \maP^{\epsilon\sigma\gd}
    \labelt{n459b}
  \end{align}
\end{subequations}
Although it is not apparent from its definition $Q^{\mn\ \ab\ \gd}$ is symmetric in all its pairs of indices, that is it is symmetric when $\mn \leftrightarrow \ab$ and $\mn \leftrightarrow \gd$ and also $\ab \leftrightarrow \gd$.
\par
The quadratic term, $G^\mn_{\chn{2}}$, gets contributions from both $G_a^\mn$ and $G_b^\mn$.
For $G_b^\mn$ we replace all instances of $g^\mn$ by $\eta^\mn$ while for $G_a^\mn$ instances of $g^\mn$ should in turn be replaced by $-h^\mn$.
Then, for the $h^2$ terms of $G_a$ and $G_b$:
\begin{subequations}
  \begin{align}
    &
    (G_{a}^\mn)_{\chn{2}}
    =
    -\Big(
    \maP^\mn_{\rho\kappa} \ h^\rho_\sigma \ \maPi^{\sigma\kappa}_\ab
    + \frac{1}{2(\stdim-2)} h^\mn \eta_\ab
    \Big)
    Q^\ab_{\gd\ \pe}
    h^{\gd,\pe}
    -\frac{1}{2} F^{\mn\ \rs}_{\gd\ \pe} h_\rs h^{\gd,\pe}
    \ ,
    \labelt{gea1}
    \\
    &
    (G_{b}^\mn)_{\chn{2}}
    =
    \frac{1}{2} Q^{\mn\ \ab\ \gd}
    \ \eta^\rs
    \ \Gamma_{\rho \ab}
    \ \Gamma_{\sigma \gd}
    \ .
    \labelt{geb1}
  \end{align}
\end{subequations}
These formulas can be used to define the three-graviton vertex.
Also, they can be used in the perturbative expansion of the classical equations of motion in Sec.~\ref{sec:PerturbativeExpansion1}.
For the \STM\ metric computation only the last term of Eq.~\eqreft{gea1} with the $F$-tensor and Eq.~\eqreft{geb1} would contribute.

\chapter{Graviton Feynman Rules}
\labelx{sec:FeynmanRules}
The Feynman rules are derived using the expansion of the action, $S$, in the graviton field, $h_\mn$ from Ch.~\ref{sec:ExpansionsAround}.
In Sec.~\ref{sec:GravitonPropagator}, we analyze the quadratic term of the gauge-fixed gravitational action from which the graviton propagator in covariant \dDo-type gauge is derived.
In Sec.~\ref{sec:ScalarGraviton} we compute the matter interactions from the scalar part of the action.
Finally, in Sec.~\ref{sec:GravitonSelf} we focus on the graviton self-interaction vertices.
Here, we will derive explicit results for the three-graviton vertex as well as expressions for the general n-graviton vertex in terms of the Einstein tensor.
\section{Covariant Gauge Graviton Propagator}
\labelx{sec:GravitonPropagator}
To derive the graviton propagator we need the quadratic term of the gauge-fixed gravitational action.
From Eq.~\eqreft{n432} we get this term:
\begin{equation}
  (S_\teEH + S_\tegf)_{\chn{2}}
  =
  \frac{1}{4}
  \int \dDx
  \ h_{\mn}^{,\rho}
  \Big(
  \delta^\rho_\sigma \mathcal{P}^\mn_\ab
  -2(1-\oov{\xi}) \mathcal{P}^{\mn}_{\rho\kappa} \mathcal{P}_{\ab}^{\sigma\kappa}
  \Big)
  h^{\ab}_{,\sigma}
  \labelt{n463}
\end{equation}
However, when we derive the Feynman rules we will use the rescaled graviton field $\hka_\mn = \frac{1}{\kappa}h_\mn$ and the rescaled action $\frac{2}{\kappa^2}(S_\teEH+S_\tegf)$.
Using the rescaled quantities, we get:
\begin{equation}
  \frac{2}{\kappa^2}(S_\teEH + S_\tegf)_{\chn{2}}
  =
  \frac{1}{2}
  \int \dDx
  \ \hka_{\mn}^{,\rho}
  \Big(
  \delta^\rho_\sigma \mathcal{P}^\mn_\ab
  -2(1-\oov{\xi}) \mathcal{P}^{\mn}_{\rho\kappa} \mathcal{P}_{\ab}^{\sigma\kappa}
  \Big)
  \hka^{\ab}_{,\sigma}
  \labelt{n464}
\end{equation}
This is the quadratic graviton action in covariant \dDo-type gauge.
For $\xi=1$ the quadratic term reduces significantly similarly to the case of quantum electrodynamics in Feynman-'t Hooft gauge.
\par
We will derive the graviton propagator in momentum space and we transform Eq.~\eqreft{n464} to momentum space and get:
\begin{equation}
  \frac{2}{\kappa^2}(S_\teEH + S_\tegf)_{\chn{2}}
  =
  \frac{1}{2}
  \int \dDp{p}
  \
  \tilde \hka_{\mn}^\dagger
  \ p^2
  \Big(
  \mathcal{P}^\mn_\ab
  -2(1-\oov{\xi})
  \mathcal{P}^{\mn}_{\rho\kappa}
  \frac{p^\rho p_\sigma}{p^2}
  \mathcal{P}_{\ab}^{\sigma\kappa}
  \Big)
  \tilde \hka^{\ab}
  \labelt{n465}
\end{equation}
We want to invert the quadratic operator between the two gravitons.
Let us analyze its tensor structure:
\begin{equation}
  \Delta^\mn_\ab =
  \mathcal{P}^\mn_\ab
  -2(1-\oov{\xi})
  \mathcal{P}^{\mn}_{\rho\kappa}
  \frac{p^\rho p_\sigma}{p^2}
  \mathcal{P}_{\ab}^{\sigma\kappa}
  \labelt{del1}
\end{equation}
It depends on the momentum of the graviton $p^\mu$ and the covariant gauge parameter $\xi$.
\par
We want to invert $\Delta^\ab_\mn$.
We can do this in several ways.
For example, in $D=4$ we can think of $\Delta^\ab_\mn$ as a matrix in 10-dimensional space.
In the inertial frame of $p_\mu$ we can then write $\Delta^\ab_\mn$ as an explicit 10-by-10 matrix and invert it with methods from linear algebra or computer algebra.
\par
Here, we will write an ansatz for the most general covariant quadratic operator depending on the graviton momentum.
There are 5 such independent operators, which we will define as follows:
\begin{subequations}
  \label{eqn:ope1}
  \begin{align}
    &I^\mn_\ab = \frac{1}{2}(\delta^\mu_\nu\delta^\alpha_\beta + \delta^\mu_\beta\delta^\nu_\alpha)
    \labelt{ope2}
    \\
    &T^\mn_\ab = \frac{1}{4} \eta^\mn \eta_\ab
    \\
    &C^\mn_\ab = \frac{1}{2}
    \Big(
    \eta^\mn \frac{p_\alpha p_\beta}{p^2}
    + \frac{p^\mu p^\nu}{p^2} \eta_\ab
    \Big)
    \\
    &\maJ^\mn_\ab = I^\mn_{\rho\kappa} \frac{p_\sigma p^\rho}{p^2} I^{\sigma\kappa}_\ab
    \\
    &K^\mn_\ab = \frac{p^\mu p^\nu}{p^2} \frac{p_\alpha p_\beta}{p^2}
  \end{align}
\end{subequations}
Any other quadratic operator built from $p^\sigma$ and $\eta^\mn$ can be written in terms of these operators.
\par
Let us now use an index free notation where matrix multiplication is understood.
Note that
\begin{equation}
  \maP = I - 2 T
  \ ,
  \labelt{n471}
\end{equation}
and:
\begin{equation}
  \Delta = \maP +2(1-\frac{1}{\xi})\maP\maJ\maP
  \ .
  \labelt{n472}
\end{equation}
We want to find $G = \Delta^{-1}$.
We write $G$ as a linear combination of the 5 operators in Eqs.~\eqreft{ope1}:
\begin{equation}
  G = \alpha_1 I + \alpha_2 T + \alpha_3 C + \alpha_4 \maJ + \alpha_5 K
  \ .
  \labelt{pro6}
\end{equation}
The coefficients $\alpha_n$ are determined from the equation
\begin{equation}
  \Delta G = I
  \ ,
  \labelt{pro4}
\end{equation}
which follows from the definition of $G$.
\par
The operators in Eqs.~\eqreft{ope1} are easily multiplied together.
During the calculations an antisymmetric operator enters as well:
\begin{equation}
  A^\mn_\ab = \frac{1}{2}
  \Big(
  \eta^\mn \frac{p_\alpha p_\beta}{p^2}
  - \frac{p^\mu p^\nu}{p^2} \eta_\ab
  \Big)
  \labelt{n475}
\end{equation}
Let us show some examples of the operators being multiplied together:
\begin{subequations}
  \label{eqn:n476}
  \begin{align}
    &T T = \frac{D}{4} T
    \labelt{n476a}
    \\
    &\maJ \maJ = \frac{1}{2} \maJ + \frac{1}{2} K
    \\
    &\maJ T \maJ = \frac{1}{4} K
    \\
    &T^\mn_\ab C^\ab_\gd =
    \frac{1}{2} T^\mn_\gd
    + \frac{D}{8} C^\mn_\gd
    +\frac{D}{8} A^\mn_\gd
    \labelt{n476d}
  \end{align}
\end{subequations}
These relations are derived by inserting the definitions of the operators from Eqs.~\eqreft{ope1} and manipulating the tensors $\eta^\mn$ and $p^\mu$.
In Eq.~\eqreft{n476d} it is necessary to show the indices since $A^\mn_\gd$ is antisymmetric.
\par
Multiplying $G$ with $\Delta$, we find:
\begin{subequations}
  \label{eqn:pro3}
  \begin{align}
    \hspace*{-1.5cm}
    \Delta^\mn_\ab G^\ab_\gd
    =\ 
    &\alpha_1
    I^\mn_\gd
    \labelt{pro3a}
    \\
    &+
    \Big(
    \alpha_1
    (-4+\frac{2}{\xi})
    +\alpha_2
    (-1+\frac{1}{2\xi})(D-2)
    -
    \alpha_3
    \frac{1}{\xi}
    \Big)
    T^\mn_\gd
    \labelt{pro3b}
    \\
    &+
    \Big(
    \alpha_1
    (2-2\frac{1}{\xi})
    +
    \alpha_2
    \frac{D-2}{4}(1-\frac{1}{\xi})
    +
    \alpha_3
    (-\frac{D-2}{2}+\frac{D}{4\xi})
    -
    (\alpha_4+\alpha_5)
    \frac{1}{2\xi}
    \Big)
    C^\mn_\gd
    \labelt{pro3c}
    \\
    &+
    \Big(
    \alpha_1
    2(-1+\frac{1}{\xi})
    +
    \alpha_4
    \frac{1}{\xi}
    \Big)
    \maJ^\mn_\gd
    \labelt{pro3d}
    \\
    &+
    \Big(
    \alpha_3
    (D-2)(\frac{1}{2}-\frac{1}{\xi})
    +
    \alpha_5
    \frac{1}{\xi}
    \Big)
    K^\mn_\gd
    \labelt{pro3e}
    \\
    &+
    \Big(
    \alpha_2
    \frac{D-2}{4}(-1+\frac{1}{\xi})
    +
    \alpha_3
    (-\frac{D-2}{2} + \frac{D-4}{4\xi})
    -
    (\alpha_4+\alpha_5)
    \frac{1}{2\xi}
    \Big)
    A^\mn_\gd
    \labelt{pro3f}
  \end{align}
\end{subequations}
This equation should be compared with Eq.~\eqreft{pro4} which is $\Delta G=I$ and we get 6 equations which determine $\alpha_n$.
They are not independent and $\alpha_n$ can e.g. be determined from the first 5 lines of Eq.~\eqreft{pro3}.
From the first line, that is Eq.~\eqreft{pro3a}, we get $\alpha_1=1$.
Then, from the fourth line, that is Eq.~\eqreft{pro3d}, we get:
\begin{align}
  2(-1+\frac{1}{\xi})
  +
  \alpha_4
  \frac{1}{\xi}
  =
  0
  \labelt{n478}
\end{align}
Hence $\alpha_4 = -2(1-\xi)$.
From Eqs.~\eqreft{pro3b}, \eqreft{pro3c} and~\eqreft{pro3e}, we can then determine $\alpha_2$, $\alpha_3$ and $\alpha_5$.
\par
In the end, the coefficients are determined to be:
\begin{subequations}
  \label{eqn:pro5}
  \begin{align}
    &\alpha_1 = 1
    \labelt{pro5a}
    \\
    &\alpha_2 = -\frac{4}{D-2}
    \\
    &\alpha_3 = 0
    \\
    &\alpha_4 = -2(1-\xi)
    \\
    &\alpha_5 = 0
  \end{align}
\end{subequations}
Inserting these values in Eq.~\eqreft{pro3} makes all terms but the first line disappear so that $\Delta G = I$.
\par
Finally, we can insert the coefficients from Eqs.~\eqreft{pro5} into the ansatz for $G^\mn_\ab$ in Eq.~\eqreft{pro6} to get the tensor structure of the graviton propagator:
\begin{equation}
  G
  =
  \maPi
  -2(1-\xi)\maJ
  \ .
  \labelt{n480}
\end{equation}
Here we have used $\maPi$ which is the inverse operator to $\maP$ and is given by:
\begin{equation}
  \maPi = I - \frac{4}{D-2} T
  \ .
  \labelt{n481}
\end{equation}
When $\xi=1$ the covariant operator reduces to $\maPi$ which is the well-known \dDo\ propagator.
In four space-time dimensions, that is $D=4$, it is even more simple, since then $\maPi=\maP$.
\par
In the end both $\Delta$ and $G$ are somewhat similar
\begin{subequations}
  \label{eqn:n482}
  \begin{align}
    &\Delta = \maP - 2(1-\frac{1}{\xi})\maP\maJ\maP
    \labelt{n482a}
    \\
    &G = \maPi
    -2(1-\xi)\maJ
    \ ,
  \end{align}
\end{subequations}
and it is the case that $G = \maPi\Delta\maPi$ when we also let $\xi\rightarrow\frac{1}{\xi}$.
\par
Let us check, that $G$ and $\Delta$ are really inverse operators to each other.
This is most easily done by splitting both $\Delta$ and $G$ into two parts according to their dependence on the covariant gauge parameter $\xi$.
In case of $\Delta$:
\begin{subequations}
  \label{eqn:pro2}
  \begin{align}
    &\Delta
    =
    \Delc + \frac{1}{\xi} \Delgf
    \labelt{pro2a}
    \\
    &\Delc
    =
    \mathcal{P}
    -2
    \mathcal{P}
    \maJ
    \mathcal{P}
    \labelt{pro2b}
    \\
    &\Delgf
    =
    2
    \mathcal{P}
    \maJ
    \mathcal{P}
    \labelt{pro2c}
  \end{align}
\end{subequations}
And in case of $G$:
\begin{subequations}
  \label{eqn:n485}
  \begin{align}
    &G = G_\tecl + \xi G_\tegf
    \labelt{n485a}
    \\
    &G_\tecl = \maPi - 2 \maJ
    \\
    &G_\tegf = 2 \maJ
  \end{align}
\end{subequations}
We will multiply $\Delta$ and $G$ together using these expressions.
The statement under Eq.~\eqreft{n482} translates to ${G_\tecl = \maPi\Delta_\tecl\maPi}$ and ${G_\tegf = \maPi\Delta_\tegf\maPi}$.
\par
First, we note the following property of $\Delta_\tecc$ and/or $Q^{\ab\ \gd\ \mn}$:
\begin{align}
  \Delc^\mn_\ab
  =
  Q^{\mn\ \sr}_\ab \frac{p_\sigma p_\rho}{p^2}
\end{align}
The $Q$-tensor was introduced in Eq.~\eqreft{n459}.
The indices $\ab$ on $Q^{\mn\ \sr}_\ab$ where lowered with the flat space metric and there should be no confusion since $Q^{\ab\ \gd\ \mn}$ is symmetric in its three ``double indices'' i.e. $\ab \leftrightarrow \gd$.
The non-trivial property of $\Delta_\tecl$ and/or $Q$ is:
\begin{align}
  p_\mu\Delc^\mn_\ab 
  =
  p_\mu Q^{\mn\ \sr}_\ab \frac{p_\sigma p_\rho}{p^2}
  = 0
  \labelt{pro1}
\end{align}
This is an identity which follows from the definitions of $\Delta_\tecl$ and/or $Q$.
It can be derived by inserting the definition of $\Delta_\tecl$
\begin{align}
  p_\mu\Delc^\mn_\ab
  =
  p_\mu\mathcal{P}^\mn_\ab
  -2
  p_\mu\mathcal{P}^{\mn}_{\rho\kappa}
  \frac{p^\rho p_\sigma}{p^2}
  \mathcal{P}_{\ab}^{\sigma\kappa}
  \ ,
  \labelt{n468}
\end{align}
and using the following relation:
\begin{align}
  2
  p_\mu\mathcal{P}^{\mn}_{\rho\kappa}
  \frac{p^\rho p_\sigma}{p^2}
  \mathcal{P}_{\ab}^{\sigma\kappa}
  =
  p_\mu \maP^\mn_\ab
  \ .
  \labelt{n469}
\end{align}
It can be verified by writing out the definition of $\maP$.
\par
Let us now multiply $\Delta$ and $G$ together using Eqs~\eqreft{pro2a} and~\eqreft{n485a}.
We get:
\begin{equation}
  \Delta G
  =
  \Delta_\tecl G_\tecl
  +
  \Delta_\tegf G_\tegf
  +
  \xi \Delta_\tecl G_\tegf
  +
  \frac{1}{\xi} \Delta_\tegf G_\tecl
  \ .
  \labelt{pro7}
\end{equation}
The last two terms, which depend on $\xi$, must necessarily vanish.
This is easily seen for $\Delta_\tecl G_\tegf$ since $G_\tegf$ contracts a momentum, $p^\mu$, to $\Delta_\tecl$ and from Eq.~\eqreft{pro1} we have that ${\Delta_\tecl}^\mn_\ab p_\mu$ vanishes.
\par
To check that $\Delta_\tegf G_\tecl$ vanishes and that the remaining terms reduce to the identity it is helpful to use the following identity:
\begin{align}
  \maJ \maP \maJ
  &=
  \maJ^2 - 2 \maJ T \maJ
  \nonumber{}
  \\
  &= \frac{1}{2} \maJ
  \labelt{jpj1}
  \ .
\end{align}
This can be checked using the relations given in Eqs.~\eqreft{n476}.
\par
Then, for $\Delta_\tegf G_\tecl$ we get
\begin{subequations}
  \label{eqn:n486}
  \begin{align}
    \Delta_\tegf G_\tecl
    &=
    2\maP\maJ\maP(\maPi-2\maJ)
    \labelt{n486a}
    \\
    &=
    2\maP(\maJ-2\maJ\maP\maJ)
    =
    0
    \ ,
  \end{align}
\end{subequations}
where we used the identity Eq.~\eqreft{jpj1} to conclude that the second line vanishes.
For the remaining $\xi$-independent terms of Eq.~\eqreft{pro7} we get:
\begin{subequations}
  \label{eqn:n487}
  \begin{align}
    \Delta_\tecl G_\tecl
    +
    \Delta_\tegf G_\tegf
    &=
    (\maP - 2\maP \maJ \maP)(\maPi - 2 \maJ)
    + 4 \maP \maJ \maP \maJ
    \labelt{n487a}
    \\
    &=
    I
    -4\maP\maJ
    +8\maP\maJ\maP\maJ
    =
    I
  \end{align}
\end{subequations}
Again, we used Eq.~\eqreft{jpj1} to conclude that the second line reduces to $I$.
\par
The major result of this section is the graviton propagator in covariant \dDo-type gauge, which we can now write down:
\begin{figure}[H]
  \centering
  \begin{tikzpicture}
    \node
    (nfd1)
        {
          \includegraphics{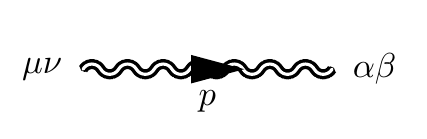}
        };
        \node
            [right=-0.5cm of nfd1]
            {
              $= \ \frac{i}{p^2+i\epsilon}G^\mn_\ab$
            };
  \end{tikzpicture}
\end{figure}
Here, $G^\mn_\ab$ is the tensor structure of the propagator and in components it is:
\begin{equation}
  G^\mn_\ab
  =
  I^\mn_\ab-\frac{1}{D-2}\eta^\mn\eta_\ab
  - 2(1-\xi)
  I^{\mn}_{\rho\kappa} \frac{\ p^\rho  p_\sigma}{p^2} I^{\kappa \sigma}_{\ab}
  \labelt{n488}
\end{equation}
We have not found this result in earlier literature.
It is a generalization of the simpler \dDo\ propagator to which the covariant propagator reduces for $\xi=1$.

\section{Matter Interaction: Scalar Graviton Vertices}
\labelx{sec:ScalarGraviton}
Our matter field is the scalar field and scalar-graviton vertices describe the interaction of gravitation and matter.
In the classical limit, the scalars can be interpreted as neutral, non-spinning point particles.
The scalar graviton vertex rules come from the expansion of the matter term of the action which from Eq.~\eqreft{nn28} is:
\begin{equation}
  S_\phi = \frac{1}{2} \int\dDx
  \sqrt{-g}
  \Big(
  g^\mn \phi_{,\mu} \phi_{,\nu} - m^2 \phi^2
  \Big)
  \labelt{ver5}
\end{equation}
In the computation of the \STM\ metric only the simplest vertex contributes, i.e. the $\phi^2 h$-vertex where a graviton interacts with a scalar.
In this section we will compute both this vertex and the next order vertex, $\phi^2 h^2$.
\par
The interactions of $\phi$ with $h_\mn$ come from the expansions of $g^\mn$ and $\sqg$.
These expansions were analyzed and computed to the necessary order in Sec.~\ref{sec:ExpansionsIn} in Eqs.~\eqreft{exp3} and~\eqreft{nn48}.
For example, the $\phi^2 h$ vertex gets one term when we replace $g^\mn$ by $-h^\mn$ and one term when we replace $\sqg$ by $\frac{1}{2}h^\mu_\mu$.
It becomes:
\begin{align}
  (S_\phi)_{\chno}
  =
  -\frac{1}{2}
  \int\dDx\ 
  h^\mn \phi_{,\mu}\phi_{,\nu}
  +\frac{1}{4}
  \int\dDx\ 
  h^\mu_\mu
  \big(
  \phi^{,\nu}\phi_{,\nu} - m^2 \phi^2
  \big)
  \labelt{ver4}
\end{align}
For the $\phi^2 h^2$ term we have to pick up quadratic terms in $h_\mn$ from $\sqg$ and $g^\mn$ and a mixed term from both $\sqg$ and $g^\mn$.
In the end, the scalar action when expanded to second order in $h_\mn$ becomes:
\begin{align}
  S_\phi
  \approx\
  \frac{1}{2} &\int\dDx
  \big(
  \eta^\mn\phi_{,\mu}\phi_{,\nu}
  -m^2\phi^2
  \big)
  \labelt{ver3}
  \\
  -\frac{1}{2}
  &\int\dDx
  \Big(
  h^\mn \phi_{,\mu}\phi_{,\nu}
  -\frac{1}{2} h^\mu_\mu
  \big(
  \phi^{,\nu}\phi_{,\nu} - m^2 \phi^2
  \big)
  \Big)
  \nonumber{}
  \\
  +\frac{1}{2}
  &\int\dDx
  \Big(
  h_\mn h^{\alpha\gamma} \maP^\mn_\ab \phi_{,\gamma}\phi^{,\beta}
  -\frac{1}{4} h_\mn h^\ab \maP^\mn_\ab
  \big(
  \phi^{,\nu}\phi_{,\nu} - m^2 \phi^2
  \big)
  \Big)
  \ .
  \nonumber{}
\end{align}
From this action we can read off the $\phi^2 h$ and the $\phi^2 h^2$ vertices as well as the scalar propagator.
Again, we should scale $h_\mn$ into $\hka_\mn=\frac{1}{\kappa}h_\mn$ before doing so.
\par
From the $\phi^2 h$ action we get a vertex rule proportional to
\begin{equation}
  \IVe_{\phi^2h}^{\mu\nu}(p,k,m)    =
  I_{\alpha\beta}^{\mu\nu} p^\alpha k^\beta - \frac{pk-m^2}{2} \eta^{\mu\nu}
  \ ,
  \labelt{ver1}
\end{equation}
where $p$ and $k$ are incoming and outgoing scalar momenta.
For example, $I^\mn_\ab p^\alpha k^\beta$ comes from the term $h^\mn \phi_{,\mu} \phi_{,\nu}$.
\par
The $\phi^2 h^2$ vertex rule can be written as $2i\kappa^2 \tau_{\phi^2h^2}^{\mu\nu\ \alpha\beta}(p,k,m)$ where again $p$ and $k$ are scalar momenta:
\begin{align}
  &\IVe_{\phi^2h^2}^{\mu\nu\ \alpha\beta}(p,k,m)    =
  \Big(
  I^{\mu\nu\ \eta}_{\hphantom{\mu\nu\ \eta}\lambda}   I^{\alpha\beta\ \lambda}_{\hphantom{\alpha\beta\ \lambda}\kappa}
  I^{\sigma\rho\ \kappa}_{\hphantom{\sigma\rho\ \kappa}\eta}
  - \frac{1}{4} \big(   \eta^{\mu\nu} I^{\alpha\beta\ \sigma\rho}
  + \eta^{\alpha\beta} I^{\mu\nu\ \sigma\rho}  \big)
  \Big)
  p_\sigma k_\rho - \frac{pk-m^2}{4} \mathcal{P}^{\mu\nu\ \alpha\beta}
  \labelt{ver2}
\end{align}
We can then write down Feynman rules for the scalar propagator and the $\kappa$ and $\kappa^2$ scalar graviton interactions.
\par
The scalar propagator is:
\begin{equation}
  \frac{i}{p^2-m^2+i\epsilon}
  \ .
\end{equation}
Here $p$ is the momentum of the scalar and $m$ its mass.
\par
The $\phi^2 h$ vertex is:
\begin{figure}[H]
  \centering
  \begin{tikzpicture}
    \node
    (nfd2)
        {
          \includegraphics{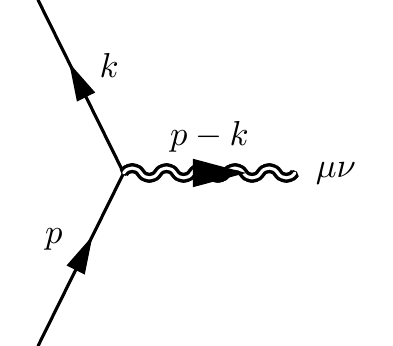}
        };
        \node
            [right=-0.5cm of nfd2]
            {
              $= \ -i\kappa \ \IVe_{\phi^2h}^{\mn}(p,k,m)$
            };
  \end{tikzpicture}
\end{figure}
The $\phi^2 h^2$ vertex is:
\begin{figure}[H]
  \centering
  \begin{tikzpicture}
    \node
    (nfd3)
        {
          \includegraphics{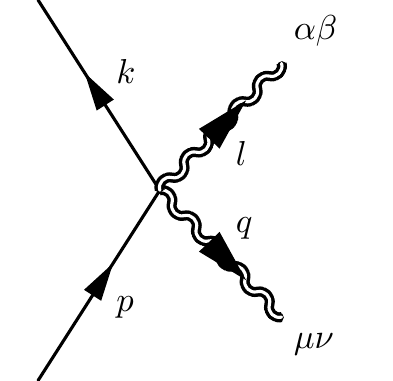}
        };
        \node
            [right=-0.5cm of nfd3]
            {
              $= \ 2i\kappa^2 \ \IVe_{\phi^2h^2}^{\mn\ \ab}(p,k,m)$
            };
  \end{tikzpicture}
\end{figure}
The tensors $\IVe_{\phi^2h}^{\mn}(p,k,m)$ and $\IVe_{\phi^2h^2}^{\mn\ \ab}(p,k,m)$ where given in Eqs.~\eqreft{ver1} and~\eqreft{ver2} respectively.
As already mentioned, only the $\phi^2 h$ vertex will contribute to the computation of the \STM\ metric.

\section{Gravitational Self-Interaction: n-Graviton Vertices}
\labelx{sec:GravitonSelf}
The graviton self-interaction vertices are important for the derivation of the \STM\ metric.
We will first derive an explicit formula for the three-graviton vertex which we will use in Ch.~\ref{sec:PerturbativeExpansion2} to compute the $(G_N)^2$ correction to the \STM\ metric in the \dDo-type gauge of Eq.~\eqreft{gau3}.
Then we will consider how the general n-graviton vertex can be written in terms of the functions $\Ghn{n}{\mn}$ and $\Hhn{n}{\mn}$ introduced in Sec.~\ref{sec:ExpansionOf}.
The general n-graviton vertex will be used in Sec.~\ref{sec:TheMetric} to relate the \STM\ metric to Feynman diagrams.
\par
The three-graviton vertex is conveniently written in terms of the $U$-tensor introduced in Sec.~\ref{sec:ActionIn}.
From Eq.~\eqreft{n432} we have the three-graviton term of the action written in terms of the $U$-tensor:
\begin{equation}
  S_{\chn{3}}
  = \frac{1}{\kappa^2} \int\ddx\  h_\mn U^{\mn\ \ab\rho\ \gd\sigma} h_{\ab,\rho}h_{\gd,\sigma}
\end{equation}
\par
As before, we introduce $\hka_\mn$ instead of $h_\mn$.
Also, this time we will explicitly transform to momentum space and follow the prescription discussed around Eq.~\eqreft{exa2}.
Hence:
\begin{equation}
  \hka_\mn = \int \dDp{l}
  \ e^{-ilx} \thka_\mn(l)
\end{equation}
Inserting this into the three-graviton action we get:
\begin{align}
  S_{\chn{3}}
  = -\kappa \int
  \dDp{l_\teon}
  \dDp{l_\tetw}
  \dDp{l_\tetr}\
  (2\pi)^D\delta^D(l_\teon+l_\tetw+l_\tetr)
  \thka_\mn^\teon
  U^{\mn\ \ab\rho\ \gd\sigma}
  \thka_{\ab}^\tetw
  \thka_{\gd}^\tetr
  l_{\tetw \rho}
  l_{\tetr \sigma}
  \ .
  \labelt{sel1}
\end{align}
Here we have e.g. written $\thka_\mn^\teon$ instead of $\thka_\mn(l_\teon)$.
\par
We now want to make Eq.~\eqreft{sel1} symmetric in $\thka_\mn^\teon$, $\thka_\mn^\tetw$ and $\thka_\mn^\tetr$.
We do this by adding three copies of Eq.~\eqreft{sel1} where we cyclically permute the graviton fields.
We get:
\begin{align}
  S_{\chn{3}}
  = -\frac{\kappa}{3} \int
  \dDp{l_\teon}
  &\dDp{l_\tetw}
  \dDp{l_\tetr}\
  (2\pi)^D\delta^D(l_\teon+l_\tetw+l_\tetr)
  \Big(
  U^{\mn\ \ab\rho\ \gd\sigma}
  l_{\tetw \rho}
  l_{\tetr \sigma}
  \nonumber{}
  \\
  &+
  U^{\ab\ \gd\rho\ \mn\sigma}
  l_{\tetr \rho}
  l_{\teon \sigma}
  +
  U^{\gd\ \mn\rho\ \ab\sigma}
  l_{\teon \rho}
  l_{\tetw \sigma}
  \Big)
  \thka_\mn^\teon
  \thka_\ab^\tetw
  \thka_\gd^\tetr
  \ .
  \labelt{sel2}
\end{align}
Now, we can read off the vertex rule from the integrand where we ignore the $\delta$-function and the $(2\pi)^D$ factors.
\par
For the three-graviton vertex, we then get:
\begin{figure}[H]
  \centering
  \begin{tikzpicture}
    \node
    (nfd8)
        {
          \includegraphics{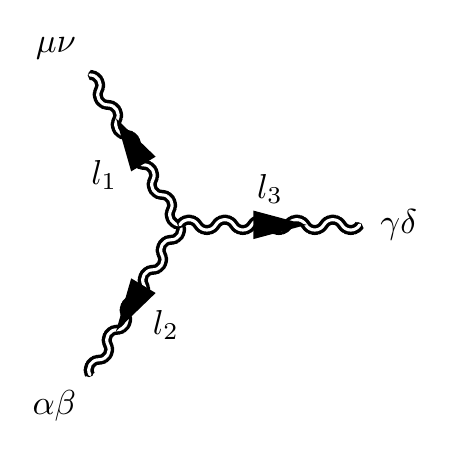}
        };
        \node
            [right=-.5cm of nfd8]
            {
              $=2i\kappa \IVe^{\mn\ \ab\ \gd}_{h^3} (l_1,l_2,l_3)$
            };
  \end{tikzpicture}
\end{figure}
Where $\IVe^{\mn\ \ab\ \gd}(l_\teon,l_\tetw,l_\tetr)$ can be written in terms of $U^{\mn\ \ab\rho\ \gd\sigma}$ as:
\begin{equation}
  \IVe^{\ab\ \gd\ \rs}_{h^3} (l_{(1)},l_{(2)},l_{(3)})
  =
  -
  \Big(
  U^{\mn\ \ab\rho\ \gd\sigma}
  l_{\tetw \rho}
  l_{\tetr \sigma}
  +
  U^{\ab\ \gd\rho\ \mn\sigma}
  l_{\tetr \rho}
  l_{\teon \sigma}
  +
  U^{\gd\ \mn\rho\ \ab\sigma}
  l_{\teon \rho}
  l_{\tetw \sigma}
  \Big)
  \labelt{be10}
\end{equation}
This, then, is the result for the three-graviton vertex.
The $U$-tensor where defined in Eqs.~\eqreft{ute1}, \eqreft{n430}, and~\eqreft{n431}.
\par
The general n-graviton vertex will now be analyzed.
Here, we use the expression for the action from Eqs.~\eqreft{n437} and~\eqreft{n438} which is:
\begin{align}
  S_\teEH
  +    S_\tegf
  =
  -
  \int \dDx
  \
  h_\mn
  \sum_{n=1..\infty} \oov{(n+1)}
  \Big(
  \Ghn{n}{\mn}(h,h,\ellipsis,h)
  +
  \frac{1}{\xi}
  \Hhn{n}{\mn}(h,h,\ellipsis,h)
  \Big)
  \labelt{sel3}
\end{align}
From Eq.~\eqreft{n440} this expression was defined to be symmetric in all factors of $h_\mn$.
We need, however, to transform to momentum space.
This requires a small development of the notation.
\par
We need to find the Fourier transform of e.g. $\Ghn{n}{\mn}(h,h,\ellipsis,h)$.
We will focus on $n=3$ from which the general case can be inferred.
We insert the definition of $h_\mn$ in terms of $\tilde h_\mn$ into $\Ghn{3}{\mn}$:
\begin{align}
  \Ghn{3}{\mn}(h_\mn,h_\mn,h_\mn)
  =
  \Ghn{3}{\mn}
  \Big(
  \int \dDp{l_\teon} e^{-ixl_\teon} \thmn_\mn^\teon
  ,
  \int \dDp{l_\tetw} e^{-ixl_\tetw} \thmn_\mn^\tetw
  ,
  \int \dDp{l_\tetr} e^{-ixl_\tetr} \thmn_\mn^\tetr
  \Big)
  \labelt{ber9}
\end{align}
Where, again, $\thmn_\mn^\teon = \thmn_\mn(l_\teon)$.
Since $\Ghn{3}{\mn}$ is linear in its arguments the integrals can be pulled out.
We get:
\begin{align}
  \Ghn{3}{\mn}(h_\mn,h_\mn,h_\mn)
  =
  \int
  \dDp{l_\teon}
  \dDp{l_\tetw}
  \dDp{l_\tetr}
  \Ghn{3}{\mn}
  \Big(
  e^{-ixl_\teon} \thmn_\mn^\teon
  ,
  e^{-ixl_\tetw} \thmn_\mn^\tetw
  ,
  e^{-ixl_\tetr} \thmn_\mn^\tetr
  \Big)
  \labelt{ber8}
\end{align}
If $\Ghn{3}{\mn}$ did not depend on the partial derivatives of $h_\mn$ we could also pull the exponential factors out.
However, we can still do so, if we replace partial derivatives, $\partial_\mu$ by $-i l_\mu$ where $l_\mu$ is the momentum of the graviton which was differentiated.
We will use a tilde on $\Ghn{n}{\mn}$ to denote that partial derivatives was replaced by $-il_\mu$.
Since there are always two derivatives in $\Ghn{n}{\mn}$ the two factors of $i$ will always cancel with each other and introduce a sign.
\par
With this notation we get:
\begin{align}
  \Ghn{3}{\mn}(h_\mn,h_\mn,h_\mn)
  =
  \int
  \dDp{l_\teon}
  \dDp{l_\tetw}
  \dDp{l_\tetr}
  e^{-ix(l_\teon+l_\tetw+l_\tetr)} 
  \tGhn{3}{\mn}
  \Big(
  \thmn_\mn^\teon
  ,
  \thmn_\mn^\tetw
  ,
  \thmn_\mn^\tetr
  \Big)
  \labelt{ber7}
\end{align}
And then for the Fourier transform:
\begin{align}
  \hspace*{-1cm}
  \int \dDx \ e^{iqx}\ 
  \Ghn{3}{\mn}(h_\mn,h_\mn,h_\mn)
  =
  \int
  \dDp{l_\teon}
  \dDp{l_\tetw}
  \dDp{l_\tetr}
  (2\pi)^D \delta^D(l_\teon+l_\tetw+l_\tetr)
  \tGhn{3}{\mn}
  \Big(
  \thmn_\mn^\teon
  ,
  \thmn_\mn^\tetw
  ,
  \thmn_\mn^\tetr
  \Big)
  \labelt{leq3}
\end{align}
The general case can then be inferred from this example.
\par
Thus, after going to momentum space, the integrand of $\frac{2}{\kappa^2}(S_\teEH+S_\tegf)$ from Eq.~\eqreft{sel3} becomes:
\begin{equation}
  \sum_{\noi}
  -\frac{2\kappa^{n-1}}{n+1} \thka_\mn^\teze
  \Big(
  \tGhn{n}{\mn}
  (\thka_\mn^\teon,..,\thka_\mn^\tenn)
  +
  \frac{1}{\xi}
  \tHhn{n}{\mn}
  (\thka_\mn^\teon,..,\thka_\mn^\tenn)
  \Big)
  \labelt{sel4}
\end{equation}
Here the momentum conserving $\delta$-function and factors of $2\pi$ was ignored in the integrand.
Due to the properties of $\Ghn{n}{\mn}$ and $\Hhn{n}{\mn}$ this expression is symmetric in all the factors of $\thka_\mn^{(i)}$ when integrated.
\par
From Eq.~\eqreft{sel4} we can read off the n-graviton self-interaction vertex by multiplying by $\big(i(n+1)!\big)$.
The ${(n+1)}$-graviton vertex rule becomes:
\begin{figure}[H]
  \centering
  \begin{tikzpicture}
    \node
    (nfd6)
        {
          \includegraphics{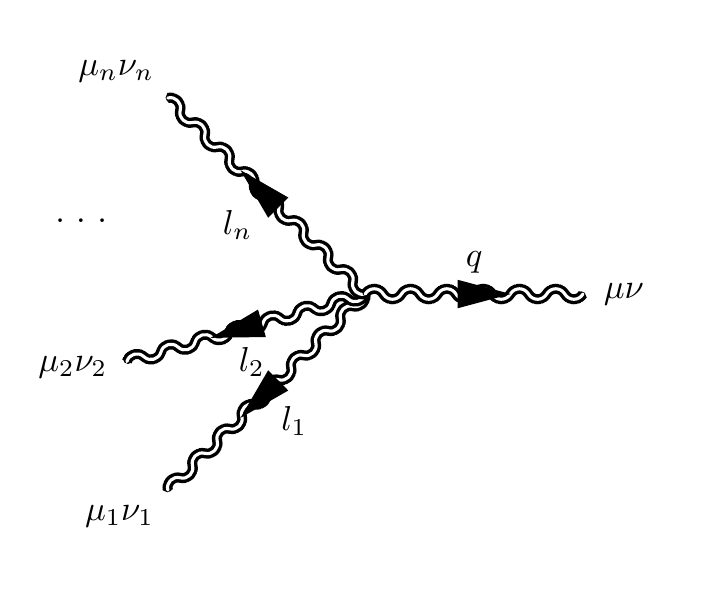}
        };
        \node
            [right=-.5cm of nfd6]
            {
              $=i\ n!\ \kappa^{n-1}\ $
              $\IVe_{h^{n+1}}^{\mn\ \mu_1\nu_1\ \mu_2\nu_2\ ...\ \mu_n\nu_n}(q,l_1,l_2,...,l_n)$
            };
  \end{tikzpicture}
\end{figure}
The tensor $\IVe_{h^{n+1}}^{\mn\ \mu_1\nu_1\ \mu_2\nu_2\ ...\ \mu_n\nu_n}(q,l_1,l_2,...,l_n)$ in terms of $\Ghn{n}{\mn}$ and $\Hhn{n}{\mn}$ becomes:
\begin{align}
  \IVe_{h^{n+1}}^{\mn\ \mu_1\nu_1\ \mu_2\nu_2\ ...\ \mu_n\nu_n}(q,l_\teon,l_\tetw,...,l_\tenn)
  \thmn_{\mu_1\nu_1}^\teon
  \thmn_{\mu_2\nu_2}^\tetw
  \ellipsis
  \thmn_{\mu_n\nu_n}^\tenn
  =
  -2
  \Big(
  &\tGhn{n}{\mn}
  \big(
  \thmn_{\mu_1\nu_1}^\teon, 
  \thmn_{\mu_2\nu_2}^\tetw, \ellipsis,
  \thmn_{\mu_n\nu_n}^\tenn
  \big)
  \nonumber{}
  \\
  +\frac{1}{\xi}
  &\tHhn{n}{\mn}
  \big(
  \thmn_{\mu_1\nu_1}^\teon, 
  \thmn_{\mu_2\nu_2}^\tetw, \ellipsis,
  \thmn_{\mu_n\nu_n}^\tenn
  \big)
  \Big)
  \labelt{ber6}
\end{align}
This, then, is the $(n+1)$-graviton vertex expressed in terms of the Einstein tensor and the analogous tensor $H^\mn$ through the tensors $\Gpz^\mn$ and $\Hpz^\mn$.
This result and the three-graviton vertex in terms of the $U$-tensor are the major results of this section.
As it stands in Eq.~\eqreft{ber6} it is not obvious that the vertex rule is symmetric in the $n+1$ graviton indices and/or momenta since e.g. it does not depend explicitly on $q^\mu$.
It is, however, symmetric in all indices and momenta due to momentum conservation which corresponds to the partial integrations used to make the original integral symmetric in Eq.~\eqreft{n440}.
When this vertex rule is used to derive the \STM\ metric, the asymmetrical role of the $q^\mu$-momentum will be an external graviton and the other momenta will be gravitons contracted to a scalar line.
\chapter{\STM\ Metric from Amplitudes}
\labelx{sec:STM}
In this chapter we will show that the \STM\ metric can be derived from the three-point vertex function of a massive scalar interacting with a graviton.
First, in Sec.~\ref{sec:GaugeFixed} we will analyze the equations of motion in covariant gauge in detail.
Then, in Sec.~\ref{sec:PerturbativeExpansion1} we will show how the classical equations of motion can be solved perturbatively in a similar fashion as Feynman expansions.
Finally in Sec.~\ref{sec:TheMetric} we will relate the metric to the three-point vertex function.
\section{Gauge-Fixed Action and Equations of Motion}
\labelx{sec:GaugeFixed}
The gauge-fixed action was discussed in Sec.~\ref{sec:GaugeFixing} and from Eq.~\eqreft{act3} it is:
\begin{align}
  S
  &=
  \frac{2}{\kappa^2}(S_{EH}+S_{gf}) + \int \dDx \sqrt{-g}\ \mathcal{L}_\phi
  \labelt{act4}
  \\
  &=
  \int \dDx \sqrt{-g}  
  \bigg(
  \frac{2R}{\kappa^2} \
  +\ 
  \mathcal{L}_\phi
  \bigg)
  +
  \int \dDx
  \frac{1}{\kappa^2 \xi} \eta^{\sigma \rho} G_\sigma G_\rho
  \nonumber
\end{align}
We want to find the classical equations of motion derived from the action by the variational principle $\delta S=0$.
These, we will refer to as the \cgE\ equations which will be similar to the Einstein equations only with a gauge breaking term.
\par
The variation of the general covariant terms of $S_\teEH$ produces the terms from the Einstein field equations and was given in Eqs.~\eqreft{ein1} and~\eqreft{n214}.
In Eq.~\eqreft{hmn1} the tensor $H^\mn$ was defined to be analogous to the Einstein tensor only derived from $S_\tegf$ instead of $S_\teEH$:
\begin{align}
  \delta S_\tegf =
  \frac{1}{\xi}
  \int
  \dDx
  \
  \eta^\rs
  G_\rho
  \delta G_\sigma
  =
  -
  \frac{1}{\xi}
  \int \dDx \
  \sqrt{-g}
  H^\mn \delta h_\mn
  \ .
  \labelt{dgs2}
\end{align}
We will now find $H^\mn$ for the generalized gauge function which was introduced in Eq.~\eqreft{gau3} and which we reprint here:
\begin{equation}
  G_\sigma =
  (1-\alpha) \ \partial_\mu (h^\mu_\sigma - \frac{1}{2} \eta^\mu_\sigma h_\nu^\nu)
  + \alpha\  g^\mn \Gamma_{\sigma\mn}
  \ .
  \labelt{gau2}
\end{equation}
The gauge function can be rewritten in a simple form using the tensors $\hat h^\mn = g^\mn-\eta^\mn$ and $\Gamma^{\sigma\ab}_{\rho\mn}$ introduced in Eqs.~\eqreft{nn45} and~\eqreft{gam2} respectively:
\begin{subequations}
  \begin{align}
    G_\sigma
    &=
    (1-\alpha) \  \eta^{\mn} \Gamma^{\rho\ab}_{\sigma\mn} h_{\ab,\rho}
    + \alpha\  g^\mn \Gamma^{\rho\ab}_{\sigma\mn} h_{\ab,\rho}
    \ .
    \labelt{gau4}
    \\
    &=
    (\eta^\mn + \alpha(g^\mn-\eta^\mn) )
    \Gamma^{\rho\ab}_{\sigma\mn} h_{\ab,\rho}
    \\
    &=
    (\eta^\mn + \alpha \hhatu{\mn})
    \Gamma^{\rho\ab}_{\sigma\mn} h_{\ab,\rho}
    \ .
    \labelt{gac1c}
  \end{align}
\end{subequations}
In the third line, we have the resulting expression for $G_\sigma$.
Here it is evident that $\alpha$ scales the non-linear terms of $G_\sigma$ since these come from $\hat h^\mn$.
\par
With Eq.~\eqreft{gac1c} it is straightforward to find $\delta G_\sigma$.
We use that
\begin{equation}
  \delta \hat h^\mn = \delta(g^\mn)=-g^{\mu\alpha}g^{\nu\beta}\delta g_\ab
  \ ,
\end{equation}
and that, naturally, $\delta g_\mn = \delta h_\mn$.
For $\delta G_\sigma$ we then get:
\begin{equation}
  \delta G_\sigma
  =
  (\eta^\mn + \alpha \hhatu{\mn})
  \Gamma^{\rho\ab}_{\sigma\mn} \delta h_{\ab,\rho}
  -
  \alpha
  \Gamma^{\rho\ab}_{\sigma\mn}  h_{\ab,\rho}
  g^{\mu\gamma}g^{\nu\delta} \delta h_{\gd}
  \ .
  \labelt{dgs1}
\end{equation}
The partial derivative on $\delta h_\mn$ in the first term makes it necessary to do a partial integration.
\par
We insert $G_\sigma$ and $\delta G_\sigma$ into $\delta S_\tegf$ from Eq.~\eqreft{dgs2} and get:
\begin{subequations}
  \begin{align}
    \delta S_{gf}
    &=
    \frac{1}{\xi}
    \int
    \dDx
    \
    \eta^\rs
    G_\rho
    \delta G_\sigma
    \\
    &=
    \frac{1}{\xi}
    \int \dDx \
    G^\sigma
    \Big(
    (\eta^\mn + \alpha \hhatu{\mn})
    \Gamma^{\rho\gd}_{\sigma\mn} \delta h_{\gd,\rho}
    -
    \alpha
    \Gamma^{\rho\ab}_{\sigma\mn}  h_{\ab,\rho}
    g^{\mu\gamma}g^{\nu\delta} \delta h_{\gd}
    \Big)
    \\
    &=
    -\frac{1}{\xi}
    \int \dDx \
    \Big(
    \Gamma^{\rho\gd}_{\sigma\mn}
    \partial_\rho
    \big(
    G^\sigma
    (\eta^\mn + \alpha \hhatu{\mn})
    \big)
    +
    \alpha
    G^\sigma
    \Gamma^{\rho\ab}_{\sigma\mn}  h_{\ab,\rho}
    g^{\mu\gamma}g^{\nu\delta}
    \Big)
    \delta h_{\gd}
  \end{align}
\end{subequations}
Here, we have treated $G_\sigma$ as a Lorentz tensor and raised its index with $\eta^\mn$.
Comparing with the expression for $S_\tegf$ in Eq.~\eqreft{dgs2} we can read off $H^\mn$ and we get:
\begin{align}
  \sqrt{-g} H^\mn
  &=
  \Hpz^\mn
  \nonumber{}
  \\
  &= \alpha G^\rho \Gamma_{\rho\ab} g^{\alpha\mu}g^{\beta\nu}
  +
  \Gamma^{\sigma\mn}_{\rho\ab}
  \partial_\sigma
  \Big(
  G^\rho
  \big(
  \eta^{\ab} + \alpha\hhatu{\ab}
  \big)
  \Big)
  \ .
  \labelt{hte1}
\end{align}
We see, that in some sense $\Hpz^\mn$ is simpler than $H^\mn$.
This is in contrast to the Einstein tensor, where $G^\mn$ is defined without a $\sqg$ while $\Gpz^\mn = \sqg G^\mn$.
\par
The tensor $H^\mn$ enters the Einstein equations as a gauge breaking term added to the Einstein tensor:
\begin{equation}
  G^{\mn} + \frac{1}{\xi} H^{\mn} = -\frac{\ \kappa^2}{4} T^{\mn}
  \ ,
  \labelt{ein3}
\end{equation}
This is the \cgE\ equation which follows from $\delta S=0$ and which we will now analyze.
\par
We expect the \cgE\ equation to describe classical general relativity.
Hence, we expect the metric to satisfy the Einstein field equations:
\begin{equation}
  G^\mn = -\frac{\ \kappa^2}{4} T^\mn
  \ .
\end{equation}
If the metric satisfies this equation, then $H^\mn$ is forced to disappear.
The conclusion is that, if we expect to get a solution from classical general relativity we must have the additional equation $H^\mn=0$.
\par
Let us see, if we can deduce that $H^\mn$ vanishes directly from the \cgE\ equation.
We find that the metric obeys an additional simple equation which follows from the \cgE\ equation by taking the covariant derivative on each side.
Since both $D_\mu G^\mn =0$ and $D_\mu T^{\mn}=0$ we get that Eq.~\eqreft{ein3} implies:
\begin{equation}
  D_\mu H^\mn = 0
  \ .
  \labelt{gau5}
\end{equation}
We interpret this additional equation as our gauge (coordinate) condition.
We will now show that in a perturbative expansion we can conclude from Eq.~\eqreft{gau5} that also $H^\mn=0$.
This is done indirectly by deducing that $G_\sigma$ must vanish, from which it follows that $H^\mn=0$.
We insert the definition of the covariant derivative from e.g. Weinberg~\cite{Weinberg:1972kfs} into Eq.~\eqreft{gau5}:
\begin{equation}
  \sqrt{-g} D_\mu H^\mn
  =
  \partial_\mu
  \big(
  \Hpz^\mn
  \big)
  +
  \Gamma^\nu_\ab \Hpz^\ab
\end{equation}
We want to expand this equation perturbatively in $G_N$.
We then know that each term in the expansion must disappear by itself.
We assume that the graviton field is at least of linear order in $G_N$ from which it follows that both $\Gamma^\nu_\ab$ and $\Hpz^\mn$ are so too.
To linear order in $G_N$ we then get:
\begin{equation}
  \sqrt{-g} D_\mu H^\mn
  \approx
  \partial_\mu \Hpz^\mn
  \ .
\end{equation}
For $\Hpz^\mn$ we find to linear order:
\begin{equation}
  H^\mn
  \approx
  \Gamma^{\sigma\mn}_{\rho\ab} \eta^\ab \partial_\sigma G^\rho
  \ .
\end{equation}
To linear order in $G_N$ we then conclude that Eq.~\eqreft{gau5} reduces to:
\begin{equation}
  D_\mu H^\mn
  \approx
  \Gamma^{\sigma\mn}_{\rho\ab} \eta^\ab \partial_\sigma \partial_\mu G^\rho
  \ .
\end{equation}
This can be simplified further by inserting the definition of $\Gamma^{\sigma\mn}_{\rho\ab}$.
We then get:
\begin{equation}
  D_\mu H^\mn
  \approx
  \frac{1}{2} \partial_\sigma\partial^\sigma G^\nu
\end{equation}
To first order in $G_N$ we get that $\partial^2 G_\sigma$ disappears.
With the boundary condition that $G_\sigma$ goes to zero at infinity we conclude that $G_\sigma=0$ to first order in $G_N$.
\par
This means that $G_\sigma$ is at least second order in $G_N$ which implies the same for $\Hpz^\mn$.
We can then follow the same line of reasoning and conclude that to second order in $G_N$ we also have $\partial^2 G_\sigma=0$.
By induction we conclude that the equation $\partial^2G_\sigma=0$ is satisfied to all orders in $G_N$ and that then $G_\sigma=0$.
This implies that $H^\mn$ also disappears.
\par
The conclusion is that the \cgE\ equation is equivalent to the Einstein field equations with the gauge/coordinate condition $G_\sigma=0$.
Hence our equations of motion in the classical limit are the Einstein field equations,
\begin{align}
  G^\mn = -\frac{\kappa^2}{4} T^\mn
  \ ,
\end{align}
and the gauge condition:
\begin{align}
  G_\sigma = 0
  \ .
  \labelt{dgs3}
\end{align}
They are combined into a single equation in the \cgE\ equation:
\begin{align}
  G^\mn + \frac{1}{\xi} H^\mn = -\frac{\kappa^2}{4} T^\mn
  \ .
\end{align}
The fact that the \cgE\ equation implies the gauge condition in Eq.~\eqreft{dgs3} is a significant result of this chapter.
\par
We will now rewrite this equation in a similar way as Weinberg~\cite{Weinberg:1972kfs} splitting it into linear and non-linear parts in $h_\mn$.
This will be the starting point of the perturbative expansion of the \cgE\ equation in the next section.
We separate $G^\mn$ and $H^\mn$ into linear and non-linear parts in $h_\mn$.
Thus as in Sec.~\ref{sec:ExpansionsIn} we have
\begin{align}
  G^\mn
  &=
  \sum_{n=1..\infty} (G^\mn)_{\chn{n}}
  \\
  &=
  G^\mn_{\chno} + G^\mn_\tenl
  \ ,
\end{align}
and similarly for $H^\mn$.
Here $G_\tenl^\mn$ and $H_\tenl^\mn$ are the non-linear parts of the expansions of $G^\mn$ and $H^\mn$ in $h_\mn$.
We then rewrite the \cgE\ equation as:
\begin{align}
  G^\mn_\chno + \frac{1}{\xi} H^\mn_\chno = -\frac{\kappa^2}{4} T^\mn-G^\mn_\tenl-\frac{1}{\xi}H^\mn_\tenl
  \labelt{cge1}
\end{align}
The part $G^\mn_\tenl$ can be interpreted as gravitational energy-momentum.
We can then introduce the total energy-momentum tensor $\tau^\mn$:
\begin{align}
  \tau^\mn = T^\mn + \frac{4}{\kappa^2} G^\mn_\tenl
  \ .
  \labelt{dgs7}
\end{align}
We will now see that this tensor is locally conserved.
\par
In terms of the energy-momentum tensor, $\tau^\mn$, the Einstein field equations become:
\begin{align}
  G^\mn_\chno = -\frac{\kappa^2}{4} \tau^\mn
  \ .
\end{align}
Inserting the formula for the linear term of $G^\mn$ from Eq.~\eqreft{n457} we get:
\begin{align}
  Q^{\mn\ \ab\ \gd} h_{\ab,\gd} = -\frac{\kappa^2}{2} \tau^\mn
\end{align}
Here, the $Q$-tensor was defined in Eq.~\eqreft{n459}.
An important property of the $Q$-tensor was discussed around Eq.~\eqreft{pro1}, namely that $Q^{\mn\ \ab\ \gd} p_\beta p_\gamma p_\delta=0$ where $p^\mu$ is any space-time vector.
Recall also, that $Q^{\mn\ \ab\ \gd}$ is symmetric in its three pairs of indices e.g. $\mn\leftrightarrow \ab$.
Using these properties we get that when the fields obey the equations of motion the energy-momentum tensor is conserved:
\begin{align}
  0 = 
  \partial_\mu Q^{\mn\ \ab\ \gd} h_{\ab,\gd}
  =
  -\frac{\kappa^2}{2} \partial_\mu \tau^\mn
\end{align}
Thus $\tau^\mn$ is locally conserved on the equations of motion.
\par
An analogous equation can be derived for $H^\mn_\tenl$.
When the fields obey the \cgE\ equation we have that $H^\mn$ vanishes.
We then get:
\begin{align}
  H^\mn_\chno = - H^\mn_\tenl
  \labelt{dgs4}
\end{align}
That is, the linear part of $H^\mn$ equals the non-linear part with a negative sign.
The linear part of $H^\mn$ is:
\begin{equation}
  H^\mn_\chno = \maP^\mn_{\sigma\kappa} \maP^{\rho\kappa}_\gd \partial^\sigma \partial_\rho h^\gd
  \ .
\end{equation}
This can e.g. be derived from the Eqs.~\eqreft{gac1c} and~\eqreft{hte1} from this section.
This should be compared to the operator $\Delta_\tegf$ from Eq.~\eqreft{pro2c}.
Similarly to Eq.~\eqreft{n486a} which is ${\Delta_\tegf G_\tecl=0}$, we get that:
\begin{equation}
  {G_\tecl}^\mn_\ab H^\ab_\chno = 0
\end{equation}
Using this and the equation of motion Eq.~\eqreft{dgs4} we conclude that the non-linear part of $H^\mn$ satisfies:
\begin{equation}
  {G_{(c)}}^\mn_\ab H^\ab_\tenl = 0
  \labelt{dgs5}
\end{equation}
This is a non-trivial equation for $H^\mn_\tenl$.
Recall, that the $G^\mn_\ab$-tensors are the tensor structure of the graviton propagator.
\par
We summarize Eq.~\eqreft{dgs5} and the conservation law of $\tau^\mn$ in the equations:
\begin{subequations}
  \label{eqn:eom1}
  \begin{align}
    &{G_\tegf}_\ab^\mn \tau^\ab = 0
    \ ,
    \labelt{eom2}
    \\
    &{G_\tecl}^\mn_\ab H^\ab_\tenl = 0
    \ .
    \labelt{eom3}
  \end{align}
\end{subequations}
These are consequences of the analysis of the \cgE\ equation.
In the next section these equations will be used to derive an expression for the metric independent of the covariant parameter $\xi$.
\par
In the final part of this section we will discuss the equations of motion in terms of the tensors $\Gpz^\mn$ and $\Hpz^\mn$ introduced in Eqs.~\eqreft{n449} and~\eqreft{n452}.
This analysis is useful since the n-graviton vertices are easily related to $\Gpz^\mn$.
Multiplying the \cgE\ equation by $\sqg$ we get:
\begin{align}
  \Gpz^\mn + \frac{1}{\xi} \Hpz^\mn = -\frac{\kappa^2}{4} \sqg \ T^\mn
  \ .
\end{align}
And we have the two simpler equations:
\begin{subequations}
  \begin{align}
    &\Gpz^\mn = -\frac{\kappa^2}{4} \sqg\ T^\mn
    \\
    &\Hpz^\mn = 0
  \end{align}
\end{subequations}
Note that the linear terms of $\Gpz^\mn$ and $G^\mn$ are equal and similarly for $\Hpz^\mn$:
\begin{subequations}
  \label{eqn:dgs6}
  \begin{align}
    &\Gpz^\mn_\chno = G^\mn_\chno
    \\
    &\Hpz^\mn_\chno = H^\mn_\chno
    \labelt{dgs6b}
  \end{align}
\end{subequations}
Thus, if we split the equations of motion in terms of $\Gpz^\mn$ and $\Hpz^\mn$ into linear and non-linear parts as in Eq.~\eqreft{cge1} we get simple relations between the non-linear parts.
\par
For example, we have
\begin{align}
  \Hpz^\mn_\chno = -\Hpz^\mn_\tenl
  \ ,
\end{align}
and due to Eqs.~\eqreft{dgs6} and Eq.~\eqreft{dgs4} we get:
\begin{align}
  \Hpz_\tenl^\mn = H_\tenl^\mn
  \labelt{hej3}
\end{align}
Similarly, using the Einstein field equations, we get an expression for the energy-momentum tensor, $\tau^\mn$, in terms of $\Gpz^\mn$:
\begin{align}
  \tau^\mn = \sqg\ T^\mn + \frac{4}{\kappa^2} \Gpz^\mn_\tenl
  \ .
\end{align}
This should be compared to Eq.~\eqreft{dgs7} which relates $\tau^\mn$ to $G^\mn$.
\par
Let us briefly remark on a special property of \dDo-gauge before moving on.
In this gauge $\alpha$ is zero and we see that $\Hpz^\mn$ from Eq.~\eqreft{hte1} is linear in the graviton field.
Hence, in this gauge $\Hpz_\tenl^\mn$ disappears and according to Eq.~\eqreft{hej3} so does $H^\mn_\tenl$.
From the quantum field theoretic point of view this is explained by the fact that $G_\sigma$ is linear in $h_\mn$ in this gauge and then the gauge dependence of the self-interaction vertices disappear.
In this case ``Landau gauge'' is possible where we let $\xi\rightarrow0$.
We can summarize this special property of \dDo-gauge by the property that in this gauge, the graviton field couples directly to the total energy-momentum tensor, $\tau^\mn$.
\section{Perturbative Expansion of the Classical Equations of Motion}
\labelx{sec:PerturbativeExpansion1}
We will now turn to the perturbative expansion of the \CEEq.
Our starting point is Eq.~\eqreft{cge1} where, after inserting the expressions for $G_\chno^\mn$ and $H_\chno^\mn$ in terms of $h_\mn$, we get:
\begin{align}
  \Big(
  \eta^\rs \mathcal{P}^{\mn\ab}
  -2(1-\oov{\xi}) \mathcal{P}^{\mn\rho\phi} \eta_\pe \mathcal{P}^{\ab\sigma\epsilon}
  \Big)
  h_{\ab,\rs}
  =
  -\frac{\kappa^2}{2} \tau^\mn
  -
  2\frac{1}{\xi}H^\mn_\tenl
  \labelt{peo1}
\end{align}
Again, in \dDo\ gauge only $\tau^\mn$ would be on the right hand side since in this gauge $H^\mn_\tenl$ disappears.
It is advantageous to transform to momentum space.
This also makes the similarity with the Feynman diagram expansion clear.
In momentum space, Eq.~\eqreft{peo1} becomes:
\begin{align}
  q^2
  \Delta^{\mn\ab}
  \tilde h_{\ab}
  =
  \frac{\kappa^2}{2} \tilde\tau^\mn
  +
  2\frac{1}{\xi} \tilde H^\mn_\tenl
  \labelt{cge2}
  \ .
\end{align}
Here, $\Delta^{\mn\ab}$ is the operator from Eq.~\eqreft{del1} which is the quadratic operator in the gauge-fixed Einstein-Hilbert action.
The momentum dependence on $q^\mu$ in Eq.~\eqreft{cge2} is hidden, although all objects including $\Delta^{\mn\ab}$ depend on the momentum variable, $q^\mu$.
Note that our perturbative expansion is in the long-range limit.
Hence, the momentum variable $q^\mu_\bot$ should be taken as small analogously to the classical limit of the Feynman diagrams.
\par
In Sec.~\ref{sec:GravitonPropagator}, the inverse to this operator was analyzed which is the graviton propagator in covariant gauge, $G^\ab_\mn$.
Multiplying with this operator on each side of Eq.~\eqreft{cge2} we get:
\begin{align}
  \tilde h_{\ab}
  =
  \frac{  G_{\ab\mn}}{q^2}
  \Big(
  \frac{\kappa^2}{2} \tilde\tau^\mn
  +
  \frac{2}{\xi} \tilde H^\mn_\tenl
  \Big)
  \ .
  \labelt{lin1}
\end{align}
Using Eqs.~\eqreft{eom1} we can rewrite it in a form independent of $\xi$:
\begin{subequations}
  \label{eqn:lin2}
  \begin{align}
    \tilde h_{\ab}
    &=
    \frac{1}{q^2}
    \Big(
    \frac{\kappa^2}{2}
         {G_\tecl}_{\ab \mn} \tilde\tau^\mn
         +
         2{G_\tegf}_{\ab \mn} \tilde H^\mn_\tenl
         \Big)
         \labelt{lin2a}
         \\
         &=
         \frac{\maPi_{\ab\mn}}{q^2}
         \Big(
         \frac{\kappa^2}{2} \tilde\tau^\mn
         +
         2 \tilde H^\mn_\tenl
         \Big)
         \ .
  \end{align}
\end{subequations}
The second line is equivalent to setting $\xi=1$ from the beginning which is fine since the metric is independent of $\xi$.
\par
We will now expand Eq.~\eqreft{lin1}, or equivalently Eq.~\eqreft{lin2}, perturbatively in $G_N$ using the techniques introduced in Ch.~\ref{sec:ExpansionsAround}.
We assume $T^\mn$ to be of zeroth order in $G_N$ and the graviton field, $h_\mn$, to be of first order.
Then $\tau^\mn$ is of zeroth order and $H_\tenl^\mn$ is of second order.
Inserting the expansions into Eq.~\eqreft{lin1} we get:
\begin{align}
  \sum_{n=1..\infty}
  \tilde h_\ab^{\cGn{n}}
  =
  \frac{G_{\ab\mn}}{q^2}
  \Big(
  \frac{\kappa^2}{2}
  \sum_{\nzi}
  \tilde\tau^\mn_{\cGn{n}}
  +
  2\frac{1}{\xi}
  \sum_{n=2..\infty}
  (\tilde H^\mn_\tenl
  )_{\cGn{n}}
  \Big)
  \ .
  \labelt{ger1}
\end{align}
Recall, that $\kappa^2=32\pi G_N$.
Equating terms of equal order in $G_N$ we get:
\begin{align}
  &\tilde h_\ab^{\cGn{1}}(q)
  =
  \frac{G_{\ab\mn}}{q^2}
  \frac{\kappa^2}{2}
  \tilde T^\mn(q)
  \labelt{leq1}
\end{align}
And for $n\geq2$ we get:
\begin{align}
  &\tilde h_\ab^{\cGn{n}}
  =
  \frac{G_{\ab\mn}}{q^2}
  \Big(
  \frac{\kappa^2}{2}
  \tilde\tau^\mn_{G^{n-1}}
  +
  \frac{2}{\xi}
  \big(
  \tilde H^\mn_\tenl
  \big)_{\cGn{n}}
  \Big)
  \ .
  \labelt{exp2}
\end{align}
In general, the material energy-momentum $T^\mn$ which is included in $\tau^\mn$ can depend in a non-trivial way on $h_\mn$ i.e. if the matter part has its own equation of motion.
In the rest of this chapter we will focus on the special case of a single inertial point particle, that is the \STM\ metric.
We will then assume that any gravitational corrections to $T^\mn$ are local and that $T^\mn$ is exact at zeroth order in the perturbative expansion.
\par
With this assumption we can rewrite Eq.~\eqreft{exp2} into:
\begin{align}
  &\tilde h_\ab^{\cGn{n}}
  =
  2\frac{G_{\ab\mn}}{q^2}
  \big(
  \tilde G^\mn_\tenl
  +
  \frac{1}{\xi}\tilde H^\mn_\tenl
  \big)_{\cGn{n}}
  \ .
  \labelt{leq2}
\end{align}
Here, we still assume $n\geq2$.
Note that this equation is equivalent to demanding:
\begin{equation}
  \big(
  G^\mn
  +
  \frac{1}{\xi}
  H^\mn
  \big)
  _{\cGn{n}} = 0
  \ .
  \labelt{ger2}
\end{equation}
This follows directly from the \cgE\ equation and the assumption that $T^\mn$ is exact at zeroth order.
\par
We will now specialize on the first terms in the expansion.
From Eq.~\eqreft{leq1}, the first order correction to $g_\mn$ is given by:
\begin{align}
  h^{\cGno}_\mn =
  \frac{\kappa^2}{2}
  \int \dDp{q}
  e^{-iqx}\
  \frac{G_{\mn\ab}}{q^2}
  \ 
  \tilde T^\ab
  \labelt{ger3}
\end{align}
This will correspond to the tree diagram in the Feynman graph expansion.
In the end of this section we will compute this simple example explicitly.
\par
Let us go to the case $n=2$ where we can use Eq.~\eqreft{leq2}:
\begin{align}
  &\tilde h_\ab^{\cGn{2}}
  =
  2\frac{G_{\ab\mn}}{q^2}
  \big(
  \tilde G^\mn_\tenl
  +
  \frac{1}{\xi}\tilde H^\mn_\tenl
  \big)_{\cGn{2}}
  \ .
  \labelt{leq4}
\end{align}
We have to find the $G_N^2$ term of $G^\mn_\tenl$ and $H^\mn_\tenl$.
We can use Eq.~\eqreft{exp1} for this and we get
\begin{align}
  (G^\mn_\tenl)_{\cGn{2}}
  =
  G_{\chn{2}}^\mn
  (
  h^{\cGno}_\mn ,
  h^{\cGno}_\mn
  )
  \ ,
  \labelt{ger4}
\end{align}
and similarly for $H_\tenl^\mn$.
These expressions, however, are in position space and it is necessary to transform them to momentum space so that we can insert them into Eq.~\eqreft{leq4}.
\par
The transformation to momentum space follows the same pattern as the arguments around Eq.~\eqreft{leq3}.
We treat the case of $G_{\chn{2}}^\mn$ explicitly and that of $H_{\chn{2}}^\mn$ follows the same line of reasoning.
We insert the expressions of $h_\mn$ in terms of $\tilde h_\mn$:
\begin{align}
  G_{\chn{2}}^\mn
  (
  h^{\cGno}_\mn ,
  h^{\cGno}_\mn
  )
  =
  G_{\chn{2}}^\mn
  (
  \int \dDp{l_1} e^{-il_1x} \tilde h^{\cGno}_\mn(l_1) ,
  \int \dDp{l_2} e^{-il_2x} \tilde h^{\cGno}_\mn(l_2)
  )
  \labelt{ger5}
\end{align}
The function $G_{\chn{2}}^\mn$ is linear in its arguments and we can pull out the integrals:
\begin{align}
  G_{\chn{2}}^\mn
  (
  h^{\cGno}_\mn ,
  h^{\cGno}_\mn
  )
  =
  \int \dDp{l_\teon} \dDp{l_\tetw}
  G_{\chn{2}}^\mn
  (
  e^{-il_1x} \tilde h^{\cGno}_\mn(l_\teon) ,
  e^{-il_2x} \tilde h^{\cGno}_\mn(l_\tetw)
  )
  \labelt{ger6}
\end{align}
We can pull out the exponential factors as well, if we substitute derivatives in $G_{\chn{2}}^\mn$ with $-il_\teon^\mu$ or $-il_\tetw^\mu$ which, as in Sec.~\ref{sec:GravitonSelf}, we symbolize with a tilde on $G_{\chn{2}}^\mn$:
\begin{align}
  G_{\chn{2}}^\mn
  (
  h^{\cGno}_\mn ,
  h^{\cGno}_\mn
  )
  &=
  \int \dDp{l_\teon} \dDp{l_\tetw}
  e^{-i(l_\teon+l_\tetw)x}
  \tilde G_{\chn{2}}^\mn
  (
  \tilde h^{\cGno}_\mn(l_\teon) ,
  \tilde h^{\cGno}_\mn(l_\tetw)
  )
  \labelt{ger7}
  \\
  &=
  \int \dDp{q}
  e^{-iqx}
  \int \dDp{l}
  \tilde G_{\chn{2}}^\mn
  (
  \tilde h^{\cGno}_\mn(l) ,
  \tilde h^{\cGno}_\mn(q-l)
  )
  \nonumber{}
\end{align}
We can now read off $G^\mn_{\chn{2}}$ in momentum space and for $(\tilde G^\mn_\tenl)_{\cGn{2}}$ we get:
\begin{align}
  (\tilde G^\mn_\tenl)_{\cGn{2}}
  =
  \int \dDp{l}
  \tilde G_{\chn{2}}^\mn
  (
  \tilde h^{\cGno}_\mn(l) ,
  \tilde h^{\cGno}_\mn(q-l)
  )
  \labelt{leq5}
\end{align}
The integral on the right-hand side looks similar to a loop integral.
\par
Inserting Eq.~\eqreft{leq5} into the expression for $\tilde h_\mn^{\cGn{2}}$ from Eq.~\eqreft{leq4} we get:
\begin{align}
  &\tilde h_\mn^{\cGn{2}}
  =
  2
  \frac{G_{\mn\ab}}{q^2}
  \int \dDp{l}
  \bigg(
  \tilde G^\ab_{\chn{2}}
  \Big(
  \tilde h^{\cGno}_\mn(l) ,
  \tilde h^{\cGno}_\mn(q-l)
  \Big)
  +
  \frac{1}{\xi} \tilde H^\ab_{\chn{2}}
  \Big(
  \tilde h^{\cGno}_\mn(l) ,
  \tilde h^{\cGno}_\mn(q-l)
  \Big)
  \bigg)
  \ .
  \labelt{lin3}
\end{align}
The second order metric is derived from a quadratic energy-momentum function of the first order metric, $G_{\chn{2}}^\mn$, and a gauge dependent term, $H_{\chn{2}}^\mn$.
Let us insert the expression for $\tilde h_\mn^\cGno$ in terms of $\tilde T^\mn$ from Eq.~\eqreft{leq1} which will make the relation to the Feynman graph expansion clear.
It is most convenient to focus only on $\tilde G^\ab_{\chn{2}}$ first:
\begin{align}
  \int \dDp{l}
  \tilde G^\ab_{\chn{2}}
  &\Big(
  \tilde h^{\cGno}_\mn(l) ,
  \tilde h^{\cGno}_\mn(q-l)
  \Big)
  =
  \int \dDp{l}
  \tilde G^\ab_{\chn{2}}
  \Big(
  \frac{\kappa^2}{2}\frac{G_{\mn\ab}\tilde T^\ab(l)}{l^2} ,
  \frac{\kappa^2}{2}\frac{G_{\mn\ab}\tilde T^\ab(l-q)}{(l-q)^2}
  \Big)
  \nonumber{}
  \\
  &=
  \frac{\kappa^4}{4}
  \int \dDp{l}
  \frac{1}{l^2(l-q)^2}
  \tilde G^\ab_{\chn{2}}
  \Big(
  G_{\mn\ab}\tilde T^\ab(l) ,
  G_{\mn\ab}\tilde T^\ab(l-q)
  \Big)
  \labelt{ger8}
\end{align}
In the second line we used that $\tilde G^\ab_{\chn{2}}$ is a linear function.
Notice that the momentum dependence of the propagators, $G_{\mn\ab}$, is not written explicitly but understood from the context.
In the second line the relevant integral is reminiscent of loop integrals from quantum field theory.
\par
In the final part of this section we will derive an expression for $h^{\cGn{3}}_\mn$.
We need the $G_N^3$ term of $G_\tenl^\mn$ and $H_\tenl^\mn$ and again, Eq.~\eqreft{exp1} can be used:
\begin{align}
  (G_\tenl^\mn)_{\cGn{3}}
  =
  2 G^\mn_{\chn{2}}
  \big(
  h^{\cGno}_\mn
  ,
  h^{\cGn{2}}_\mn
  \big)
  +
  G^\mn_{\chn{3}}
  \big(
  h^{\cGno}_\mn
  ,
  h^{\cGno}_\mn
  ,
  h^{\cGno}_\mn
  \big)
  \labelt{ger9}
\end{align}
Again, an analogous equation holds for $H_\tenl^\mn$.
The Fourier transformation to momentum space is similar to the case of $h^{\cGn{2}}_\mn$ above and the case discussed around Eq.~\eqreft{leq3}:
\begin{align}
  G^\mn_{\chn{2}}
  \big(
  h^{\cGno}_\mn
  ,
  h^{\cGn{2}}_\mn
  \big)
  =
  \int \dDp{q}
  e^{-iqx}
  \int \dDp{l}
  \tilde G_{\chn{2}}^\mn
  \Big(
  \tilde h^{\cGno}_\mn(l) ,
  \tilde h^{\cGn{2}}_\mn(q-l)
  \Big)
  \labelt{ge10}
\end{align}
And:
\begin{align}
  &G^\mn_{\chn{3}}
  \big(
  h^{\cGno}_\mn
  ,
  h^{\cGno}_\mn
  ,
  h^{\cGno}_\mn
  \big)
  =
  \int \dDp{l_\teon} \dDp{l_\tetw} \dDp{l_\tetr}
  e^{-i(l_\teon + l_\tetw + l_\tetr)x}
  \tilde G^\mn_{\chn{3}}
  \Big(
  \tilde h^{\cGno}_\mn (l_\teon)
  ,
  \tilde h^{\cGno}_\mn (l_\tetw)
  ,
  \tilde h^{\cGno}_\mn (l_\tetr)
  \Big)
  \nonumber{}
  \\
  &\qquad\qquad
  =
  \int  \dDp{q}
  e^{-iqx}
  \int \dDp{l_\teon} \dDp{l_\tetw}
  \tilde G^\mn_{\chn{3}}
  \Big(
  \tilde h^{\cGno}_\mn (l_\teon)
  ,
  \tilde h^{\cGno}_\mn (l_\tetw-l_\teon)
  ,
  \tilde h^{\cGno}_\mn (q-l_\tetw)
  \Big)
  \labelt{ge11}
\end{align}
So that we can now write down the equation for $h_\ab^{\cGn{3}}$:
\begin{align}
  &\tilde h_\ab^{\cGn{3}}
  =
  2
  \frac{G_{\ab\mn}}{q^2}
  \Big(
  (\tilde G^\mn_\tenl)_{\cGn{3}}
  +
  \frac{1}{\xi}
  (\tilde H^\mn_\tenl)_{\cGn{3}}
  \Big)
  \ .
  \labelt{ge12}
\end{align}
where
\begin{align}
  (\tilde G^\mn_\tenl)_{\cGn{3}}
  =
  &\ 2\int \dDp{l}
  \tilde G_{\chn{2}}^\mn
  \Big(
  \tilde h^{\cGno}_\mn(l) ,
  \tilde h^{\cGn{2}}_\mn(q-l)
  \Big)
  \labelt{lin4}
  \\
  +
  &\int \dDp{l_\teon} \dDp{l_\tetw}
  \tilde G^\mn_{\chn{3}}
  \Big(
  \tilde h^{\cGno}_\mn (l_\teon)
  ,
  \tilde h^{\cGno}_\mn (l_\tetw-l_\teon)
  ,
  \tilde h^{\cGno}_\mn (q-l_\tetw)
  \Big)
  \nonumber{}
\end{align}
And an equivalent equation holds for $(\tilde H^\mn_\tenl)_{\cGn{3}}$.
These formulas are similar to two-loop graphs.
\par
In the next section, Sec.~\ref{sec:TheMetric}, the equation for $\tilde h_\ab^{\cGn{3}}$ and that for $\tilde h_\ab^{\cGn{2}}$ will be compared to explicit Feynman diagram expansions.
First, as a concrete example to show how the formulas of this section work, we will use the simple formula for $\tilde h_\ab^{\cGn{1}}$ to derive the first order contribution to the \STM\ metric.
\subsection{Newton Potential in Arbitrary Dimension}
\labelx{sec:NewtonPotential}
As an example we will apply Eq.~\eqreft{leq1} to derive the first order correction to the metric in arbitrary dimensions.
As $T^\mn$ we will take the point particle energy-momentum tensor of special relativity (see e.g.~\cite{Weinberg:1972kfs}).
We take a point particle of momentum $k^\mu$ and mass $m$ and use the covariant notation introduced in Eqs.~\ref{nn32}:
\begin{subequations}
  \begin{align}
    T^\mn
    &=
    \frac{k^\mu k^\nu}{m} \delta^{D-1}(x_\bot)
    \ ,
    \\
    &=
    m
    \ \eta_\prl^\mn
    \ \delta^{D-1}(x_\bot)
    \ .
    \labelt{ber1}
  \end{align}
\end{subequations}
That this tensor describes an inertial particle is easily verified in the reference frame of $k^\mu$.
In momentum space this becomes:
\begin{align}
  \tilde T^\mn (q)
  =
  \int \dDx\ 
  e^{iqx}\ 
  T^\mn
  =
  2\pi\delta(q_\prl)\ 
  m\ 
  \eta_\prl^\mn
  \ .
  \labelt{ber2}
\end{align}
In this expression it is more easily seen that the tensor is Lorentz covariant.
The energy-momentum tensor $T^\mn$ is conserved since $q_\bot^\mu \eta_\mn^\prl$ vanishes.
Then, using Eq.~\eqreft{leq1} we get:
\begin{subequations}
  \label{eqn:lin5}
  \begin{align}
    \tilde h_\ab^{\cGn{1}}(q)
    &=
    \frac{G_{\ab\mn}}{q^2}
    \frac{\kappa^2}{2}
    2\pi\delta(q_\prl)\
    m\ 
    \eta_\prl^\mn
    \ ,
    \labelt{lin5a}
    \\
    &=
    \frac{\kappa^2m}{2}
    \frac{2\pi\delta(q_\prl)}{q_\bot^2}
    G_{\ab\mn}\ \eta^\mn_\prl
    \ .
  \end{align}
\end{subequations}
Since $T^\mn$ is conserved we get:
\begin{align}
  G_{\ab\mn}\ \tilde T^\mn = \maPi_{\ab\mn} \tilde T^\mn
  \ .
  \labelt{ber3}
\end{align}
The tensor structure of $\tilde h_\ab^{\cGn{1}}(q)$ becomes:
\begin{subequations}
  \label{eqn:ber4}
  \begin{align}
    \maPi_{\ab\mn} \tilde \eta^\mn_\prl
    &=
    \eta^\mn_\prl - \frac{1}{D-2}\eta^\mn
    \ ,
    \labelt{ber4a}
    \\
    &=
    \frac{D-3}{D-2}
    \big(
    \eta^\prl_\mn - \frac{1}{D-3} \eta^\bot_\mn
    \big)
    \ .
  \end{align}
\end{subequations}
This should be inserted in Eq.~\eqreft{lin5}.
\par
We have now the final expression for $\tilde h_\mn^\cGno$:
\begin{align}
  \tilde h_\mn^\cGno
  =
  \frac{\kappa^2m}{2}
  \frac{2\pi\delta(q_\prl)}{q_\bot^2}
  \frac{D-3}{D-2}
  \big(
  \eta^\prl_\mn - \frac{1}{D-3} \eta^\bot_\mn
  \big)
  \ .
  \labelt{ber5}
\end{align}
Using a Fourier integral introduced later in Eq.~\eqreft{vv34} we can go to position space.
The leading order contribution to $h_\mn$ becomes:
\begin{align}
  h^{\cGno}_\mn 
  = -\frac{\mu}{\sqrt{-x_\bot^2}^{D-3}}
  \big(
  \etat{\mn} - \frac{1}{D-3} \etar{\mn}
  \big)
  \ .
  \labelt{me32}
\end{align}
Here, we used the \STM\ parameter from Eq.~\eqreft{mud1}.
This result is in agreement with Refs.~\cite{Collado:2018isu,Emparan:2008eg}.
It is independent of the gauge parameter $\alpha$ since this parameter only enters in the graviton self-interaction vertices.
It obeys the first order gauge condition $(G_\sigma)^{\cGno}=0$ where:
\begin{align}
  (G_\sigma)^{\cGno} = G_\sigma^{\chno}(h^{\cGno}_\mn)
  = \partial^\rho \maP^\mn_\rs h^{\cGno}_\mn
\end{align}

\section{The Metric from the Three-Point Vertex Function}
\begin{figure}[h]
  \centering
  \captionsetup{width=.8\linewidth}
  \includegraphics[width=5cm]{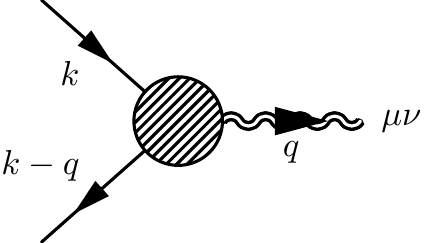}
  \caption{
    A massive scalar interacts with a graviton.
    The diagram represents the exact three-point vertex function, $i\maM_\teve^\mn$.
    \flabel{amp2}
    \ffig{fd9}
  }
  \label{fig:amp2}
\end{figure}

\labelx{sec:TheMetric}
It is now the goal to derive the \STM\ metric from a Feynman diagram expansion.
The relevant amplitude will be the exact three-point vertex function of a massive scalar interacting with a graviton which is shown in Fig.~\ref{fig:amp2}.
Here the scalar momentum $k^\mu$ is on-shell $k^2=m^2$ but the graviton momentum, $q^\mu$, is considered as arbitrary.
\par
\begin{figure}[h]
  \centering
  \captionsetup{width=0.8\linewidth}
  \begin{tikzpicture}
    \node(fd110)
         {
           $i\maM_{\text{2-loop}}^\mn=$
         };
         \node(fd11)[right=0cm of fd110]
              {
                \includegraphics[width=4.4cm]{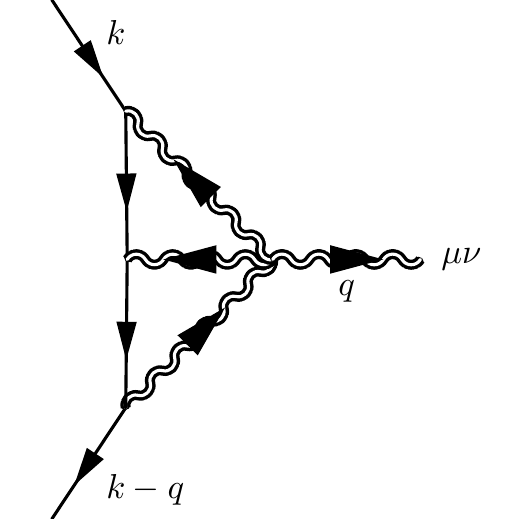}
              };
              \node[right=0cm of fd11]
                   {
                     $+$
                   };
                   \node(fd12)[right=.5cm of fd11]
                        {
                          \includegraphics[width=4.4cm]{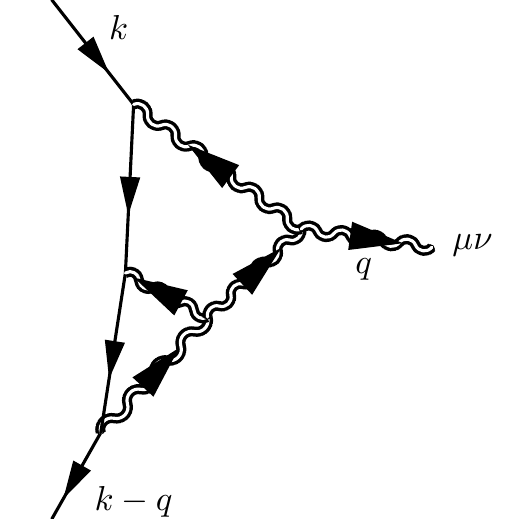}
                        };
                        \node[right=0cm of fd12]
                             {
                               $+\ .\ .\ .$
                             };
  \end{tikzpicture}
  \caption{
    $\maM_{\text{$2$-loop}}^\mn$ is expanded in Feynman diagrams.
    The ellipsis denotes two permutations of the second diagram.
    Other 2-loop diagrams exist which do not contribute in the classical limit.
    \flabel{amp3}
    \ffig{fd11,fd12}
  }
  \label{fig:amp3}
\end{figure}

We will use the ideas of Refs.~\cite{Bjerrum-Bohr:2018xdl,Galusha:cand} to reduce the n-loop integrals in the expansion of $\maM_\teve^\mn$ to simple couplings to classical sources.
Essentially, the idea is that in the classical limit only triangle graphs contribute.
Two-loop examples of such diagrams are found in Fig.~\ref{fig:amp3}.
A property of such triangle graphs is that, if the scalar line is removed, the remaining graph is a graviton tree diagram.
The result for the n-loop integrals in the classical limit is that they reduce to a convolution integral which contracts $(n+1)$ classical sources to the graviton tree diagram which remains after the scalar line is removed.
\par
We have not found exact formulas for how the reduction is done in the general case since, also, both of the Refs.~\cite{Bjerrum-Bohr:2018xdl,Galusha:cand} work only in $D=4$.
We will then not focus on exact equalities but instead indicate how the \STM\ metric can be derived from the three-point function given that the integrals reduce as explained.
To this aim, our Feynman rules for the general n-graviton vertex in terms of $\Gpz^\mn$ and $\Hpz^\mn$ will be helpful.
\par
In this section, we will analyze the one-loop and two-loop cases.
From this the general case can be worked out.
In the next chapter we will then compute the one-loop case explicitly using the formulas developed in this section and the three-graviton vertex from Sec.~\ref{sec:GravitonSelf}.
\par
Let us first discuss how, in the end, the \STM\ metric will be related to the three-point function.
Then afterwards we will show how this relation can be derived.
The three-point function $\maM_\teve^\mn$ is interpreted as the source of the graviton field.
We relate $\maM_\teve^\mn$ to the tensors introduced in Sec.~\ref{sec:GaugeFixed}:
\begin{equation}
  2\pi\delta(kq) \maM^\mn_{\text{vertex}}
  =
  -\kappa
  \tilde \tau^\mn
  -\frac{4}{\kappa\xi}
  \tilde H_\tenl^\mn
  \ .
  \labelt{ver6}
\end{equation}
This is exactly the right-hand side of Eq.~\eqreft{cge1} (up to a factor).
It is then straightforward to relate the amplitude to the metric:
\begin{equation}
  g_\mn = \eta_\mn
  - \frac{\kappa}{2}
  \int \frac{d^Dq\ \delta(kq)\ e^{-iq x}}{(2\pi)^{D-1}}
  \frac{G_{\mn\ab}}{q^2}
  \mathcal{M}_{\text{vertex}}^\ab
  \ .
  \labelt{ext2}
\end{equation}
This is an exciting formula relating the Lorentz covariant three-point function from quantum field theory to the general covariant all order metric from general relativity.
Following similar arguments as those that led to Eq.~\eqreft{lin2} we can get an equation independent of $\xi$ relating $g_\mn$ and $\maM_\teve^\mn$.
We will write it in terms of $\tilde h_\mn$ instead of $g_\mn$:
\begin{equation}
  \tilde h_\mn = \frac{\mathcal{P}^{-1}_{\mn\ab}}{q^2}
  \Big(
  \frac{\kappa^2}{2}
  \tilde \tau^\ab
  +2
  \tilde H_{\text{non-linear}}^\ab
  \Big)
  \ .
  \labelt{ver7}
\end{equation}
And, of course, $g_\mn = \eta_\mn + h_\mn$.
\par
\begin{figure}[h]
  \centering
  \captionsetup{width=.8\linewidth}
  \includegraphics[width=5cm]{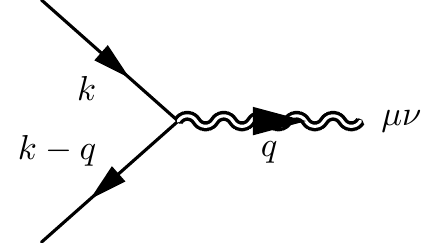}
  \caption{
    The tree-level contribution to the three-point vertex function.
    In the classical limit, this becomes the point particle energy-momentum tensor.
    \flabel{amp6}
    \ffig{fd16}
  }
  \label{fig:amp6}
\end{figure}

Let us now look into the derivation of these formulas.
First, we will look at the tree-level contribution to the three-point function which comes from the diagram in Fig.~\ref{fig:amp6}.
This will show how the formulas are used in a concrete example.
Also the exact constant of proportionality between the three-point amplitude and $\tilde \tau^\mn$ is easily determined from this example.
\par
The scalar graviton vertex rules were derived in Sec.~\ref{sec:ScalarGraviton}.
The diagram in Fig.~\ref{fig:amp6} will be proportional to the $\phi^2 h$ vertex rule which from Eq.~\eqreft{ver1} is:
\begin{align}
  \IVe_{\phi^2h}^{\mu\nu}(k,k-q,m)
  &=
  I_{\alpha\beta}^{\mu\nu} k^\alpha (k-q)^\beta - \frac{k(k-q)-m^2}{2} \eta^{\mu\nu}
  \labelt{ver8}
  \\
  &=
  I_{\alpha\beta}^{\mu\nu} k^\alpha (k-q)^\beta + \frac{kq}{2} \eta^{\mu\nu}
  \ .
  \nonumber{}
\end{align}
We have $\maM_\teee^\mn = -\kappa\IVe_{\phi^2h}^{\mu\nu}(k,k-q,m)$.
We can simplify the amplitude in the classical limit where we can neglect factors of the graviton momentum, $q^\mu$, in comparison to the particle momentum, $k^\mu$.
In the classical limit the amplitude then reduces to:
\begin{align}
  \maM_\teee^\mn = -\kappa k^\mu k^\nu
  = -\kappa m^2 \eta_\prl^\mn
\end{align}
According to Eq.~\eqreft{ver6} we then get $\tilde \tau^\mn$ as:
\begin{subequations}
  \begin{align}
    \tilde \tau^\mn
    &=
    -\frac{2\pi\delta(kq)}{\kappa} \maM_\teee^\mn
    \\
    &=
    2\pi\delta(q_\prl)
    \
    m
    \ \eta_\prl^\mn
  \end{align}
\end{subequations}
This is exactly the energy-momentum tensor from Eq.~\eqreft{ber2} which describes an inertial point particle.
Thus, Eq.~\eqreft{ver6} is correct at tree-level.
\par
\begin{figure}[h]
  \centering
  \captionsetup{width=.8\linewidth}
  \includegraphics[width=5cm]{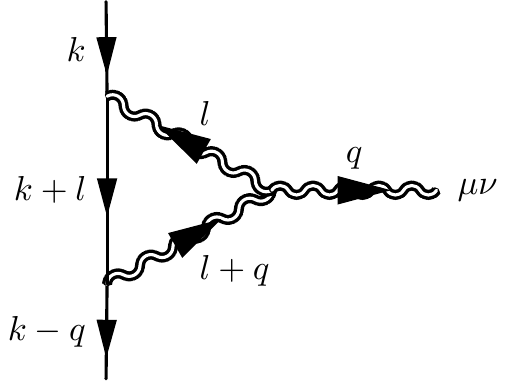}
  \caption{
    In position space, this diagram is reduced to a local quadratic function of $h^\cGno_\mn$.
    \flabel{amp7}\ffig{fd7}
  }
  \label{fig:amp7}
\end{figure}

Let us turn to the one-loop contribution to the metric.
We will compute this exactly in the next chapter.
Here, we will only show how it can be related to the perturbative expansion of the classical equations of motion.
For the one-loop case only one diagram, the one in Fig.~\ref{fig:amp7}, contributes in the classical limit.
The idea is, that the massive propagator with momentum $k+l$ can be integrated away with $l_\prl$.
Then we can think of the amplitude, $\maM_\teol^\mn$, as two tree-diagrams of the type of Fig.~\ref{fig:amp6} contracted to the three-graviton vertex with a convolution integral.
\par
In the next chapter we will look at these integrals explicitly.
For now, however, the arguments will only be suggestive.
We take the three-graviton vertex rule from Eq.~\eqreft{ber6}:
\begin{equation}
  i2\kappa
  \IVe_{h^{3}}^{\mn\ \ab\ \gd}(q,l,l+q)
  \tfmn_{\ab}(l)
  \tfmn_{\gd}(l+q)
  =
  -i4\kappa
  \Big(
  \tGhn{2}{\mn}
  \big(
  \tfmn_{\ab}^{(l)}, 
  \tfmn_{\gd}^{(l+q)}
  \big)
  +\frac{1}{\xi}
  \tHhn{2}{\mn}
  \big(
  \tfmn_{\ab}^{(l)},
  \tfmn_{\gd}^{(l+q)}
  \big)
  \Big)
  \labelt{vv43}
\end{equation}
Here $\tilde f_\ab(l)$ and $\tilde f_\ab(l+q)$ are the tree-diagrams from Fig.~\ref{fig:amp6} after they have been propagated with the graviton propagator.
Thus, they are the first order correction to the metric, $\thmn^{\cGno}_\mn$, from Eq.~\eqreft{ber5}.
The one-loop amplitude then becomes something like:
\begin{align}
  \maM_\teol^\mn
  \sim
  \int \dDp{l}
  \IVe_{h^{3}}^{\mn\ \ab\ \gd}(q,l,l+q)
  \thmn^\cGno_{\ab}(l)
  \thmn^\cGno_{\gd}(l+q)
  \labelt{vv44}
\end{align}
Inserting Eq.~\eqreft{vv43} into Eq.~\eqreft{vv44} we recognize this equation as the source term of $\tilde h^{\cGn{2}}_\mn$ from Eq.~\eqreft{lin3}.
Thus, if the integrals reduce as discussed, this amplitude produces the correct second order contribution to the metric when propagated with the graviton propagator as in Eq.~\eqreft{ext2}.
\par
\begin{figure}[h]
  \centering
  \captionsetup{width=0.8\linewidth}
  \begin{tikzpicture}
    \node(fd13)
         {
           \includegraphics{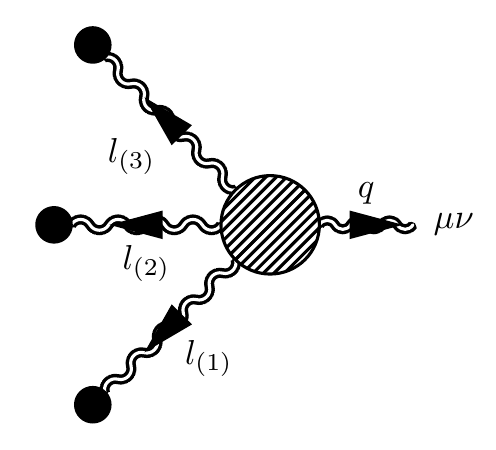}
         };
         \node(fd130)[right=-0.3cm of fd13]
              {
                $=$
              };
              \node(fd14)[right=-0.3cm of fd130]
                   {
                     \includegraphics{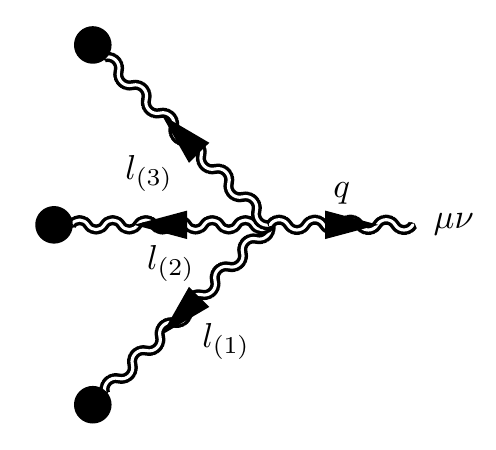}
                   };
                   \node(fd140)[right=-0.3cm of fd14]
                        {
                          $+\quad3\times$
                        };
                        \node(fd15)[right=-0.3cm of fd140]
                             {
                               \includegraphics{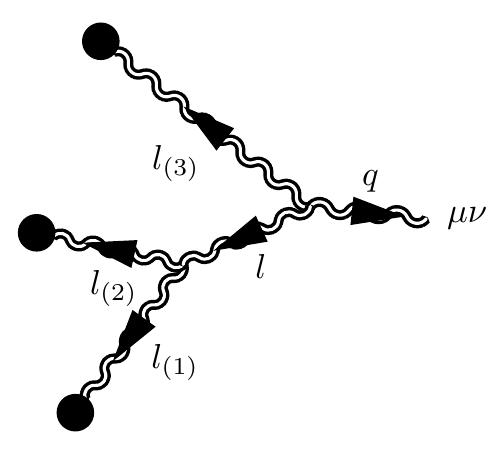}
                             };
  \end{tikzpicture}
  \caption{
    The four-graviton tree amplitude contracted with three classical sources.
    \flabel{amp5}
    \ffig{fd13,fd14,fd15}
  }
  \label{fig:amp5}
\end{figure}

We will now analyze the two-loop case.
The two-loop triangle graphs which contribute in the classical limit where shown in Fig.~\ref{fig:amp3}.
There are four such graphs, three of them being permutations of each other.
Assuming that the integrals reduce as discussed in the classical limit we can cut the scalar line into three pieces, cutting it where massive propagators are.
The result is that three $\phi^2 h$ tree diagrams from Fig.~\ref{fig:amp6} are contracted to the 4-point graviton tree amplitude as depicted in Fig.~\ref{fig:amp5}.
\par
In Fig.~\ref{fig:amp5} the shaded blob on the left-hand side is the graviton 4-point tree amplitude.
It is contracted to 3 symmetric classical sources which are depicted as smaller solid blobs.
On the right-hand side of Fig.~\ref{fig:amp5} the graviton tree amplitude is expanded in graviton self-interaction vertices.
The three distinct Feynman graphs of Fig.~\ref{fig:amp3} which were permutations of each other have now been reduced to three equivalent terms due to the three classical sources being contracted in a symmetric way.
\par
The terms on the right-hand side of Fig.~\ref{fig:amp5} should be compared to Eqs.~\eqreft{ge11} and~\eqreft{ge10} from the perturbative expansion of the classical equations of motion determining $\thmn^{\cGn{3}}_\mn$.
For example, in the last term two classical sources are contracted to a three-graviton vertex which gives $h^{\cGn{2}}_\mn$.
The three-graviton vertex together with a classical source is then contracted to another three-graviton vertex.
We thus relate the last term of Fig.~\ref{fig:amp5} to the first term of Eq.~\eqreft{lin4} which is:
\begin{equation}
  \int \dDp{l}
  \Big(
  \tilde G_{\chn{2}}^\mn
  \big(
  \tilde h^{\cGno}_\mn(l) ,
  \tilde h^{\cGn{2}}_\mn(q-l)
  \big)
  +
  \frac{1}{\xi}
  \tilde H_{\chn{2}}^\mn
  \big(
  \tilde h^{\cGno}_\mn(l) ,
  \tilde h^{\cGn{2}}_\mn(q-l)
  \big)
  \Big)
  \labelt{ge13}
\end{equation}
Similarly we relate the first term of Fig.~\ref{fig:amp5} to the last term of Eq.~\eqreft{lin4} which is:
\begin{equation}
  \int \dDp{l_\teon} \dDp{l_\tetw}
  \tilde G^\mn_{\chn{3}}
  \Big(
  \tilde h^{\cGno}_\mn (l_\teon)
  ,
  \tilde h^{\cGno}_\mn (l_\tetw-l_\teon)
  ,
  \tilde h^{\cGno}_\mn (q-l_\tetw)
  \Big)
  \labelt{ge14}
\end{equation}
This expression is thus related to the diagram where 3 classical sources meet the 4-graviton self-interaction vertex.
\par
In Eq.~\eqreft{lin4} there is a factor 2 in front of Eq.~\eqreft{ge13} and factor unity in front of Eq.~\eqreft{ge14}.
In Fig.~\ref{fig:amp5} it looks as though there is a factor 3 in front of Eq.~\eqreft{ge13} and factor unity in front of Eq.~\eqreft{ge14}.
We should, however, remember that the graviton self-interaction vertices are related to $\Gpz^\mn$ and $\Hpz^\mn$ with a numerical factor which depends on the number of gravitons.
The numerical factor is $(n!)$ for the ${(n+1)}$-graviton vertex.
Thus, instead, in Fig.~\ref{fig:amp5} Eq.~\eqreft{ge13} comes with a factor 12 and Eq.~\eqreft{ge14} comes with a factor 6 (along with other common factors which we disregard).
We conclude that also in Fig.~\ref{fig:amp5} the terms come with a correct relative factor of 2.
\par
There is also the difference between Fig.~\ref{fig:amp5} and Eq.~\eqreft{lin4} that the n-graviton vertices are related to $\Gpz^\mn$ and $\Hpz^\mn$ while Eq.~\eqreft{lin4} is in terms of $G^\mn$ and $H^\mn$.
However, in the final part of Sec.~\ref{sec:GaugeFixed} we discussed that $\Gpz^\mn$ and $G^\mn$ can be related to each other and similarly for $\Hpz^\mn$ and $H^\mn$.
\par
\begin{figure}[h]
  \centering
  \captionsetup{width=0.8\linewidth}
  \begin{tikzpicture}
    \node(fd100)
         {
           $i\maM_\tenp^\mn\sim$
         };
         \node(fd10)[right=0cm of fd100]
              {
                \includegraphics{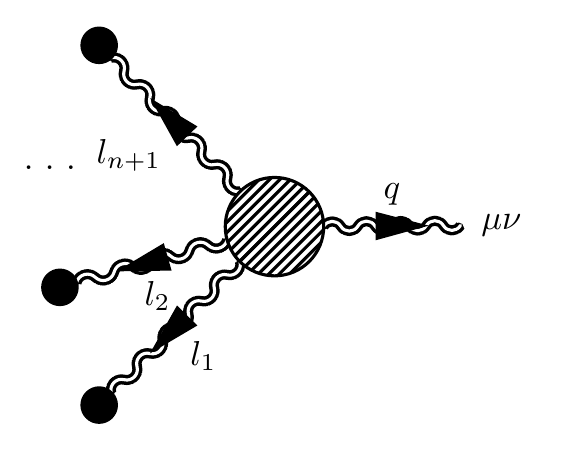}
              };
  \end{tikzpicture}
  \caption{
    The shaded blob represents the $(n+2)$-graviton tree amplitude and the smaller solid blobs represent the $\phi^2h$-tree diagrams i.e. the point particle energy-momentum tensor.
    \flabel{amp4}
    \ffig{fd10}
  }
  \label{fig:amp4}
\end{figure}

The general case can be deduced from these examples and is depicted in Fig.~\ref{fig:amp4}.
First, it requires consideration of the general combinatorics of comparing the graviton tree-amplitudes to the terms of the classical expansion of the \cgE\ equation.
Here, the general formula for the n-graviton vertex turns out to be suitable for the inductive structure of the triangle diagrams.
Also, the n-loop integrals have to be reduced in the classical limit, so that the assumptions made in this section are fulfilled.
This is left for future investigations which would allow the general n-loop diagram to be reduced explicitly to the corresponding terms in the perturbative classical expansion.
\chapter{\STM\ Metric at One-Loop Order}
\labelx{sec:PerturbativeExpansion2}
We will now compute the one-loop correction to the \STM\ metric in \dDo-type gauge.
We will use Eq.~\eqreft{ext2} which relates the three-point vertex function to the metric.
First, in Sec.~\ref{sec:OneLoop} we go through the technical details of the Feynman diagram computation.
This includes evaluation of the one-loop triangle integrals in the classical limit and working out the tensor contractions of the three-graviton vertex.
In Sec.~\ref{sec:SecondOrder} we will then use Eq.~\eqreft{ext2} to derive the metric contribution in position space from the amplitude.
This requires us contract the amplitude with the graviton propagator and make a Fourier transform to position space.
\par
In space-time dimension $D=5$ a logarithm appears in position space.
This is analyzed in Sec.~\ref{sec:AppearanceOf} where the triangle integrals are treated more carefully and divergencies are removed.
The logarithm is connected to a redundant gauge freedom, which also occurs in $D=4$.
Finally, in Sec.~\ref{sec:ClassicalDerivation} we present a derivation of the second order contribution to the \STM\ metric using only methods from classical general relativity.
The results of this computation verify the amplitude computation including the appearance of logarithms in $D=5$.

\section{Feynman Diagram Computation}
\begin{figure}[h]
  \centering
  \captionsetup{width=0.8\linewidth}
  \includegraphics[width=5cm]{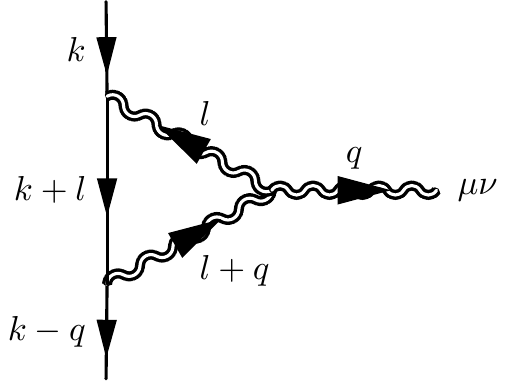}
  \caption{Feynman triangle diagram. The solid line is a massive scalar and wiggly lines are gravitons.\flabel{amp1}\ffig{fd7}}
  \label{fig:amp1}
\end{figure}

\labelx{sec:OneLoop}
The diagram in Fig.~\ref{fig:amp1} is the only one that contributes to the three-point vertex function at one-loop order in the classical limit.
Other diagrams do not have the required triangle structure.
In this section we will compute the corresponding amplitude, $\maM_\teol^\mn$, in detail.
\par
The Feynman graph in Fig.~\ref{fig:amp1} was already discussed in Sec.~\ref{sec:TheMetric} around Fig.~\ref{fig:amp7}.
Here the one-loop triangle integrals where assumed to reduce to a simple convolution integral.
In Sec.~\ref{sec:TriangleIntegrals} we will evaluate these integrals explicitly.
We find that only the spatial part of the integrals conform to the assumptions made in Sec.~\ref{sec:TheMetric}.
However, in Sec.~\ref{sec:Tensor} we see during the calculation, that only the space-components of the integrals contribute to the amplitude so that in the end, the assumptions made in Sec.~\ref{sec:TheMetric} are satisfied at one-loop order.
This is seen using the three-graviton vertex in terms of the $U^{\mn\ \ab\rho\ \gd\sigma}$ tensor.
It would be interesting if another argument could show this immediately, which would make the Feynman rule in terms of $\Gpz^\mn$ and $\Hpz^\mn$ advantageous.
\subsection{Triangle Loop Integrals}
\labelx{sec:TriangleIntegrals}
The triangle one-loop integrals relevant for the graph in Fig.~\ref{fig:amp1} are well known in the literature in $D=4$ \cite{Bjerrum-Bohr:cand,BjerrumBohr:2002kt,Donoghue:1994dn} and also in arbitrary dimensions \cite{Cristofoli:2020uzm}.
\par
Our derivation will be slightly different from \cite{Cristofoli:2020uzm}, although the results agree.
Let us first consider the simplest case of the scalar integral:
\begin{equation}
  I = \int \dDp{l}
  \frac{1}{l^2+i\epsilon}
  \frac{1}{(l+q_\bot)^2+i\epsilon}
  \frac{1}{(l+k)^2-m^2+i\epsilon}
  \labelt{vvv2}
\end{equation}
In the literature, they have been solved with the implicit constraint on $q^\mu$ that $kq\approx0$.
In our work we treat $q^\mu$ as an arbitrary momentum variable.
We have then added the subscript on $q_\bot^\mu$ in the triangle integrals so that we still have the similar equation that $q_\bot k=0$.
\par
The classical limit allows us to simplify the massive propagator:
\begin{subequations}
  \label{eqn:vvv1}
  \begin{align}
    \frac{1}{(l+k)^2-m^2+i\epsilon}
    &\approx
    \frac{1}{2kl+i\epsilon}
    \\
    &=
    \frac{1}{2m}
    \frac{1}{l_\prl+i\epsilon}
    \labelt{vvv1b}
  \end{align}
\end{subequations}
In the first line we used that $l^2 \ll kl$ and in the second line, that $kl=ml_\prl$.
\par
We rewrite Eq.~\eqreft{vvv1b} using the following formula:
\begin{align}
  \frac{1}{l_\prl+i\epsilon}
  =
  \frac{1}{l_\prl}
  -
  i\pi\delta(l_\prl)
  \labelt{rel2}
\end{align}
When we insert this equation into the scalar triangle integral of Eq.~\eqreft{vvv2} we can neglect the first term of Eq.~\eqreft{rel2}.
This is so since the two graviton propagators are even in $l_\prl$.
In the classical limit the scalar triangle integral is then reduced to:
\begin{equation}
  I =
  -\frac{i}{4m}
  \int \dDp{l}
  2\pi\delta(l_\prl)
  \frac{1}{l_\bot^2}
  \frac{1}{(l_\bot+q_\bot)^2}
  \ .
  \labelt{vvv3}
\end{equation}
Here we have neglected $i\epsilon$ since both $l_\bot$ and $q_\bot$ are space-like.
The remaining integral is a bubble integral, which is e.g. solved in Srednicki~\cite{Srednicki:2007qs}.
\par
The integral in Eq.~\eqreft{vvv3} is similar to the integrals from the expansion of the classical equations of motion.
We will symbolize it as $N_{D-1}$:
\begin{subequations}
  \label{eqn:vvv4}
  \begin{align}
    N_{D-1}
    &=
    \int \dDp{l}
    2\pi\delta(l_\prl)
    \frac{1}{l_\bot^2}
    \frac{1}{(l_\bot+q_\bot)^2}
    \labelt{vvv4a}
    \\
    &=
    \int \ddp{l_\bot}
    \frac{1}{l_\bot^2}
    \frac{1}{(l_\bot+q_\bot)^2}
  \end{align}
\end{subequations}
And we find that
\begin{equation}
  N_d = -\frac{\Omega_{d-2}\sqrt{-q^2_\bot}^{d-4}}{4(4\pi)^{d-2}\ \sin(\frac{\pi}{2}d)}
  \labelt{vvv5}
\end{equation}
where $\Omega_{d-1}$ is the surface area of a sphere in d-dimensional space which was defined in Eq.~\eqreft{n217}.
The integral $N_{D-1}$ from Eq.~\eqreft{vvv4} is special since it conforms exactly to the analysis from Sec.~\ref{sec:TheMetric}.
It is a convolution integral which in position space corresponds to simple multiplication.
\par
The tensor triangle integrals can be treated analogously.
Only the integrals corresponding to $I^\mu$ and $I^\mn$ contribute in the classical limit:
\begin{subequations}
  \label{eqn:vvv7}
  \begin{align}
    I^\mu &= \int \dDp{l}
    \frac{l^\mu}{\big(l^2+i\epsilon\big)
      \big((l+q_\bot)^2+i\epsilon\big)
      \big((l+k)^2-m^2+i\epsilon\big)}
    \labelt{vvv7a}
    \\
    I^\mn &= \int \dDp{l}
    \frac{l^\mu l^\nu}{\big(l^2+i\epsilon\big)
      \big((l+q_\bot)^2+i\epsilon\big)
      \big((l+k)^2-m^2+i\epsilon\big)}
    \labelt{vvv7b}
  \end{align}
\end{subequations}
Integrals with more loop momenta in the numerator does not contribute in the classical limit.
This is due to the fact that the three-graviton vertex adds two graviton momenta to the numerator and terms with extra graviton momenta are neglected in comparison.
\par
The tensor integrals can be solved algebraically by writing an ansatz.
Let us work out $I^\mu$ as an example:
\begin{equation}
  I^\mu =  A q^\mu_\bot + B k^\mu
  \labelt{vvv8}
\end{equation}
The coefficients $A$ and $B$ are found using the equations $q_\mu^\bot I^\mu = q_\bot^2 A$ and $k_\mu I^\mu = m^2 B$.
Then we can use relations such as $2 k_\mu l^\mu = ((l+k)^2-m^2)-l^2$ to reduce the numerators so that we again have scalar integrals without any loop momenta in the numerator.
We will do this for the $I_\prl$ and $I_\bot^\mu$ parts separately.
\par
For the $I_\bot^\mu$ we get a result similar to the scalar triangle integral:
\begin{subequations}
  \label{eqn:vvv9}
  \begin{align}
    I_\bot^\mu
    &= A q^\mu_\bot
    \labelt{vvv9a}
    \\
    &=
    -\frac{i}{4m}
    \int \dDp{l}
    2\pi\delta(l_\prl)
    \frac{l^\mu_\bot}{l_\bot^2
      (l_\bot+q_\bot)^2}
    \labelt{vvv9b}
  \end{align}
\end{subequations}
In the first line, we inserted the ansatz from Eq.~\eqreft{vvv8} and then, in the second line, the definition from Eq.~\eqreft{vvv7a} using the simplifications of the massive propagator in Eq.~\eqreft{rel2}.
The integral is similar to the $N_{D-1}$-integral defined in Eq.~\eqreft{vvv4} and we define:
\begin{align}
  N_{D-1}^\mu
  &=
  \int \dDp{l}
  2\pi\delta(l_\prl)
  \frac{l^\mu_\bot}{l_\bot^2
    (l_\bot+q_\bot)^2}
  \labelt{vv10}
\end{align}
To solve it algebraically we dot with $q_\bot$ and use $2l_\bot q_\bot = (l_\bot+q_\bot)^2-l_\bot^2-q_\bot^2$.
We get:
\begin{align}
  \hspace*{-.5cm}
  2q^\bot_\mu
  N_{D-1}^\mu
  &=
  \int \dDp{l}
  2\pi\delta(l_\prl)
  \frac{1}{l_\bot^2}
  -
  \int \dDp{l}
  2\pi\delta(l_\prl)
  \frac{1}{(l_\bot+q_\bot)^2}
  -
  q^2_\bot
  \int \dDp{l}
  2\pi\delta(l_\prl)
  \frac{1}{l_\bot^2
    (l_\bot+q_\bot)^2}
  \nonumber{}
  \\
  &=
  -q_\bot^2
  N_{D-1}
  \labelt{vvv6}
\end{align}
The two first integrals of the first line can be neglected in the classical limit and the third one is recognized as $N_{D-1}$.
The expression for $N^\mu_{D-1}$ becomes:
\begin{align}
  N^\mu_{D-1}
  =
  -\frac{1}{2}q_\bot^\mu
  N_{D-1}
  \labelt{vv11}
\end{align}
Again, $N_{D-1}^\mu$ has the same simple interpretation as $N_{D-1}$.
\par
We will now compute the parallel part of $I^\mu$:
\begin{subequations}
  \label{eqn:vv12}
  \begin{align}
    I_\prl &= mB
    \labelt{vv12a}
    \\
    &=
    \frac{1}{2m}
    \int \dDp{l}
    \frac{1}{l^2
      (l+q_\bot)^2}
    \frac{l_\prl}{l_\prl+i\epsilon}
    \labelt{vv12b}
  \end{align}
\end{subequations}
Again, in the first line we inserted the ansatz from Eq.~\eqreft{vvv8} and then in the second line the definition from Eq.~\eqreft{vvv7a}.
This time, however, the simplification of the massive propagator is different since the remaining integral is now odd in $l_\prl$.
Thus instead of a $\delta$-function from the factor $1/(l_\prl+i\epsilon)$ it cancels with the $l_\prl$ in the numerator.
Thus the final factor of Eq.~\eqreft{vv12b}, $l_\prl/(l_\prl+i\epsilon)$, is unity.
After a Wick rotation the remaining integral is of the same type as $N_D$ and we get:
\begin{align}
  I_\prl =
  i\frac{N_D}{2m}
\end{align}
The final expression for $I^\mu$ becomes:
\begin{subequations}
  \begin{align}
    I^\mu
    &= -\frac{i}{4m} N_{D-1}^\mu
    + \frac{i}{2m^2} N_D k^\mu
    \\
    &=
    \frac{i}{8m} N_{D-1}\ q_\bot^\mu
    +\frac{i}{2m^2} N_D\ k^\mu
  \end{align}
\end{subequations}
We can treat $I^\mn$ in the same fashion and the result is:
\begin{align}
  I^\mn
  &=
  -\frac{i}{4m} N^\mn_{D-1}
  -\frac{i}{4m^2}
  N_D \ 
  (q^\mu_\bot k^\nu + k^\mu q^\nu_\bot)
\end{align}
Where $N^\mn_{D-1}$ is:
\begin{subequations}
  \begin{align}
    N^\mn_{D-1}
    &=
    \int \dDp{l}
    2\pi\delta(l_\prl)
    \frac{l^\mu_\bot l^\nu_\bot}{l_\bot^2
      (l_\bot+q_\bot)^2}
    \\
    &=
    \frac{q_\bot^2}{4(D-2)} N_{D-1}
    \Big(
    (D-1)\qbmnu{\mu}{\nu}
    -
    \eta^\mn_\bot
    \Big)
  \end{align}
\end{subequations}
We have now computed all the triangle integrals that contribute to our amplitude in the classical limit.
The results agree with Ref.~\cite{Cristofoli:2020uzm} the only difference being that we have expressed them in terms of $N_{D-1}$ and $N_D$.
\par
As already mentioned, the $N_{D-1}$-integrals have the simple interpretation in position space of two classical sources multiplied together and contracted to the three-graviton vertex.
This is not the case for the factors of $N_D$ in $I^\mu$ and $I^\mn$.
We can summarize it as follows.
The space components of the triangle integrals are given by the $N_{D-1}$-integrals which have the interpretation of a convolution between two classical sources.
The time and mixed components of the triangle integrals are given by the $N_D$ integral.
\par
According to the ideas in Sec.~\ref{sec:TheMetric}, only the $N_{D-1}$ integrals ought to contribute to the amplitude.
This is indeed the case as we will see.
We would like to have a simpler argument for this fact, however.
It seems that the $N_D$ part of the integral would describe mixed components of the metric, that is components with one time and one space index.
In $D=4$ the standard expression for the Schwarzschild metric in harmonic gauge found in e.g.~\cite{Weinberg:1972kfs} does not have such components.
However, there exist other expressions for the metric in harmonic gauge which have mixed components~\cite{Petrov:2020abu}.
This is also expected to be the case for the \STM\ metric in arbitrary dimensions.
It would be interesting if the expressions for the metric with mixed components could also be derived from the amplitude in which case the $N_D$ part of the triangle integrals would possibly play a role.
\par
Finally, we should comment on the fact that the expression for $N_d$ in Eq.~\eqreft{vvv4} diverges when $d$ is even.
This divergence should be treated with the methods of dimensional regularization.
This means that for even $d$ we should expand $N_d$ in a Laurent series.
The divergent pole will be an analytic function in $q_\bot$ which describes local corrections to the metric.
The pole should be subtracted from $N_{d}$ and the non-analytic finite term is the relevant part of $N_{d}$.
We will analyze this in Sec.~\ref{sec:AppearanceOf}.
By using regularized integrals, it turns out that we can ignore the divergence in all space-time dimensions except $D=5$.
In Sec.~\ref{sec:SecondOrder} we will then derive the metric contribution in all dimensions $D\neq5$ and specialize on $D=5$ in Sec.~\ref{sec:AppearanceOf}.
\subsection{Three-Graviton Vertex and Tensor Manipulations}
\labelx{sec:Tensor}
We will now go through the tensor manipulations of the amplitude computation.
We use the three-graviton vertex rule in terms of the $U$-tensor.
We expand the Feynman diagram in Fig.~\ref{fig:amp1} as:
\begin{subequations}
  \label{eqn:vv13}
  \begin{align}
    2\pi\delta(kq) i\maM^\mn_\teol
    &=
    -i\frac{4}{\kappa}
    \Big(
    \tilde G_\teol+\frac{1}{\xi} \tilde H_\teol
    \Big)
    \labelt{vv13a}
    \\
    &
    =
    2\pi\delta(kq)m^4(2i\kappa)(-i\kappa)^2i^3
    \int \dDp{l}
    \frac{1}{l^2(l+q_\bot)^2
      \big((l+k)^2-m^2+i\epsilon\big)}
    \nonumber{}
    \\
    &\qquad\qquad\qquad\qquad\qquad
    \times
    f_\ab f_\gd
    \IVe^{\ab\ \gd\ \mn}_{h^3}(l,-l-q,q)
    \ .
    \labelt{vv13b}
  \end{align}
\end{subequations}
In the front, we have gathered factors of $i$, $\kappa$ and $m$.
Afterwards we have the loop integration and the propagators which make the triangle integrals.
Then, in the end of Eq.~\eqreft{vv13} we have the tensor structure including the three-graviton vertex $\IVe_{h^3}$ and tensors $f_\ab$ defined as:
\begin{align}
  f_\ab = \frac{k^\mu k^\nu}{m^2} \maPi_{\mn\ab}
  = \eta_\prl^\mn \maPi_{\mn\ab}
  \ .
\end{align}
The tensors $f_\ab$ describe the tensor structure of the $\phi^2 h$ vertices contracted to the graviton propagator.
Here, we have already made use of two simplifications due to the classical limit.
First, we have neglected factors of $l$ and $q$ in the scalar-graviton vertex.
Second, we have neglected the momentum dependent part of the graviton propagator.
The first simplification is similar to the simplification of the tree-level diagram.
Here, the graviton momenta are neglected in comparison to the scalar momenta.
The momentum dependent part of the propagator is less obvious since graviton momenta are introduced both in the denominator and numerator.
\par
There are two graviton propagators with momenta $l^\mu$ and $(l+q)^\mu$.
The momentum dependent part of the graviton propagator with e.g. momentum $l_\mu$ is:
\begin{align}
  - 2(1-\xi)
  I^{\mn}_{\rho\kappa} \frac{\ l^\rho  l_\sigma}{l^2} I^{\kappa \sigma}_{\ab}
  \ .
\end{align}
Thus there are also integrals with more loop momenta squared in the denominator than the triangle integrals.
However, in these integrals one of the graviton momenta $l^\mu$ or $q^\mu$ is always contracted to the scalar vertex.
A simple explanation for neglecting the momentum part of the propagator is then, that since the scalar graviton vertex represents the point particle energy-momentum tensor, it is conserved.
When the graviton momentum is contracted with the conserved energy-momentum tensor we then get zero.
\par
Let us analyze the integral where the graviton propagator with momentum $l^\mu$ meets the scalar graviton vertex:
\begin{subequations}
  \begin{align}
    \int \dDp{l}
    \frac{lk}{l^2 l^2 (l+q)^2 \big((l+k)^2-m^2+i\epsilon\big)}
    =
    \frac{1}{2}
    \int \dDp{l}
    \frac{\big((l+k)^2-m^2\big)-l^2}{
      l^2 l^2 (l+q)^2 \big((l+k)^2-m^2+i\epsilon\big)}
    \\
    =
    \frac{1}{2}
    \int \dDp{l}
    \frac{1}{
      l^2 l^2 (l+q)^2 }
    -
    \frac{1}{2}
    \int \dDp{l}
    \frac{1}{
      l^2 (l+q)^2 \big((l+k)^2-m^2+i\epsilon\big)}
  \end{align}
\end{subequations}
In the second line, the first term does not have the right structure for the classical limit.
The second term has the structure of the triangle integrals.
However, there will be three factors of graviton momenta ($l^\mu$ or $q^\mu$) in the numerator, one remaining from the propagator and two from the three-graviton vertex.
This is one extra factor of graviton momentum compared to the terms from the momentum independent part of the propagator.
\par
In the classical limit, we can thus make the simplifications in the definition of $f_\ab$.
Note that $f_\ab$ is the tensor structure of the first order metric contribution, $h_\mn^\cGno$, from Eq.~\eqreft{ber4a}:
\begin{equation}
  f_\mn = \frac{D-3}{D-2} \eta_\mn^\prl - \frac{1}{D-2} \eta_\mn^\bot
  \ ,
  \labelt{vv22}
\end{equation}
Focus now on the tensor part of the integral with the three-graviton vertex and two factors of $f_\ab$.
We use the vertex rule in terms of $U^{\mn\ \ab\rho\ \gd\sigma}$ from Eq.~\eqreft{be10}:
\begin{align}
  \hspace*{-1cm}
  f_\ab f_\gd
  \tau^{\ab\ \gd\ \mn}_{h^3}(l,-l-q,q)
  =
  f_\ab f_\gd
  \Big( U^{\mn\ \ab\rho\ \gd\sigma} l_\rho (l+q)_\sigma
  + U^{\ab\ \gd\rho\ \mn\sigma} (l+q)_\rho q_\sigma
  - U^{\gd\ \mn\rho\ \ab\sigma} q_\rho l_\sigma
  \Big)
\end{align}
The final two terms of this equation can be simplified by expanding $(l+q)$ and using the symmetries of $U^{\mn\ \ab\rho\ \gd\sigma}$.
We get
\begin{align}
  f_\ab f_\gd
  \tau^{\ab\ \gd\ \mn}_{h^3}(l,-l-q,q)
  =
  f_\ab f_\gd
  \Big( U^{\mn\ \ab\rho\ \gd\sigma} l_\rho (l+q)_\sigma
  + U^{\ab\ \gd\rho\ \mn\sigma} q_\rho q_\sigma
  \Big)
  \ ,
\end{align}
which we insert in the expression for the amplitude in Eq.~\eqreft{vv13}.
Inserting the triangle integrals as well, we get:
\begin{align}
  2\pi \delta(kq)
  i\mathcal{M}_{\text{1-loop}}^\mn
  =
  -
  2\pi\delta(kq)
  2 m^4 \kappa^3
  f_\ab f_\gd
  \Big( U^{\mn\ \ab\rho\ \gd\sigma} (I_\rs+I_\rho q_\sigma)
  + U^{\ab\ \gd\rho\ \mn\sigma} q_\rho q_\sigma I
  \Big)
  \ .
\end{align}
All the tensors of this expression are known and it is now the task to compute all the contractions.
\par
The integral $I_\rs+I_\rho q_\sigma$ is given by:
\begin{align}
  I_\rs+I_\rho q_\sigma
  =
  i
  \frac{q^2\ N_{D-1}}{16(D-2)m}
  \Big(
  (D-3)\qbmnd{\rho}{\sigma}
  + \eta^\bot_\rs
  \Big)
  +
  i
  \frac{N_D}{4m^2}
  \big(
  k_\rho q^\bot_\sigma
  -
  q^\bot_\rho k_\sigma
  \big)
  \labelt{vv14}
\end{align}
Because of the symmetries of $f_\ab f_\gd U^{\mn\ \ab\rho\ \gd\sigma}$, only the symmetric part of the integral contributes.
This is exactly what we want, since the symmetric part is given in terms of the $N_{D-1}$-integrals which follow the ideas of Sec.~\ref{sec:TheMetric}.
We can then conclude that the $N_D$ integral does not contribute to the one-loop diagram in Fig.~\ref{fig:amp1} in the classical limit.
\par
Inserting the integral from Eq.~\eqreft{vv14} in the amplitude we get
\begin{align}
  \hspace*{-.6cm}
  2\pi\delta(kq) \mathcal{M}_{\text{1-loop}}^\mn
  &=
  2\pi\delta(q_\prl) \frac{m^2 \kappa^3  q_\bot^2 N_{D-1}}{2}
  f_\ab f_\gd
  \Big(
  -  \frac{1}{4(D-2)}
  U^{\mn\ \ab\rho\ \gd\sigma}
  M_\rs^\bot
  + U^{\ab\ \gd\rho\ \mn\sigma} \qbmnd{\rho}{\sigma}
  \Big)
  \ ,
  \labelt{vv15}
\end{align}
where we have defined:
\begin{align}
  M_\rs^\bot
  =
  (D-3)\qbmnd{\rho}{\sigma}+\eta^\bot_\rs
  \ .
\end{align}
The superscript on $M^\bot_\rs$ makes it clear that it has only space-like components.
\par
We will compute most of the contractions in Eq.~\eqreft{vv15} in detail.
Let us start with the contribution to $\tHmn^\mn_\teol$ which is simpler than the one to $\tGmn^\mn_\teol$.
Thus we compute the part which comes from $U_\tegf$.
The first term of Eq.~\eqreft{vv15} has the tensor structure:
\begin{align}
  U^{\mn\ \ab\rho\ \gd\sigma}_{gf} f_\ab f_\gd M_\rs^\bot
  \ .
\end{align}
To evaluate this it is convenient to use the expression for $U_\tegf$ in Eq.~\eqreft{n430b} which is in terms of $h_{\mn,\rho}$.
Instances of $h_\mn$ should be replaced by $f_\mn$ and partial derivatives by $M_\rs^\bot$.
For example:
\begin{align}
  -\mathcal{P}^\rs_\ab
  h^\ab_{,\sigma}
  h_{\rho}^{\mu,\nu}  
  \rightarrow
  -\mathcal{P}^\rs_\ab
  f^\ab
  f^{\mu}_\rho
  \Mbo^\nu_\sigma
  \ .
  \labelt{vv20}
\end{align}
We then get:
\begin{align}
  U^{\mn\ \ab\rho\ \gd\sigma}_{gf} f_\ab f_\gd M_\rs^\bot
  &=
  \alpha
  \mathcal{P}^\rs_\ab
  f^\ab
  \Big(
  - f_{\rho}^\mu  \Mbo^{\nu}_\sigma
  - f_{\rho}^\nu  \Mbo^{\mu}_\sigma
  + f^\mn M^\bot_\rs
  \Big)
  \\
  &=
  \alpha
  \eta_\prl^\rs
  \Big(
  - f_{\rho}^\mu  \Mbo^{\nu}_\sigma
  - f_{\rho}^\nu  \Mbo^{\mu}_\sigma
  + f^\mn M^\bot_\rs
  \Big)
  =
  0
\end{align}
Going from the first to the second line we used that $\maP f = \eta_\prl$.
The second line is easily seen to disappear since $M^\bot_\rs$ and $\eta_\prl^\mn$ are orthogonal.
\par
The second term of Eq.~\eqreft{vv14} has the tensor structure:
\begin{align}
  U^{\ab\ \gd\rho\ \mn\sigma}_{gf} f_\ab f_\gd \qbmnd{\rho}{\sigma}
  \ .
  \labelt{vv17}
\end{align}
Here, the two factors of $f$ are contracted in an asymmetrical way.
Then Eq.~\eqreft{n430b} for $U_\tegf$ is less useful and we use instead Eq.~\eqreft{n436}.
\par
For the first term of Eq.~\eqreft{n436} we get:
\begin{subequations}
  \label{eqn:vv16}
  \begin{align}
    \hspace*{-1cm}
    -\alpha
    \Big(
    I^{\ab\rho\kappa}
    I^\gd_{\kappa\lambda}
    \maP^{\lambda\sigma\mn}
    +
    I^{\ab\sigma\kappa}
    I^\mn_{\kappa\lambda}
    \maP^{\lambda\rho\gd}
    \Big)
    f_\ab f_\gd \qbmnd{\rho}{\sigma}
    &=
    -\alpha
    \Big(
    f^{\rho\kappa}
    f_{\kappa\lambda}
    \maP^{\lambda\sigma\mn}
    \qbmnd{\rho}{\sigma}
    +
    f^{\sigma\kappa}
    I^\mn_{\kappa\lambda}
    \eta_\prl^{\lambda\rho}
    \qbmnd{\rho}{\sigma}
    \Big)
    \labelt{vv16a}
    \\
    &=
    -\alpha\frac{1}{(D-2)^2}\maP^{\rho\sigma\mn}
    \qbmnd{\rho}{\sigma}
    \labelt{vv16b}
  \end{align}
\end{subequations}
On the right-hand side of Eq.~\eqreft{vv16a} the second term vanishes and the first term simplifies to the one in the second line.
\par
For the second term of Eq.~\eqreft{n436} we get:
\begin{subequations}
  \begin{align}
    \frac{1}{2}
    \alpha
    \Big(
    \maP^{\gd\rs} I^{\mn\ab}
    +
    \maP^{\mn\rs} I^{\gd\ab}
    \Big)
    f_\ab f_\gd \qbmnd{\rho}{\sigma}
    &=
    \frac{1}{2}
    \alpha
    \Big(
    \eta_\prl^{\rs} f^{\mn}\qbmnd{\rho}{\sigma}
    +
    \maP^{\mn\rs} f_\ab f^\ab \qbmnd{\rho}{\sigma}
    \Big)
    \\
    &=
    \frac{1}{2}
    \alpha
    f_\ab f^\ab
    \maP^{\mn\rs} \qbmnd{\rho}{\sigma}
  \end{align}
\end{subequations}
Here, only the second term of the first line is non-zero.
It simplifies to the one in the second line.
\par
We have now computed the tensor structure in Eq.~\eqreft{vv17}:
\begin{subequations}
  \begin{align}
    U^{\ab\ \gd\rho\ \mn\sigma}_{gf} f_\ab f_\gd \qbmnd{\rho}{\sigma}
    &=
    \alpha
    \Big(
    -\frac{1}{(D-2)^2}
    +
    \frac{1}{2}
    f_\ab f^\ab
    \Big)
    \maP^{\mn\rs} \qbmnd{\rho}{\sigma}
    \\
    &=
    \alpha
    \frac{D-3}{2(D-2)}
    \maP^{\mn\rs} \qbmnd{\rho}{\sigma}
    \\
    &=
    \alpha\frac{D-3}{4(D-2)}
    (2\qbmnu{\mu}{\nu}-\eta^\mn)
  \end{align}
\end{subequations}
This allows us to find $\tHmn^\mn_\teol$ defined in Eq.~\eqreft{vv13a}:
\begin{align}
  &\tilde H_\teol^\mn =
  -\alpha
  \ 2\pi\delta(q_\prl)
  \ \frac{\kappa^4 m^2 q^2 N_{D-1}}{
    16}
  \frac{D-3}{D-2}
  \maP^{\mn\rs} \qbmnd{\rho}{\sigma}
  \ .
  \labelt{vv27}
\end{align}
This is the gauge-dependent part of Eq.~\eqreft{vv15}.
\par
We will now compute the part of Eq.~\eqreft{vv15} which contributes to $\tGmn^\mn$.
There are two tensor structures to be computed, namely
\begin{align}
  U_\tecl^{\mn\ \ab\rho\ \gd\sigma} f_\ab f_\gd M_\rs^\bot
  \ ,
  \labelt{vv18}
\end{align}
and:
\begin{align}
  U_\tecl^{\ab\ \gd\rho\ \mn\sigma} f_\ab f_\gd \qbmnd{\rho}{\sigma}
  \ .
  \labelt{vv19}
\end{align}
Both computations are similar to the ones for $U_\tegf$ only more complicated due to $U_\tecl$ being more complicated than $U_\tegf$.
We will do Eq.~\eqreft{vv18} in detail and only quote the result for Eq.~\eqreft{vv19}.
\par
For the computation of Eq.~\eqreft{vv18}, we use Eq.~\eqreft{ute1} for the definition of $U_\tecl$.
As in Eq.~\eqreft{vv20} instances of $h_\ab$ should be replaced with $f_\ab$ and the derivatives on $h_\ab$ should be replaced with $M_\rs$.
Let us reprint Eq.~\eqreft{ute1}:
\begin{align}
  U_\tecl^{\mn\ \ab\rho\ \gd\sigma} \ h_{\ab,\rho} h_{\gd,\sigma}
  =
  &
  2 I^\mn_\pe \maP^\ab_\rs \maP^{\sigma\phi}_{\gamma\delta} h^{,\epsilon}_\ab h^{\gd,\rho}
  - \maP^{\mu\rho}_\ab \maP^{\nu\sigma}_\gd \eta_\rs h^\ab_{,\kappa} h^{\gd,\kappa}
  \labelt{vv21}
  \\
  &
  + \maP^\mn_\rs
  \Big(
  h^{\rho\alpha}_{,\beta} h^{\sigma\beta}_{,\alpha}
  -\frac{1}{2} h^{\alpha,\rho}_{\beta} h^{\beta,\sigma}_{\alpha}
  - h^\rs_{,\alpha} h^\ab_{,\beta}
  \Big)
  \nonumber
  \ .
\end{align}
We will consider each term by itself.
\par
The first term of $U_\tecl$ from Eq.~\eqreft{vv21} becomes:
\begin{align}
  2 I^\mn_\pe \maP^\ab_\rs \maP^{\sigma\phi}_{\gamma\delta} f_\ab f^{\gd} M^{\epsilon\rho}_\bot
  =
  2 I^\mn_\pe \eta^\prl_\rs \eta^{\sigma\phi}_\prl M^{\epsilon\rho}_\bot
  =
  0
\end{align}
It is straightforward to apply $\maP$ to $f$.
We then get $\eta_\prl$ which is orthogonal to $M^\bot$ so that this term vanishes.
\par
Next term of Eq.~\eqreft{vv21}:
\begin{align}
  - \maP^{\mu\rho}_\ab \maP^{\nu\sigma}_\gd \eta_\rs f^\ab f^\gd \Mbo^\kappa_\kappa
  =
  - \eta^{\mu\rho}_\prl \eta^{\nu\sigma}_\prl \eta_\rs \Mbo^\kappa_\kappa
  =
  -\eta^\mn_\prl \Mbo^\kappa_\kappa
\end{align}
Again, $\maP f$ is simply $\eta_\prl$.
We need to compute the trace of $\Mbo$:
\begin{align}
  \Mbo^\kappa_\kappa = D-3+D-1 = 2(D-2)
  \labelt{vv23}
\end{align}
So from the second term of Eq.~\eqreft{vv21} we get:
\begin{equation}
  - \maP^{\mu\rho}_\ab \maP^{\nu\sigma}_\gd \eta_\rs f^\ab f^\gd \Mbo^\kappa_\kappa
  =
  -2(D-2)\eta_\prl^\mn
  \ .
\end{equation}
\par
The next three terms of Eq.~\eqreft{vv21} are contracted with $\maP^\mn_\rs$.
We will compute the three terms in the brackets first and then do the contraction.
\par
The first term in the brackets:
\begin{align}
  f^{\rho\alpha} f^{\sigma\beta} M_\ab^\bot
  =
  \frac{1}{(D-2)^2} M^\rs_\bot
\end{align}
Again, we used that $M_\ab^\bot$ is orthogonal to $k^\mu$ and the explicit structure of $f_\ab$ from Eq.~\eqreft{vv22}.
\par
The second term in the brackets:
\begin{align}
  -\frac{1}{2} f^\ab f_\ab M^\rs_\bot
  =
  -\frac{1}{2} \frac{(D-3)^2+D-1}{(D-2)^2} M^\rs_\bot
\end{align}
Here, we inserted $f^\ab f_\ab$ directly.
\par
The third term in the brackets:
\begin{align}
  -f^\rs f^\ab M_\ab^\bot
  =
  \frac{1}{D-2} f^\rs \Mbo^\kappa_\kappa
  =
  2 f^\rs
\end{align}
The trace, $\Mbo^\kappa_\kappa$, was given in Eq.~\eqreft{vv23}.
\par
The sum of these three terms become:
\begin{align}
  f^{\rho\alpha} f^{\sigma\beta} M_\ab
  -\frac{1}{2} f^\ab f_\ab M^\rs
  -f^\rs f^\ab M_\ab
  =
  -\frac{1}{2}\frac{D-3}{D-2} M^\rs
  +2 f^\rs
  \ .
  \labelt{vv24}
\end{align}
These are the three terms in the brackets from Eq.~\eqreft{vv21}.
They should be propagated with $\maP^\mn_\rs$.
For ${\maP^\mn_\rs M^\rs}$, we get:
\begin{align}
  \maP^\mn_\rs M^\rs = -(D-2) \eta_\prl^\mn
  +(D-3)
  \Big(\qbmnu{\mu}{\nu}-\eta_\bot^\mn
  \Big)
\end{align}
The three terms in Eq.~\eqreft{vv24} propagated with $\maP^\mn_\rs$ become:
\begin{align}
  \maP^\mn_\rs
  \Big(
  f^{\rho\alpha} f^{\sigma\beta} M_\ab
  -\frac{1}{2} f^\ab f_\ab M^\rs
  -f^\rs f^\ab M_\ab
  \Big)
  =
  \frac{D+1}{2} \eta_\prl^\mn
  -
  \frac{1}{2}\frac{(D-3)^2}{D-2}
  \big(
  \qbmnu{\mu}{\nu}-\eta_\bot^\mn
  \big)
  \ .
\end{align}
We have now computed all the terms of Eq.~\eqreft{vv21}.
\par
Thus, we have computed the contribution to $\tGmn^\mn$ in Eq.~\eqreft{vv18}:
\begin{align}
  U^{\mn\ \ab\rho\ \gd\sigma} f_\ab f_\gd M_\rs
  =
  -\frac{3(D-3)}{2} \eta_\prl^\mn
  -
  \frac{1}{2}\frac{(D-3)^2}{D-2}
  \big(
  \qbmnu{\mu}{\nu}-\eta_\bot^\mn
  \big)
  \labelt{vv25}
\end{align}
We will not do the detailed computation of the  other contribution to $\tGmn^\mn$ from Eq.~\eqreft{vv19}.
The result is:
\begin{align}
  U^{\ab\ \gd\rho\ \mn\sigma} f_\ab f_\gd \qbmnd{\rho}{\sigma}
  =
  \frac{1}{2}\frac{D-4}{D-2} \eta_\prl^\mn
  -
  \frac{1}{2} \frac{D-1}{D-2}
  \big(
  \eta_\bot^\mn - \qbmnu{\mu}{\nu}
  \big)
  \labelt{vv26}
\end{align}
With Eqs.~\eqreft{vv25} and~\eqreft{vv26} we can compute $\tGmn^\mn$.
We combine Eqs.~\eqreft{vv13a} and~\eqreft{vv15}:
\begin{align}
  &\tilde G_\teol^\mn
  =
  2\pi\delta(q_\prl)
  \ \frac{\kappa^4 m^2 q^2 N_{D-1}}{
    64
  }
  \bigg(
  \frac{D-7}{D-2} \eta^\mn_\parallel
  - \frac{(D-3)(3D-5)}{(D-2)^2}
  \Big(
  \eta_\bot^\mn - \frac{\ q^\mu q^\nu}{q^2}
  \Big)
  \bigg)
  \ .
  \labelt{amp1}
\end{align}
We have now computed the amplitude in Fig.~\ref{fig:amp1}.
The major results are Eqs.~\eqreft{vv27} and~\eqreft{amp1} for $\tHmn^\mn_\teol$ and $\tGmn^\mn_\teol$ respectively in terms of which the amplitude is defined in Eq.~\eqreft{vv13a}.

\section{Second Order Correction to the Metric}
\labelx{sec:SecondOrder}
With the result for the one-loop contribution to the three-point function we can derive the second order contribution to the \STM\ metric.
Let us summarize the results of Sec.~\ref{sec:OneLoop} for the amplitude in Fig.~\ref{fig:amp1}:
\begin{subequations}
  \label{eqn:vv28}
  \begin{align}
    &2\pi\delta(kq) \maM^\mn_\teol
    =
    -\frac{4}{\kappa}
    \Big(
    \tilde G_\teol^\mn
    +\frac{1}{\xi}
    \tilde H_\teol^\mn
    \Big)
    \ ,
    \labelt{vv28a}
    \\
    &\tilde G_\teol^\mn
    =
    2\pi\delta(q_\prl)
    \ \frac{\kappa^4 m^2 q^2 N_{D-1}}{
      64
    }
    \bigg(
    \frac{D-7}{D-2} \eta^\mn_\parallel
    - \frac{(D-3)(3D-5)}{(D-2)^2}
    \Big(
    \eta_\bot^\mn - \frac{\ q^\mu q^\nu}{q^2}
    \Big)
    \bigg)
    \ ,
    \labelt{vv28b}
    \\
    &\tilde H_\teol^\mn =
    -\alpha
    \ 2\pi\delta(q_\prl)
    \ \frac{\kappa^4 m^2 q^2 N_{D-1}}{
      16}
    \frac{D-3}{D-2}
    \maP^{\mn\rs} \qbmnd{\rho}{\sigma}
    \ .
    \labelt{vv28c}
  \end{align}
\end{subequations}
Note that $\ttau^\mn_\teol=\frac{4}{\kappa^2}\tGmn_\teol^\mn$.
We have chosen to work with $\tGmn_\teol^\mn$ instead of $\ttau^\mn_\teol$ due to it being analogous to $\tHmn_\teol^\mn$.
The functional dependence on $q^\mu$ is included in the integral $N_{D-1}$ which was defined in Eq.~\eqreft{vvv5}.
\par
We will find the metric and go to position space according to Eqs.~\eqreft{ext2} and~\eqreft{ver7}.
First let us briefly check if the results in Eqs.~\eqreft{vv28} obey the relations discussed in Eqs.~\eqreft{eom1}.
These where the conservation law of $\ttau^\mn$ and the analogous equation for $\tHmn^\mn$:
\begin{equation}
  {G_\tecl}^\mn_\ab \tHmn^\ab_\tenl = 0
  \ .
  \labelt{vv29}
\end{equation}
It is easily checked that the one-loop contribution to $\ttau^\mn$ in Eq.~\eqreft{vv28b} is conserved.
That is $q_\mu \tGmn^\mn_\teol$ disappears since $\tGmn^\mn_\teol$ depends on the combination $(\eta_\bot^\mn - \frac{\ q^\mu q^\nu}{q^2})$.
It is less apparent that Eq.~\eqreft{vv29} is satisfied for the one-loop contribution to $\tHmn^\mn$.
\par
Let us verify that ${G_{(c)}}^\mn_\ab \tHmn_\teol^\ab$ vanishes.
We insert the tensor structure of $\tHmn_\teol^\ab$ and then that of ${G_{(c)}}^\mn_\ab$:
\begin{align}
  {G_c}_\mn^\ab \maP^{\mn\rs} \qbmnd{\rho}{\sigma}
  &=
  \qbmnd{\mu}{\nu}
  -2 \maJ_\mn^\ab \maP_\ab^\rs \qbmnd{\rho}{\sigma}
  \labelt{vv30}
\end{align}
We then use Eq.~\eqreft{jpj1} to reduce the second term on the right-hand side:
\begin{subequations}
  \label{eqn:vv31}
  \begin{align}
    \maJ_\mn^\ab \maP_\ab^\rs \qbmnd{\rho}{\sigma}
    &=
    \maJ_\mn^\ab \maP_\ab^\rs \maJ_\rs^\gd \eta_\gd
    \labelt{vv31a}
    \\
    &=
    \frac{1}{2} \maJ_\mn^\gd \eta_\gd
    \\
    &=
    \frac{1}{2}
    \qbmnd{\mu}{\nu}
  \end{align}
\end{subequations}
Combining Eqs.~\eqreft{vv30} and~\eqreft{vv31} we see that ${G_{(c)}}^\mn_\ab \tHmn_\teol^\ab=0$ is satisfied.
These equations for $\tHmn^\mn_\teol$ and $\tGmn^\mn_\teol$ secure that the second order contribution to the metric is independent of $\xi$.
\par
We need to contract $\maM_\teol^\mn$ with the graviton propagator to get the metric.
We can use Eq.~\eqreft{ver7} or equivalently:
\begin{align}
  \thmn^{\cGn{2}}_\mn
  =
  2 \frac{G^\tecl_{\mn\ab}}{q^2} \tGmn^\ab_\teol
  +
  2 \frac{G^\tegf_{\mn\ab}}{q^2} \tHmn^\ab_\teol
  \labelt{vv32}
\end{align}
If we exchange $G^\tecl$ and $G^\tegf$ in Eq.~\eqreft{vv32} with $\maPi$ we get Eq.~\eqreft{ver7}.
We can do this due to Eq.~\eqreft{vv29} and the conservation law of $\ttau^\mn$.
After propagating with $G^\tecl$ and $G^\tegf$ we get:
\begin{subequations}
  \label{eqn:vv33}
  \begin{align}
    &\frac{G^\tecl_{\mn\ab}}{q^2} 2 \tGmn^\ab_\teol
    =
    2\pi\delta(q_\prl)
    \frac{\kappa^4 m^2 N_{D-1}}{32}
    \bigg(
    4\frac{(D-3)^2}{(D-2)^2} \eta^\prl_\mn
    + \frac{(D-3)(3D-5)}{(D-2)^2} \frac{\ q_\mu q_\nu}{q^2}
    - \frac{D-7}{(D-2)^2} \eta^\bot_\mn
    \bigg)
    \labelt{vv33a}
    \\
    &\frac{G^\tegf_{\mn\ab}}{q^2} 2 \tHmn^\ab_\teol
    =
    -
    2\pi\delta(q_\prl)
    \alpha
    \frac{\kappa^4 m^2 N_{D-1}}{8} \frac{D-3}{D-2} \frac{\ q_\mu q_\nu}{q^2}
    \labelt{vv33b}
  \end{align}
\end{subequations}
It is especially simple to compute Eq.~\eqreft{vv33b}.
The tensor structure of $\tHmn_\teol^\mn$ is
\begin{align}
  \maP^{\mn\rs} \qbmnd{\rho}{\sigma}
  \ ,
\end{align}
and it is thus straightforward to contract it with $\maPi_{\mn\ab}$.
In case of Eq.~\eqreft{vv33a} we applied $\maPi$ to each term of Eq.~\eqreft{vv28b}.
\par
We can now write down an expression for $h^{\cGn{2}}_\mn$ in momentum space:
\begin{align}
  \thmn^{\cGn{2}}_\mn
  =
  2\pi\delta(q_\prl)
  \frac{\kappa^4 m^2 N_{D-1}}{32}
  \frac{(D-3)^2}{(D-2)^2}
  \bigg(
  4 \eta^\prl_\mn
  +
  \Big(
  \frac{3D-5}{D-3}
  -4\alpha\frac{D-2}{D-3}
  \Big)\frac{\ q_\mu q_\nu}{q^2}
  - \frac{D-7}{(D-3)^2} \eta^\bot_\mn
  \bigg)
  \labelt{vv38}
\end{align}
We recall the definition of $N_{D-1}$:
\begin{align}
  N_{D-1}
  =
  \frac{
    \Omega_{D-3} \sqrt{-q^2_\bot}^{D-5}
  }{
    4(4\pi)^{D-3} \cos(\frac{\pi}{2}D)
  }
  \labelt{vv39}
\end{align}
Again, this expression diverges in odd space-times.
For now, we will ignore this and continue, and in Sec.~\ref{sec:AppearanceOf} we analyze it in detail.
\par
To transform $\thmn_\mn^{\cGn{2}}$ to position space we need to transform $({-q^2_\bot})^{\frac{D-5}{2}}$ and the same function multiplied by $\frac{\ q_\mu q_\nu}{q^2}$.
For this purpose we can use the formula:
\begin{equation}
  \int \frac{d^dq_\bot}{(2\pi)^d} \ e^{-ix_\bot q_\bot}
  (-q_\bot^2)^{\frac{n}{2}} =
  \frac{2^n}{\sqrt{\pi}^d}
  \frac{\Gamma(\frac{d+n}{2})}{\Gamma(-\frac{n}{2})}
  \frac{1}{(-x_\bot^2)^\frac{d+n}{2}}
  \ ,
  \labelt{vv34}
\end{equation}
The same integral is found in Ref.~\cite{Collado:2018isu}.
It transforms $({-q^2_\bot})^{\frac{n}{2}}$ into a similar function of $x_\bot^2$.
We could have used the dimensions of the integral to make an ansatz for the result.
Note, that when $\frac{n}{2}$ is an integer the integral vanishes since we divide by $\Gamma(-\frac{n}{2})$.
This is so, since when $\frac{n}{2}$ is an integer, then $(-q_\bot^2)^{\frac{n}{2}}$ is an analytic function in $q_\bot^2$.
Such a function does not contribute in the classical limit.
We will see, however, that the vanishing of this integral cancels the divergence of $N_{D-1}$.
We will understand this better in Sec.~\ref{sec:AppearanceOf}.
\par
Using the integral in Eq.~\eqreft{vv34} we transform $N_{D-1}$ and get:
\begin{align}
  \int
  \frac{d^{D-1}q_\bot\ e^{-iq_\bot x_\bot}\ }{(2\pi)^{D-1}}
  N_{D-1}
  =
  \Bigg(
  \frac{
    1
  }{
    \Omega_{D-2}\ (D-3)\ 
    \sqrt{-x_\bot^2}^{D-3}
  }
  \Bigg)^2
  \ .
  \labelt{vv36}
\end{align}
It is clear that the divergence in $N_{D-1}$ has disappeared in position space after using the integral in Eq.~\eqreft{vv34}.
The Fourier transform of $\frac{\ q_\mu q_\nu}{q^2}$ can also be derived directly from Eq.~\eqreft{vv34}.
We differentiate it with respect to $x^\mu$ twice:
\begin{align}
  \frac{\partial}{\partial x_\bot^\mu}
  \frac{\partial}{\partial x_\bot^\nu}
    \int \frac{d^dq_\bot}{(2\pi)^d} \ e^{-ix_\bot q_\bot}\ 
    (-q_\bot^2)^{\frac{n}{2}}
    =
    -  \int \frac{d^dq_\bot}{(2\pi)^d} \ e^{-ix_\bot q_\bot}\ 
    q^\bot_\mu q^\bot_\nu (-q_\bot^2)^{\frac{n}{2}}
    \labelt{vv48}
\end{align}
Using this method we get the Fourier transform of $N_{D-1} \frac{q^\bot_\mu q^\bot_\nu}{q^2_\bot}$:
\begin{equation}
  \int \frac{d^{D-1}q_\bot \ e^{-iq_\bot x_\bot}}{(2\pi)^{D-1}}
  N_{D-1}
  \frac{q^\bot_\mu q^\bot_\nu}{q^2_\bot}
  =
    \Bigg(
    \frac{
      1
  }{
    \Omega_{D-2}\ (D-3)\ 
    \sqrt{-x_\bot^2}^{D-3}
  }
  \Bigg)^2
  \frac{1}{D-5}
  \Big(
  2(D-3)\frac{x^\bot_\mu x^\bot_\nu}{x_\bot^2}
  -\etar{\mn}
  \Big)
  \ .
  \labelt{vv35}
\end{equation}
This integral is seen to diverge in space-time $D=5$ and the integral is then only valid in $D\neq5$.
This is due to the fact that we have not subtracted the divergent term in $N_{D-1}$.
It only becomes problematic in $D=5$ where a logarithmic dependence on the radial coordinate appears in the metric.
\par
Using the integrals from Eqs.~\eqreft{vv36} and~\eqreft{vv35} we can transform $\thmn_\mn^{\cGn{2}}$ to position space for all space-time dimensions $D\neq5$.
We will use the \STM\ parameter, $\mu$, defined in Eq.~\eqreft{mud1} and we get:
\begin{align}
  \hspace*{-.5cm}
  h^{\cGn{2}}_\mn =
  \frac{\mu^2}{r^{2(D-3)}}
  \Bigg(
  \oov{2} \etat{\mn}
  -
  \frac{
    (4\alpha-3)D
    -8\alpha + 5
  }{
    4(D-5)
  }
  \Xmnb{\mu}{\nu}
  -
  \frac{
    2(1-\alpha)D^2
    -(13-10\alpha)D
    +25-12\alpha
  }{
    4(D-3)^2(D-5)
  }
  \etar{\mn}
  \Bigg)
  \ .
  \labelt{vv37}
\end{align}
Here, we have introduced the variable $r$ defined by $r^2=-x_\bot^2$.
In the reference frame of $k^\mu$ we have $r=\absvec{x}$.
The equation is valid for all $D\neq5$ and for all values of the gauge parameter $\alpha$.
The special choice of $\alpha=\frac{5}{6}$ removes the pole in $D=5$ and for this special choice Eq.~\eqreft{vv37} can also be used in $D=5$.
This is easy to see in momentum space in Eq.~\eqreft{vv38} where the coefficient of $\frac{\ q_\mu q_\nu}{q^2}$ disappears in $D=5$ for $\alpha=\frac{5}{6}$.
The same conclusion holds for any choice of $\alpha$ which depends on $D$ in such a way that $\alpha=\frac{5}{6}$ when $D=5$.
We would expect higher order corrections to the metric to depend on arbitrary powers of $\alpha$.
\par
In the literature similar expansions are found in Refs.~\cite{Collado:2018isu,BjerrumBohr:2002ks,Goldberger:2004jt}.
Both Refs.~\cite{BjerrumBohr:2002ks,Goldberger:2004jt} work in $D=4$ while Ref.~\cite{Collado:2018isu} derives the second order contribution to the metric in arbitrary dimensions in \dDo-gauge.
We find that their result agrees with our Eq.~\eqreft{vv37} when we put $\alpha=0$.
Note, however, that there is a misprint in their article.
The relevant equation in their article~\cite{Collado:2018isu} is Eq.~(5.34).
In the fourth line of this equation the square on $(D-p-3)^2$ should be removed, so that instead it reads $(D-p-3)$.
We thank Paolo Di Vecchia for confirming this.
Note, also, that in their equation $p$ should be set to zero and $R_p$ should be related to $\mu$ appropriately.
The divergence of Eq.~\eqreft{vv37} in $D=5$ is not discussed in Ref.~\cite{Collado:2018isu}.
\par
We have also checked that Eq.~\eqreft{vv37} in $D=4$ for $\alpha=1$ agrees with the standard formula for the Schwarzschild metric in harmonic gauge from e.g. Ref.~\cite{Weinberg:1972kfs}.
For general $\alpha$ and $D\neq5$ we have compared it to the classical derivation in Sec.~\ref{sec:ClassicalDerivation} and we find agreement.
\par
As expected the metric in Eq.~\eqreft{vv37} satisfies the gauge condition $G_\sigma=0$ which at second order in $G_N$ reads:
\begin{align}
  G^{\cGn{2}}_\sigma =
  \maP^\mn_\rs {h_{\cGn{2}}}_\mn^{,\rho}
  - \alpha \Gamma_{\sigma\ab}^{\rho\mn} \ h_\cGno^\ab \ h^{\cGno}_{\mn,\rho}
  \ .
\end{align}
Here, $h_\cGno^\mn$ should be taken from Eq.~\eqreft{me32}.
\section{Appearance of Logarithms in the Metric}
\labelx{sec:AppearanceOf}
What happened to the divergence in $N_{D-1}$ and why is Eq.~\eqreft{vv37} for $h^{\cGn{2}}_\mn$ not valid in $D=5$?
First, we will consider the $N_{D-1}$ integral again and remove the divergent part.
Then we will use the regularized expression for $N_{D-1}$ to derive the metric in $D=5$.
Here, it turns out that a logarithmic dependence on the radial coordinate appears in position space.
We will then analyze this phenomenon in terms of gauge transformations and show that besides $D=5$ we only expect logarithms in $D=4$.
\par
The divergent factor in the expression for $N_{D-1}$ in Eq.~\eqreft{vv39} is $1/\cos(\frac{\pi}{2}D)$.
This is so since $\cos(\frac{\pi}{2}D)$ vanishes when $D$ is an odd integer.
It is then convenient to analyze $N_{D-1}$ for the cases of even and odd $D$ separately.
\par
When $D$ is even the expression in Eq.~\eqreft{vv39} is finite and reduces to
\begin{align}
  N_{D-1} = \frac{(-1)^n}{8(16\pi)^n n!} \frac{(-q^2_\bot)^n}{\sqrt{-q^2_\bot}}
  \ ,
\end{align}
where $n = \frac{1}{2}(D-4)$ is an integer and $n\geq0$.
This expression is finite and non-analytic in $q^2$ and it presents no difficulties when it is Fourier transformed to position space with Eq.~\eqreft{vv34}.
\par
Next let us consider odd $D$.
In this case, the expression for $N_{D-1}$ in Eq.~\eqreft{vv39} is divergent.
Instead, we expand $N_{D-1}$ in a Laurent series in a small deviation from the odd dimension.
We write $D=5+2n+2\epsilon$ and expand $N_{D-1}$ in $\epsilon$:
\begin{align}
  N_{D-1}
  =
  \frac{1}{\epsilon} \frac{(-1)^{n+1}\Omega_{2+2n} (-q^2_\bot)^n}{(4\pi)^{3+2n}}
  +
  \frac{(-1)^{n+1} n!}{16\pi^2 (4\pi)^n (1+2n)!}
  (-q^2_\bot)^n
  \ln(-q^2_\bot r_0^2)
  + \ellipsis
\end{align}
There are other finite terms in the expansion but the one shown is the only non-analytic term.
The pole term is an analytic function, namely an integer power of $q^2$.
The arbitrary length scale, $r_0$, comes from the dimensional dependence of $G_N$.
We conclude that the relevant part of $N_{D-1}$ for the classical limit is
\begin{align}
  N_{D-1}
  =
    \frac{(-1)^{n+1} n!}{16\pi^2 (4\pi)^n (1+2n)!}
  (-q^2_\bot)^n
    \ln(-q^2_\bot r_0^2)
    \ ,
    \labelt{vv40}
\end{align}
where still $D=5+2n$.
The Fourier transform to position space of this expression for $N_{D-1}$ is equivalent to the transformation of the formula for $N_{D-1}$ in Eq.~\eqreft{vv39}.
We can use Eq.~\eqreft{vv34} which gives the Fourier transform of $(-q^2)^{\frac{n}{2}}$ if we use the following expression for the logarithm:
\begin{align}
    \ln(-q^2)
  = \frac{1}{\epsilon}
  \Big(
  (-q^2)^{\epsilon}-1
  \Big)
  \labelt{vv41}
\end{align}
Here, $\epsilon$ is an infinitesimal parameter.
It is now clear, that the Fourier transforms of both expressions for $N_{D-1}$ are the same.
This is so, since when we insert Eq.~\eqreft{vv41} into Eq.~\eqreft{vv40} the constant term in Eq.~\eqreft{vv41}, $(-\frac{1}{\epsilon})$ will be transformed to zero because of the property of Eq.~\eqreft{vv34} discussed below that equation.
The $q$-dependent term in Eq.~\eqreft{vv41} then makes the expression for $N_{D-1}$ in Eq.~\eqreft{vv40} equivalent to the one in Eq.~\eqreft{vv39}.
\par
However, this is not the case when we transform $N_{D-1} \frac{q^\bot_\mu q^\bot_\nu}{q^2_\bot}$.
Here, the transformed expression in position space is free of divergence when we use Eq.~\eqreft{vv40} but as we have seen it has a divergence in $D=5$ when we use Eq.~\eqreft{vv39}.
Let us indicate how Eqs.~\eqreft{vv41} and~\eqreft{vv34} can be used to transform ${N_{D-1} \frac{q^\bot_\mu q^\bot_\nu}{q^2_\bot}}$ to momentum space when $D$ is odd using Eq.~\eqreft{vv40} for the definition of $N_{D-1}$.
This will show how formulas which are all continuous in the dimension of space-time, $D$, can produce such different results in $D\neq5$ and $D=5$.
\par
First, we find:
\begin{align}
  \hspace*{-1cm}
  \int \frac{d^{D-1}q_\bot \ e^{-iq_\bot x_\bot}}{(2\pi)^{D-1}}
  \frac{  N_{D-1}}{q^2_\bot}
  &=
  \frac{(-1)^{n} n!}{16\pi^2 (4\pi)^n (1+2n)!}
  \int \frac{d^{D-1}q_\bot \ e^{-iq_\bot x_\bot}}{(2\pi)^{D-1}}
  (-q^2_\bot)^{n-1}
  \ln(-q^2_\bot )
  \labelt{47}
  \\
  &=
  \frac{(-1)^{n} n!}{16\pi^2 (4\pi)^n (1+2n)!}
  \int \frac{d^{D-1}q_\bot \ e^{-iq_\bot x_\bot}}{(2\pi)^{D-1}}
  \frac{1}{\epsilon}
  \Big(
  (-q^2_\bot)^{n-1+\epsilon}
  -
  (-q^2_\bot)^{n-1}
  \Big)
  \nonumber{}
\end{align}
In the first line we inserted the definition of $N_{D-1}$ from Eq.~\eqreft{vv40} and in the second line we used Eq.~\eqreft{vv41} to rewrite the logarithm function.
We have put the arbitrary scale to unity, $r_0=1$, and $n$ is still defined by $D=5+2n$.
We can now use Eq.~\eqreft{vv34} to evaluate the integral:
\begin{align}
  \hspace*{-1cm}
  \int \frac{d^{D-1}q_\bot \ e^{-iq_\bot x_\bot}}{(2\pi)^{D-1}}
  \frac{  N_{D-1}}{q^2_\bot}
  =
  \frac{(-1)^{n} n!}{16\pi^2 (4\pi)^n (1+2n)!}
  \frac{4^{n-1}}{\pi^{2+n} (-x_\bot^2)^{2n+1}}
  \frac{1}{\epsilon}
  \Big(
  \frac{4^\epsilon \Gamma(2n+1+\epsilon)}{\Gamma(1-n-\epsilon)(-x^2_\bot)^\epsilon}
  -
  \frac{\Gamma(2n+1)}{\Gamma(1-n)}
  \Big)
  \labelt{vv46}
\end{align}
All dependence on the infinitesimal parameter $\epsilon$ has been gathered in the final factor.
Of course, $\epsilon$ is only introduced as an intermediate parameter and disappears from the final expression.
Focus on the last factor:
\begin{align}
  \frac{1}{\epsilon}
  \Big(
  \frac{4^\epsilon\Gamma(2n+1+\epsilon)}{\Gamma(1-n-\epsilon)(-x^2_\bot)^\epsilon}
  -
  \frac{\Gamma(2n+1)}{\Gamma(1-n)}
  \Big)
  =
  \Gamma(2n+1)
  \bigg(
  \frac{\psi(1-n) }{\Gamma(1-n)}
  +
  \frac{\psi(2n+1)
  -
  \ln(-\frac{x^2_\bot}{4})}{\Gamma(1-n)}
  \bigg)
  \labelt{vv45}
\end{align}
On the right-hand side, $\epsilon$ has disappeared again.
The function, $\psi(z)$, is the digamma function which is used to differentiate the $\Gamma$-function using $\Gamma'(z) = \Gamma(z) \psi(z)$.
To arrive at the Fourier transform of $N_{D-1} \frac{q^\bot_\mu q^\bot_\nu}{q^2_\bot}$ we would now insert Eq.~\eqreft{vv45} into Eq.~\eqreft{vv46} and then differentiate Eq.~\eqreft{vv46} twice with respect to $x_\bot^\mu$ as in Eq.~\eqreft{vv48}.
However, we can already draw important conclusions without doing the final steps in detail.
\par
Focus on the right-hand side of Eq.~\eqreft{vv45} where there are two terms in the brackets.
The first term, ${\psi(1-n)/\Gamma(1-n)}$ is finite and non-zero for all integers $n\geq0$ (for $n\geq1$ both numerator and denominator diverge and the limit is taken).
However, for integer $n\geq0$ the second term is non-zero only for $n=0$.
This is so since the numerator of the second term is finite for all $n\geq0$ while the denominator $\Gamma(1-n)$ diverges when $n$ is an integer $\geq1$.
Hence, the second term only plays a role in $D=5$ and in this way $D=5$ is so different from other dimensions.
\par
If we proceed with the computations in Eqs.~\eqreft{vv46} and~\eqreft{vv45}, the first term on the right hand side of Eq.~\eqreft{vv45} will produce the same results in all odd $D\neq5$ as the integral in Eq.~\eqreft{vv35} which was valid for all $D\neq5$.
In $D=5$, i.e. $n=0$, both terms in Eq.~\eqreft{vv45} are finite and the resulting integral would then also be finite.
Let us put $n=0$ in Eq.~\eqreft{vv45}:
\begin{align}
  \frac{1}{\epsilon}
  \Big(
  \frac{4^\epsilon\Gamma(1+\epsilon)}{\Gamma(1-\epsilon)(-x^2_\bot)^\epsilon}
  -
  1
  \Big)
  =
  2\psi(1)
  -
  \ln(-\frac{x^2_\bot}{4})
  \ .
  \labelt{hej1}
\end{align}
Using that $\psi(1)=-\gamma$ where $\gamma$ is the Euler-Mascheroni constant we get that this simplifies to:
\begin{align}
  \frac{1}{\epsilon}
  \Big(
  \frac{4^\epsilon\Gamma(1+\epsilon)r_0^{2\epsilon}}{\Gamma(1-\epsilon)(-x^2_\bot)^\epsilon}
  -
  1
  \Big)
  =
  -
  \ln(-\frac{x^2_\bot e^{2\gamma}}{4r_0^2})
  \ .
  \labelt{hej2}
\end{align}
Here, we reintroduced the arbitrary scale, $r_0$.
This expression should be inserted in Eq.~\eqreft{vv46} which when differentiated twice with respect to $x_\bot^\mu$ would give the desired integral.
It is then clear that the result will depend on the combination $\ln(-\frac{x^2_\bot e^{2\gamma}}{4r_0^2})$.
\par
We compute the final steps and find the result for $N_{4} \frac{q^\bot_\mu q^\bot_\nu}{q^2_\bot}$ in position space:
\begin{align}
  \hspace*{-.5cm}
  \int \frac{d^5q \ \delta(q_\prl) \ e^{-iqx}}{(2\pi)^4}
  N_4
  \frac{q_\mu q_\nu}{q^2}
  =
  -\frac{1}{32\pi^4\sqrt{-x_\bot^2}^{4}}
  \bigg(
  \etar{\mn}
  -6 \Xmnb{\mu}{\nu}
  -\Big(
  \etar{\mn} - 4\Xmnb{\mu}{\nu}
  \Big)
  \ln(-\frac{x_\bot^2 e^{2\gamma}}{4r_0^2})
  \bigg)
  \ .
  \labelt{vv49}
\end{align}
Again, $\gamma$ is the Euler-Mascheroni constant which can be removed by redefining the scale $r_0$.
\par
Using Eq.~\eqreft{vv49} we compute the second order metric in $D=5$.
After a redefinition of $r_0$ we get:
\begin{eqnarray}
  h^{(2)}_\mn =
  \frac{\mu^2}{r^{4}}
  \Big(
  \oov{2} \etat{\mn}
  - \frac{2(6\alpha-5)\ln\frac{r}{r_0}-1}{16} \etar{\mn}
  + \frac{(6\alpha-5)(4\ln\frac{r}{r_0}-1)}{8} \Xmnb{\mu}{\nu}
  \Big)
  \ .
  \labelt{vv50}
\end{eqnarray}
Again, $r^2 = -x_\bot^2$.
We have not found this result in earlier literature, although a similar situation occurs in $D=4$ at third order in $G_N$ in \dDo\  gauge \cite{Goldberger:2004jt}.
Note that we can make the logarithm disappear with the special choice $\alpha=\frac{5}{6}$.
The arbitrary scale, however, would in principle still be there.
In analogy, in $D=4$ we know that for $\alpha=1$ in harmonic gauge there are no logarithms.
\par
How can it be, that the metric in $D=5$ depends on the arbitrary scale $r_0$?
It will now be seen that the freedom in choosing this scale corresponds to a redundant gauge freedom.
Thus, there is a certain gauge transformation of $h_\mn$ which leaves $h_\mn$ in the gauge $G_\sigma=0$.
This case is similar to the case of linearized gravity in \dDo-gauge where the coordinates can be translated by a harmonic function while the metric stays in \dDo-gauge.
The exact gauge transformations of the graviton field where discussed in Sec.~\ref{sec:GaugeTransformations}.
\par
Let us transform our coordinates according to:
\begin{equation}
  x^\mu
  \rightarrow
  x^\mu
  + \beta
  \frac{\mu^2}{r^4}
  x_\bot^\mu
  + \ ...
  \labelt{cor1}
\end{equation}
This is of the form $x^\mu\rightarrow x^\mu+\epsilon^\mu(x)$.
The ellipsis indicates higher order corrections to this transformation.
To second order in $G_N$ only the second order metric, $h^{\cGn{2}}_\mn$, will be changed by the transformation in Eq.~\eqreft{cor1}.
The change will be
\begin{align}
  h^{\cGn{2}}_\mn
  \rightarrow
  h^{\cGn{2}}_\mn
  -
  \epsilon_{\mu,\nu}
  -
  \epsilon_{\nu,\mu}
  \ ,
  \labelt{me25}
\end{align}
where $\epsilon_\mu=\beta\frac{\mu^2x^\bot_\mu}{r^4}$.
Since the first order metric is unchanged only the second order gauge condition will be changed.
We find that:
\begin{align}
  G_\sigma^{\cGn{2}}
  \rightarrow
  G_\sigma^{\cGn{2}}
  +\partial^2 \epsilon_\sigma
  \labelt{me26}
\end{align}
Thus, if $\epsilon_\sigma$ is a harmonic function, we will stay in the \dDo-type gauge $G_\sigma=0$.
The situation is thus completely analogous to the case of linearized gravity.
\par
The choice of $\epsilon^\mu=\beta\frac{\mu^2x_\bot^\mu}{r^4}$ makes $\epsilon^\mu$ a harmonic function.
This can be seen from the following argument.
We start from the potential from a point source which obeys:
\begin{align}
  \partial^2 \frac{1}{r^{D-3}}
  \propto
  \delta^{D-1}(x_\bot)
  \labelt{me27}
\end{align}
Then we take the partial derivative on each side:
\begin{align}
  \partial^2 \frac{x_\mu^\bot}{r^{{D-1}}}
  \propto
  \partial_\mu \partial^2 \frac{1}{r^{D-3}}
  \propto
  \partial_\mu \delta^{D-1}(x_\bot)
  \labelt{me28}
\end{align}
Putting $D=5$ we get the result that $\epsilon^\mu$ is harmonic when $r\neq0$.
\par
The conclusion is that the metric stays in the \dDo-type gauge when we make the coordinate transformation in Eq.~\eqreft{cor1}.
This will introduce the coefficient $\beta$ in Eq.~\eqreft{cor1} as a free parameter in the metric.
How does the metric change?
We use Eq.~\eqreft{me25}:
\begin{equation}
  h^{\cGn{2}}_\mn
  \rightarrow
  h^{\cGn{2}}_\mn
  - 2\beta
  \frac{\mu^2}{r^4}
  \Big(
  -4 \Xmnb{\mu}{\nu}
  +
  \eta^\bot_\mn
  \Big)
\end{equation}
Let us compare this to a rescaling of the arbitrary length scale $r_0$ in $h^{\cGn{2}}_\mn$ in Eq.~\eqreft{vv50}.
We find that when we let $r_0 \rightarrow \gamma r_0$ then $h^{\cGn{2}}_\mn$ changes by:
\begin{equation}
  h^{\cGn{2}}_{\mn}
  \rightarrow
  h^{\cGn{2}}_{\mn}
  +
  \ln(\gamma)
  \frac{6\alpha-5}{8}
  \frac{\mu^2}{r^4}
  \Big(
  -4 \Xmnb{\mu}{\nu} + \delrb{\mn}
  \Big)
\end{equation}
We see that the freedom to change coordinates according to Eq.~\eqreft{cor1} exactly corresponds to the freedom in choosing the arbitrary scale $r_0$.
\par
What happens in space-times different than $D=5$?
Here, the equivalent transformation would be
\begin{align}
  \epsilon^\sigma = \beta \frac{x_\bot^\sigma}{r^{D-1}}
  \ ,
  \labelt{me29}
\end{align}
which is seen to be a harmonic function in Eq.~\eqreft{me28}.
In contrast to $\beta$ in the transformation from Eq.~\eqreft{cor1}, $\beta$ in Eq.~\eqreft{me29} is dimensionful.
\par
We can introduce an arbitrary parameter in the metric in any dimension if we translate our coordinates with $\epsilon^\sigma$ from Eq.~\eqreft{me29}.
However, we only expect logarithms to appear in $D=4$ besides $D=5$.
The reason is to be found in the dimension of $\beta$.
Only in $D=4$ and $D=5$ is it comparable to the dimension of the \STM\ parameter $\mu$.
The mass dimension of $\beta$ is found to be $(-D+1)$ and that of $\mu$ is $(-D+3)$.
When can $\beta$ be expressed as an integer power of $\mu$?
This requires
\begin{align}
  -D+1 = m(-D+3)
  \ ,
  \labelt{me30}
\end{align}
for some integer $m$.
For $D=4$ we find $m=3$ while for $D=5$ we get $m=2$.
For other values of $D\geq4$ we find that $m$ is not an integer.
The interpretation is that in $D=5$ a logarithm appears at second order, while in $D=4$ it appears at third order.
This conclusion is in agreement with the results in this chapter as well as Ref.~\cite{Goldberger:2004jt}.
In Ref.~\cite{Goldberger:2004jt} the Schwarzschild metric in $D=4$ was computed perturbatively to third order in \dDo-gauge and it was found that a logarithm appears in the metric at this order.

\section{Classical Derivation of the \STM\ Metric}
\labelx{sec:ClassicalDerivation}

In this section we derive the \STM\ metric with methods from classical general relativity.
The method is analogous to that in Weinberg~\cite{Weinberg:1972kfs} where the Schwarzschild metric is derived in harmonic coordinates in $D=4$ by a coordinate transformation from the Schwarzschild metric in spherical coordinates.
Here we work in arbitrary dimensions $D\geq4$ with the generalized \dDo-type gauge from Sec.~\ref{sec:GaugeFixing}.
That is, we will derive the \STM\ metric perturbatively to second order in $G_N$ in coordinates which satisfy $G_\sigma=0$ where $G_\sigma$ is the function in Eq.~\eqreft{gau3}.
\par
The \STM\ metric was given in Eq.~\eqreft{n215} and is:
\begin{align}
  &d\tau^2 =
  (1-\frac{\mu}{R^\dmt}) dt^2
  -\frac{1}{1-\frac{\mu}{R^\dmt}} dR^2
  - R^2 d\Omega^2_{D-2}
  \labelt{met1}
\end{align}
The \STM\ parameter $\mu$ was defined in Eq.~\eqreft{mud1} and is proportional to $mG_N$.
Also, we use $n=D-3$.
In Eq.~\eqreft{met1} we have used $R$ as the radial coordinate.
The goal is to change to new coordinates in terms of a new radial coordinate, $r$.
The old radial coordinate $R$ is a function of the new one, $r$, and we expand $R(r)$ perturbatively in $r$ and determine the coefficients of this expansion from the gauge condition $G_\sigma=0$.
\par
The new coordinates, $x^\mu$, should be Cartesian-like and are defined in terms of $r$ by passing from the spherical coordinates $r$ and angle variables to Cartesian coordinates:
\begin{align}
  &x^0 = t
  \labelt{met2}
  \\
  &x^1 = r \cos(\theta_1) \nonumber \\
  &x^2 = r \sin(\theta_1)\cos(\theta_2) \nonumber \\
  &x^3 = r \sin(\theta_1)\sin(\theta_2)\cos(\theta_3) \nonumber \\
  &... \nonumber \\
  &x^{D-1} = r \sin(\theta_1)\sin(\theta_2) ... \sin(\theta_{D-2})
  \nonumber{}
\end{align}
Here $x_i$ are the Cartesian-like coordinates and $\theta_i$ are the angle coordinates where $i=1\ ...\ (D-2)$.
The angle coordinates are included in the \STM\ metric in $d\Omega^2_{D-2}$.
The angles $\theta_i$ run from $0$ to $\pi$ when $i=1\ ...\ (D-3)$ and $\theta_{D-2}$ runs from $0$ to $2\pi$.
\par
The new coordinates $x^\mu$ are interpreted as a Lorentz covariant vector and indices on $x^\mu$ are then raised and lowered with the flat space metric.
We will also use the same Lorentz covariant notation as introduced in Sec.~\ref{sec:GeneralRelativity2}.
We want to write the \STM\ metric in terms of the new coordinates.
We use the following relations:
\begin{subequations}
  \label{eqn:met3}
\begin{align}
  &dR^2 = \frac{dR^2}{dr^2}(\frac{x_\bot dx_\bot}{r})^2
  \ ,
  \labelt{met3a}
  \\
  &d\Omega^2_{D-2} = -
  \frac{1}{r^2}
  \big(
  dx_\bot^2 + (\frac{x_\bot dx_\bot}{r})^2
  \big)
  \ .
  \labelt{met3b}
\end{align}
\end{subequations}
Here we have already generalized to the Lorentz covariant notation.
Then the radial variable $r$ is defined as $r^2 = -x_\bot^2$.
Thus, although the \STM\ metric in Eq.~\eqreft{met1} describes a stationary particle the equations in terms of the Cartesian-like coordinates are easily generalized to describe a particle with any momentum.
\par
We insert Eqs.~\eqreft{met3} in the \STM\ metric from Eq.~\eqreft{met1}:
\begin{align}
  d\tau^2 =
  (1-\fcR) dt^2
  -\frac{1}{1-\fcR} \frac{dR^2}{dr^2} (\frac{x_\bot dx_\bot}{r})^2
  + \frac{R^2}{r^2} \Big(  dx_\bot^2 + (\frac{x_\bot dx_\bot}{r})^2  \Big)
  \labelt{met4}
\end{align}
We can read off the metric, $g_\mn$:
\begin{align}
  &g_{\mn} =
  (1-\fcR) \eta^\prl_\mn
  +\frac{
    1
  }{
    1-\fcR
  }
  \frac{dR^2}{dr^2}
  \Xmnb{\mu}{\nu}
  +
  \frac{R^2}{r^2}
  (\eta^\bot_\mn - \Xmnb{\mu}{\nu})
  \ .
  \labelt{met5}
\end{align}
We will expand this expression for the metric in $r$.
\par
We expand $R$ in the dimensionless quantity $\fcr$:
\begin{align}
  R = r(1 + a\fcr + b \ \Big(a\frac{\mu}{r^\dmt}\Big)^2 + ...)
  \labelt{met6}
\end{align}
The coefficient, $a$, has been included in the $\mu^2$ term of the expansion for convenience.
Determining $a$ gives the first order correction to the metric while $b$ gives the second.
Likewise $a$ is determined from the first order gauge condition $G_\sigma^{\cGno}=0$ and $b$ from the second order gauge condition $G_\sigma^{\cGn{2}}=0$.
\par
Expanding the metric from Eq.~\eqreft{met5} to first order, we find:
\begin{align}
  &g_{\mn} \approx
  \eta_\mn
  -\fcr \eta^\prl_\mn
  +2a\fcr \eta^\bot_{\mn}
  - \big( 2na-1 \big) \fcr \Xmnb{\mu}{\nu}
  \labelt{met7}
\end{align}
We read off the first order correction:
\begin{align}
  h^\cGno_\mn =
  -\fcr
  \Big(
  \eta^\prl_\mn
  -2a \eta^\bot_{\mn}
  + \big( 2na-1 \big) \Xmnb{\mu}{\nu}
  \Big)
  \ .
  \labelt{met8}
\end{align}
From Eq.~\eqreft{n419} the first order gauge condition is:
\begin{align}
  \maP^\mn_\rs {h^{\cGno}}_\mn^{,\rho} = 0
  \labelt{met9}
\end{align}
We want to insert $h^\cGno_\mn$ from Eq.~\eqreft{met8} into the gauge condition from Eq.~\eqreft{met9}.
It is thus necessary to compute the partial derivative of $h_\mn^\cGno$.
We get:
\begin{align}
  \partial_\sigma
  h^\cGno_\mn
  =
  -n\frac{\mu}{r^{n+1}}
  \Xrhob{\sigma}
  \Big(
  \eta^\prl_{\mn}
  - 2a\eta^\bot_{\mn}
  + \big(2na-1\big) \Xmnb{\mu}{\nu} \Big)
  -\oov{r^{n+1}} (2na-1) \lambda_{\sigma \mn}
  \labelt{me10}
\end{align}
Here $\lambda_{\sigma \mn}$ is a convenient tensor defined as:
\begin{subequations}
  \label{eqn:me11}
\begin{align}
  \lambda_{\sigma\mn}
  &=
  r\ \partial_\sigma
  \Xmnb{\mu}{\nu}
  \labelt{me11a}
  \\
  &=
  \Big( -\eta^\bot_{\sigma\mu} + \Xmnb{\sigma}{\mu} \Big) \Xrhob{\nu}
  + \Big( -\eta^\bot_{\sigma\nu} + \Xmnb{\sigma}{\nu} \Big) \Xrhob{\mu}
  \label{lam1e}
\end{align}
\end{subequations}
It is symmetric in $\mu$ and $\nu$.
\par
We insert $\partial_\sigma h^{\cGno}_\mn$ into the gauge condition and get:
\begin{align}
  G^\cGno_\sigma = \frac{\mu}{r^{n+1}}(2na-1)\Xrhob{\sigma}
  \labelt{me12}
\end{align}
In this way $a$ is determined to be $a=\frac{1}{2n}$.
Inserting this value for $a$ in the expression for $h^\cGno_\mn$ from Eq.~\eqreft{met8} produces the same result as we computed in Sec.~\ref{sec:NewtonPotential}.
\par
We will now derive the next term in the expansion of the \STM\ metric which is determined by $b$.
This is the second order term, which we want to compare with the amplitude computation.
We expand the coefficients of the metric from Eq.~\eqreft{met5} which are $(1-\frac{\mu}{R^{n}})$, $\frac{R^2}{r^2}$, and $1/(1-\frac{\mu}{R^{n}}) \frac{dR^2}{dr^2}$:
\begin{subequations}
  \label{eqn:me13}
\begin{align}
  &1-\fcR
  = 1 - \fcr + \frac{1}{2} \Big(\fcr\Big)^2 + ...
  \labelt{me13a}
  \\
  &\frac{R^2}{r^2} =
  1 + \frac{1}{\dmt} \fcr
  +(2b+1)\frac{1}{4\dmt^2} \Big(\fcr\Big)^2 + ...
  \\
  &\frac{dR^2}{dr^2}
  = 1 - \frac{\dmt-1}{\dmt} \fcr
  - \Big(
  2(2\dmt-1)b - (\dmt-1)^2
  \Big) \frac{1}{4\dmt^2} \Big(\fcr\Big)^2 +...
  \\
  &\frac{1}{1-\fcR}
  = 1 + \fcr + \frac{1}{2} \Big(\fcr\Big)^2 + ...
  \\
  &\frac{1}{1-\fcR} \frac{dR^2}{dr^2} =
  1 + \frac{1}{\dmt} \fcr
  +\Big(  1-\dmt(\dmt-2) - 2(2\dmt-1)b   \Big)
  \frac{1}{4\dmt^2} \Big(\fcr\Big)^2 + ...
  \labelt{me13e}
\end{align}
\end{subequations}
Inserting these expansions into the metric we get its expansion to second order in $G_N$:
\begin{align}
  g_\mn \approx
  \eta_\mn
  - \frac{\mu}{r^\dmt}
  \big(&\deltb{\mn} - \oov{\dmt} \delrb{\mn} \big)
  +\fcrs
  \big(\oov{2} \deltb{\mn}
  + \frac{2b+1}{4\dmt^2} \delrb{\mn}
  -\frac{4b+\dmt-2}{4\dmt} \Xmnb{\mu}{\nu}   \big)
  \labelt{me14}
\end{align}
And we read off $h^{\cGn{2}}_\mn$:
\begin{align}
  h^{\cGn{2}}_\mn
  =
  \frac{\mu^2}{r^{2\dmt}}
  \big(\oov{2} \deltb{\mn}
  + \frac{2b+1}{4\dmt^2} \delrb{\mn}
  - \frac{4b+\dmt-2}{4\dmt} \Xmnb{\mu}{\nu}  \big)
  \labelt{me15}
\end{align}
The coefficient $b$ is determined from the second order term of the gauge condition.
\par
From Eq.~\eqreft{n419} the second order gauge condition is found to be:
\begin{align}
  G^{\cGn{2}}_\sigma =
  \maP^\mn_\rs {h^{\cGn{2}}}_\mn^{,\rho}
  - \alpha \Gamma_{\sigma\ab}^{\rho\mn} \ h_\cGno^\ab \ h^{\cGno}_{\mn,\rho}
  \labelt{me16}
\end{align}
We insert $h^\cGno_\mn$ from Eqs.~\eqreft{met8} and~\eqreft{me10} and $h^{\cGn{2}}_\mn$ from Eq.~\eqreft{me15}.
The partial derivative of $h^{\cGn{2}}_\mn$ is:
\begin{align}
  \partial_\rho
  h^{\cGn{2}}_\mn
  =
  \frac{2\dmt \mu^2}{r^{2\dmt+1}}
  \Xrhob{\rho}
  \Big(
  \oov{2} \deltb{\mn}
  + \frac{2b+1}{4\dmt^2} \delrb{\mn}
  - \frac{4b+\dmt-2}{4\dmt} \Xmnb{\mu}{\nu}
  \Big)
  -\frac{\mu^2}{r^{2\dmt+1}} \frac{4b+\dmt-2}{4\dmt} \lambda_{\rho\mn}
  \labelt{me17}
\end{align}
The tensor, $\lambda_{\rho\mn}$, was defined in Eq.~\eqreft{me11}.
We compute the second order gauge condition in two steps.
First, $\alpha \Gamma_{\sigma\ab}^{\rho\mn} \ h_\cGno^\ab \ h^{\cGno}_{\mn,\rho}$, where we get:
\begin{align}
  \alpha \Gamma_{\sigma\ab}^{\rho\mn} \ h_\cGno^\ab \ h^{\cGno}_{\mn,\rho}
  =
  - \alpha
  \frac{\mu^2}{r^{2\dmt+1}}
  \frac{\dmt+1}{2} \Xrhob{\sigma}
  \labelt{me18}
\end{align}
Then, $\maP^\mn_\rs {h^{\cGn{2}}}_\mn^{,\rho}$, where we get:
\begin{align}
  \maP^\mn_\rs {h^{\cGn{2}}}_\mn^{,\rho}
  =
  -
  \frac{\mu^2}{r^{2\dmt+1}}
  \frac{\dmt^2+1+(\dmt-2)b}{2\dmt}
  \Xrhob{\sigma}
  \labelt{me19}
\end{align}
The coefficient $b$ is determined from these expressions.
\par
Combining Eqs.~\eqreft{me16}, \eqreft{me18} and~\eqreft{me19}, we get:
\begin{align}
  G_\sigma^{\cGn{2}}
  =
  -\frac{\mu^2}{2nr^{2n+1}}
  \Xrhob{\sigma}
  \Big(
  n^2+1+(n-2)b
  -
  \alpha
  n(n+1)
  \Big)
  \labelt{me21}
\end{align}
We determine $b$ from $G^{\cGn{2}}_\sigma=0$ to be:
\begin{align}
  b = \frac{-(1-\alpha)\dmt^2+\alpha\dmt -1}{\dmt-2}
  \labelt{me20}
\end{align}
This value of $b$ diverges when $n=2$, that is $D=5$.
This signals that the chosen expansion of $R(r)$ does not work in $D=5$ which is expected, since logarithms appear in $D=5$.
The value of $b$ is inserted in Eq.~\eqreft{me15} to get the second order contribution to the metric in $D\neq5$.
We find that Eq.~\eqreft{me15} with the value of $b$ in Eq.~\eqreft{me20} agrees with the amplitude computation from Eq.~\eqreft{vv37}.
\par
We will now specialize on $D=5$.
It is clear, that we cannot choose $b$ to satisfy Eq.~\eqreft{me21} when $n=2$ since $b$ disappears from the equation.
However, when $\alpha=\frac{5}{6}$ the equation is identically satisfied, and in this case $b$ can be chosen arbitrarily.
This is in agreement with the results from the amplitude computation.
\par
To find the metric in $D=5$ for general $\alpha$ it is necessary to modify the expansion of $R(r)$ in $r$.
We generalize the coefficient $b$ from the expansion in Eq.~\eqreft{met6} by letting $b\rightarrow b_0 + b_1 \ln(\frac{r}{r_0})$.
The expansion of $R(r)$ is then of the form:
\begin{align}
  R = r
  \bigg(
  1 + \frac{\mu}{2\dmt r^\dmt} +
  (b_0+b_1\ln(\frac{r}{r_0}))
  \Big(\frac{\mu}{2\dmt r^\dmt}\Big)^2 + ...
  \bigg)
  \labelt{me22}
\end{align}
Where $b_0$ and $b_1$ are coefficients to be determined and $r_0$ is a length scale.
If we define $b=b_0+b_1\ln(\frac{r}{r_0})$ then Eq.~\eqreft{me22} is similar to the earlier expansion in Eq.~\eqreft{met6} only that $b$ now includes a logarithmic dependence.
\par
The metric in Eq.~\eqreft{met5} is expanded with the new definition of $b$.
For example, the expansion of the coefficient in Eq.~\eqreft{me13e} is changed to:
\begin{align}
  \frac{1}{1-\fcR} \frac{dR^2}{dr^2} = (...) + 2b_1 \Big(\frac{\mu}{2\dmt r^\dmt}\Big)^2
  \ .
\end{align}
The ellipsis indicate the result in Eq.~\eqreft{me13e} to which the term linear in $b_1$ should be added.
In case of $h^{\cGn{2}}_\mn$ and its derivative we get:
\begin{subequations}
\begin{align}
  &h^{\cGn{2}}_\mn = (...) + 2 b_1 \Xmnb{\mu}{\nu} \Big(\frac{\mu}{2\dmt r^\dmt}\Big)^2
  \\
  &\partial_\rho h^{\cGn{2}}_\mn = (...) + b_1\frac{\mu^2}{4\dmt^2 r^{2\dmt+1}}(
  8\dmt \Xrhob{\sigma}\Xmnb{\mu}{\nu}
  -2 \Xrhob{\sigma}\delrb{\mn}
  +2 \lambda_{\sigma\mn}
  \Big)
\end{align}
\end{subequations}
Again, the ellipsis indicate that the results from Eqs.~\eqreft{me15} and~\eqreft{me17} should be inserted.
For example:
\begin{align}
    h^{\cGn{2}}_\mn
  =
  \frac{\mu^2}{r^{2\dmt}}
  \bigg(
  \oov{2} \deltb{\mn}
  + \frac{2b+1}{4\dmt^2} \delrb{\mn}
  +
  \big(
  -
  \frac{4b+\dmt-2}{4\dmt}
  +\frac{b_1}{2n^2}
  \big)
  \Xmnb{\mu}{\nu}
  \bigg)
  \labelt{me24}
\end{align}
As before, we insert $h^{\cGn{2}}_{\mn,\sigma}$ into the gauge condition from Eq.~\eqreft{me16}.
\par
For the term, which depends on $h^{\cGn{2}}_\mn$ in the gauge condition at second order we get:
\begin{align}
  \maP^\mn_\rs {h^{\cGn{2}}}_\mn^{,\rho}
  =
  \frac{\mu^2}{r^{2\dmt+1}}
  \Big(
  -\frac{\dmt^2+1+(\dmt-2)b}{2\dmt}
  +\frac{3\dmt-2}{4\dmt^2}b_1
  \Big)
  \Xrhob{\sigma}
\end{align}
Then, with the inclusion of the logarithmic dependence in $b$ the gauge condition ${G^{\cGn{2}}_\sigma=0}$ from Eq.~\eqreft{me16} becomes:
\begin{align}
  G^{\cGn{2}}_\sigma
  =
  -\frac{\mu^2}{2nr^{2n+1}}
  \Big(
  (\dmt-2)b-\frac{3\dmt-2}{2\dmt} b_1 - \alpha \dmt (\dmt+1) + \dmt^2 + 1
  \Big)
\end{align}
Recall the definition of $b=b_0+b_1\ln(\frac{r}{r_0})$.
When $n\neq2$ the logarithmic dependence in $b$ is forced to vanish and we get the same result as in Eq.~\eqreft{me20}.
Instead, when $n=2$ the coefficient of $b$ disappears and we get an equation which determines $b_1$ to be:
\begin{align}
  b_1 = 5-6\alpha
  \ .
  \labelt{me23}
\end{align}
As expected $b_1$ disappears for $\alpha=\frac{5}{6}$.
When $b_1\neq0$ the roles of $b_0$ and $r_0$ are similar and we put $b_0=0$.
Inserting $b_1$ from Eq.~\eqreft{me23} into the expression for $h^{\cGn{2}}_\mn$ in Eq.~\eqreft{me24} we get $h^{\cGn{2}}_\mn$ in $D=5$.
We find that this expression for $h^{\cGn{2}}_\mn$ agrees with the one from the amplitude computation in Eq.~\eqreft{vv50}.

\chapter{Concluding Remarks and Perspectives}
\labelx{sec:Conclusion}
We have analyzed several aspects of classical general relativity treated as a quantum field theory.
These include gauge theory of spin-2 gravitons, expansion of general relativity around flat space-time and Feynman rules of gravity.
Gravitational interactions between particles are carried by gravitons.
The gauge symmetry of the gravitons is the quantum field theoretic manifestation of general covariance from classical general relativity and in the long-range, classical limit gravitons can be interpreted like other quantum fields in the framework of special relativity.
\par
The use of covariant gauge introduced the arbitrary covariant gauge parameter $\xi$ which clearly shows how different quantities such as the graviton propagator or the three-point vertex function depend on the quantum gauge-fixing procedure.
Also, the role of the gauge-fixing function, $G_\sigma$, was analyzed as well as the freedom in choosing this function.
Here, we introduced a novel gauge-fixing function which combines harmonic and \dDo\ gauge in an entire family of gauge choices depending on the gauge parameter $\alpha$.
Our results were thus very general depending on the two gauge parameters $\xi$ and $\alpha$ as well as the arbitrary space-time dimension, $D$.
\par
The expansion of objects from general relativity such as the \EHA\ action around flat space-time requires manipulation of numerous tensors with several indices.
We discussed two distinct expansions, namely in the graviton field and in the gravitational constant, and showed how to relate them to each other.
An important result is the expansion of the \EHA\ action in $h_\mn$ in terms of the tensors $\Gpz^\mn$ and $\Hpz^\mn$ related to the Einstein tensor and the analogous gauge-breaking tensor $H^\mn$ respectively.
This made it possible to relate the n-graviton vertex rule to the Einstein tensor and $H^\mn$.
\par
Using the results of our expansions in $h_\mn$ we were able to derive Feynman rules for quantum gravity.
Here, we were in particular interested in the gravitational part of the action from which the graviton propagator and n-graviton self-interaction vertices where derived.
The graviton propagator in covariant \dDo\ gauge as well as the n-graviton vertices in terms of the Einstein tensor and $H^\mn$ are novel results, that we have not found in earlier literature.
In addition we expanded the gravitational action in detail to third order in $h_\mn$ which allowed us to derive an explicit result for the three-graviton vertex.
\par
Using the Feynman rules for the n-graviton vertex we were able to relate the exact three-point vertex function of a massive scalar interacting with a graviton to the \STM\ metric.
This was done by comparing the Feynman diagram expansion to a perturbative expansion of the classical equations of motion derived from the gauge-fixed action by $\delta S=0$.
Here, it was assumed that the n-loop triangle integrals can be reduced to certain convolution integrals in the classical limit which has been shown explicitly only in $D=4$~\cite{Bjerrum-Bohr:2018xdl,Galusha:cand}.
The metric thus derived is independent of $\xi$ and satisfies the gauge/coordinate condition $G_\sigma=0$.
It is an exciting result that the \STM\ metric can be derived from the Lorentz covariant three-point amplitude to all orders in $G_N$ which deserves further analysis in the future.
\par
Using the formulas relating the three-point amplitude and the \STM\ metric to each other, we computed the one-loop contribution to the metric with the generalized \dDo-type gauge function depending on the arbitrary parameter $\alpha$.
This, in particular, required use of the three-graviton vertex and evaluation of triangle one-loop integrals in arbitrary dimension.
For specific values of the parameters $\alpha$ and $D$ we were able to compare the results with the literature.
This was the case for $\alpha=0$ and arbitrary $D$ where we found agreement with \cite{Collado:2018isu} and for $D=4$ and $\alpha=1$ where we found agreement with the standard result for the Schwarzschild metric in harmonic coordinates.
\par
The introduction of generalized Lorentz covariant gauge-fixing functions is an interesting consequence of the quantum field theoretic approach to general relativity and it would be exciting to examine if it is possible to derive exact, all-order results with such coordinate-conditions in classical general relativity.
Here we think of the \dDo\ gauge condition from Eq.~\eqreft{vv42} or the generalized gauge-function from our work Eq.~\eqreft{gau3}.
In this work, we used methods from classical general relativity to compute the \STM\ metric perturbatively in the \dDo-type gauge depending on $\alpha$ to second order in $G_N$.
The results of this classical computation agreed with the amplitude method.
\par
In $D=5$, at second order in $G_N$, a logarithmic dependence on the radial coordinate appeared in the metric.
In the amplitude computation it was necessary to carefully remove divergent terms from the triangle integrals in order to arrive at the correct metric in $D=5$.
The appearance of logarithms was analyzed in terms of redundant gauge freedom.
From this analysis it is expected that besides the case of $D=5$ at second order, logarithms will appear only in $D=4$ at third order which explains the logarithms in Ref.~\cite{Goldberger:2004jt} which appear exactly at this order in $D=4$.
\par
The reduction of n-loop triangle integrals in the classical limit is still to be done in detail.
In this thesis, we computed the one-loop integrals but only after going through several calculations, we were able to show that just the space-components of these integrals contribute which are exactly the part that has a simple interpretation in the classical limit.
Further investigations would be to include spinning and charged particles which would produce the Kerr-Newman metric.
Another direction would be to study diagrams with two distinct massive scalars (i.e. four external scalar lines) and an external graviton.
Such diagrams would describe binary systems and possibly a metric for the two-body system could be defined.
A third possibility would be to analyze the $\hbar$ correction to the \STM\ metric as was done already in Ref.~\cite{BjerrumBohr:2002ks}.
Is it possible to derive this correction in a sensible way to all orders in $G_N$?
In general, the study of general relativity from quantum field theory promises to provide many exciting results in the future.

\blankpage
\chapter*{Acknowledgments}
I would like to thank my advisors Poul Henrik Damgaard and Emil Bjerrum-Bohr for many helpful discussions during the work on this thesis.
\blankpage

\addcontentsline{toc}{chapter}{Bibliography}
\bibliographystyle{unsrt}
\bibliography{refs}

\end{document}